\documentclass[prl,aps,twocolumn,showpacs]{revtex4}

\usepackage{ifpdf}
\ifpdf
\usepackage{subfigure}
\usepackage[pdftex]{graphicx}
\else
\usepackage{graphicx}
\fi

\usepackage{setspace}
\usepackage{psfrag}
\usepackage{epstopdf}
\usepackage{graphicx}
\DeclareGraphicsExtensions{.eps}

\usepackage{amsmath,graphicx,epsfig,bm,color}
\usepackage{amssymb,amsfonts}
\usepackage{rotating}
\usepackage{graphics}
\usepackage{setspace}

\begin{document}
	
\title{\bf Phase diagrams and Thomas-Fermi estimates for spin-orbit coupled Bose-Einstein Condensates under rotation}
\author{Amandine Aftalion${}^{1}$ \& Peter Mason${}^{2}$}
\affiliation {${}^1$CNRS \& Universit\'e Versailles-Saint-Quentin-en-Yvelines,
Laboratoire de Math\'ematiques de Versailles, CNRS UMR 8100, 45
avenue des \'Etats-Unis, 78035 Versailles C\'edex, France \\
${}^2$Joint Quantum Centre (JQC) Durham-Newcastle, Department of Physics, Durham University, Durham DH1 3LE, United Kingdom.}
\date{\today}
\begin{abstract}
We provide complete phase diagrams describing the ground state of a trapped spinor BEC under the combined effects of rotation and a Rashba
 spin-orbit coupling. The interplay between the different parameters (magnitude of rotation, strength of the spin-orbit coupling and interaction) leads to a rich ground state physics that we classify. We explain some features analytically in the Thomas-Fermi approximation, writing the problem in terms of the total density, total phase and spin. In particular, we analyze the giant skyrmion, and find
 that it is of degree 1 in the strong segregation case. In some regions of the phase
  diagrams, we relate the patterns to a ferromagnetic energy.
\end{abstract}
\maketitle

\section{I. Introduction}

Bose Einstein condensates (BEC's) provide a unique experimental and theoretical testing ground for many
macroscopic quantum phenomena. One such area that has recently attracted a lot of attention is the engineering of synthetic non-Abelian gauge potentials coupled to neutral atoms \cite{jacob,juz,campbell,DGJO} to create a spin-orbit coupled Bose-Einstein condensate \cite{rashba1,rashba2}, where the internal spin states and the orbital momentum of the atoms are coupled. These spin-orbit coupled condensates support a variety of ground state density profiles; for instance in the most straightforward case of a spin-$1/2$ condensate \cite{zhai,OB,galitski,WuMZ,HZ,HLL,yip,ZMZ,HRPL,OB2}, the density either displays a plane wave or a striped wave. The transition between the two depends on the interaction parameters. However, if one is to consider a spin-$1$ or spin-$2$ condensate then more exotic ground state profiles, based on the helical modulation of the order parameter, can be created \cite{KMM,XKYU,spin-2,CZ}.

In addition to the various ground state density profiles, one can look to the basic elementary excitations created in these spin-orbit condensates, like the vortex \cite{ram,JZ,subh}, dark soliton \cite{fialko}, or bright soliton \cite{xu} in spin-$1/2$ condensates, or the skyrmion in spin-$1$ and spin-$2$ condensates \cite{su,KMNM,RHM,LL}. In contrast to a single component or two-component condensate, where the appearance and energetic stability of the elementary excitations are dependent on a rotation of the system to impart angular momentum, spin-orbit coupled condensates naturally impart momentum through the coupling of the internal spin and orbital momentum of the atoms. But when combining both  the spin-orbit coupling and the rotation, various novel features have been predicted to occur \cite{LL,LFZWL,zzwu,xuhan,radic}. Through a suitable control of the condensate, an experimental scheme for rotating spin-orbit coupled condensates has been proposed in \cite{radic}.

 The aim of this paper is to study the combined effect of a Rashba spin-orbit coupling and a rotation on spinor BEC's for spin-$1/2$ condensates. The interplay between trap energy, spin-orbit coupling and interaction leads to a rich ground state physics: stripe phases, half vortices or vortex lattices with some behaviours reminiscent of vortex lattices appearing for fast rotating condensates \cite{cornell,MHo}. We provide a complete phase diagram according to the magnitude of rotation, spin-orbit coupling and interaction. Some features have been analyzed by Subhasis et al. \cite{subh}, Zhou et al. \cite{zzwu} and Xu \& Han \cite{xuhan}, but here we want to investigate a full phase diagram behaviour.

Our paper is organised as follows. In Section II we introduce the energy functional in terms of individual wave functions before making the transformation to the non-linear Sigma model where the energy is instead written in terms of the total density and a spin density. In Section III we provide numerically determined phase space diagrams for the ground states of the condensate as functions of the rotation, spin-orbit coupling and interaction. We explain some features analytically by using a Thomas-Fermi approximation in Section IV.


\section{II. Problem Statement and Energy Functional}

We are interested in a two-dimensional $(x,y)$ rotating spin-coupled Bose-Einstein condensate. This has the following non-dimensional energy functional in terms of the wave functions $\psi_1$ and $\psi_2$:
\begin{equation}
	\label{er}
	\begin{split}
	E&=\int\sum_{k=1,2}\Bigg(\frac{1}{2}|\nabla\psi_k|^2+\frac{1}{2}r^2|\psi_k|^2
-{\Omega}\psi^*_k L_z \psi_k\\
&\qquad+\frac{g_k}{2}|\psi_k|^4-\kappa\psi_k^*\left[i\frac{\partial\psi_{3-k}}{ \partial x}+(-1)^{3-k}\frac{\partial\psi_{3-k}}{\partial y}\right]\Bigg)\\
&\qquad+g_{12}|\psi_1|^2|\psi_2|^2\qquad d^2{r},
\end{split}
\end{equation}
under the constraint that $\int |\psi_1|^2+|\psi_2|^2=N$, $N$ being the number of atoms. Here, $g_k$ is the self interaction of each component (intracomponent coupling) that we will later take to be equal for both components, $g_{12}$ measures the effect of interaction between the two components (intercomponent coupling) and $\Omega$ is the rotational velocity, applied equally to both components, with $L_z=-i(x\partial_y-y\partial_x)$ the angular momentum operator acting in the $z$ direction. We consider a Rashba spin-orbit interaction strength, $\kappa$, being equal in both the $x$ and $y$ direction.

One of the key ingredients in the analysis will be to use the nonlinear Sigma
 model introduced for two component condensates in the absence of a spin-orbit coupling \cite{ueda,am2,ktu}; that is to write the energy in terms of the total density $\rho$,
\begin{equation}\label{rhot}
    \rho=|\psi_1|^2+|\psi_2|^2,
\end{equation}
and a normalised complex-valued spinor
$\bm{\chi}=[{\chi_1},{\chi_2}]^T$: the
wave functions can be decomposed as $\psi_1=\sqrt{\rho}\chi_1$,
$\psi_2=\sqrt{\rho}\chi_2$ where $|\chi_1|^2+|\chi_2|^2=1$.
 We define the spin density $\bm{S}=\bar{\bm{\chi}}\bm{\sigma}\bm{\chi}$, where
$\bm{\sigma}=(\sigma_x,\sigma_y,\sigma_z)$ are the Pauli matrices, with the components of $\bm{S}$ following as
\begin{subequations}
    \label{seqs}
\begin{align}
    \label{sx}
    S_x=&\chi^*_1\chi_2+\chi_2^*\chi_1,\\
    \label{sy}
    S_y=&-i(\chi^*_1\chi_2-\chi_2^*\chi_1),\\
    S_z=&|\chi_1|^2-|\chi_2|^2,
    \label{sz}
\end{align}
\end{subequations}
such that $|\bm{S}|^2 = 1$ everywhere. For a rotating condensate, it is natural to introduce \begin{equation}
    \bm{v}_{\text{eff}}=\frac{\nabla\Theta}{2}+\frac{\bm{R}S_z}{2(1-S_z^2)}=\frac 12 S_z \nabla (\Theta_1 -\Theta_2),
\end{equation} where $\Theta=\Theta_1+\Theta_2$, $\Theta_k$ is the phase of $\psi_k$, that is $\psi_k=\sqrt \rho |\chi_k|e^{i\Theta_k}$ and $\bm{R}=S_y\nabla S_x-S_x\nabla
S_y$. This allows us to rewrite the energy functional (\ref{er}) as
\begin{equation}
		\label{ener2}
\begin{split}
 E=&\int\frac{1}{2}\left(\nabla\sqrt{\rho}\right)^2+\frac{\rho}{8}\left(\nabla\bm{S}\right)^2+\frac{\rho}{2}\left(\bm{v}_{\text{eff}}-\bm{\Omega}\times\bm{r}\right)^2\\
		&+\rho\kappa\left(\bm{S_{\perp}}\cdot \bm{v}_{\text{eff}}+\frac{1}2 \bm{S}\cdot \nabla \times \bm{S} \right)\\
		&+\frac{\rho}{2}(1-\Omega^2)r^2+(c_0+c_1S_z+c_2S_z^2)\frac{\rho^2}{2}\qquad d^2{r},
\end{split}
\end{equation}
where $\bm{S_{\perp}}=(S_x,S_y)$ and
\begin{subequations}
\begin{align}
\label{c00}
{c}_0=&\frac{1}{4}(g_1+g_2+2 g_{12}),\\
    {c}_1=&\frac{1}{2}(g_1-g_2),\\
\label{c22}
    {c}_2=&\frac{1}{4}(g_1+g_2-2 g_{12}).
\end{align}
\end{subequations}  A derivation to this form of the energy from (\ref{er}) and to other forms is given in the Appendix.

Note that our choice of the effective spin-orbit Hamiltonian [Eq. (\ref{er})] is assumed to remain stationary in the rotating frame. This is in contrast to the experimental schemes proposed in \cite{radic}, where  there is a time dependence inherent in the Hamiltonian. We justify our assumption and use of a time-independent Hamiltonian on two fronts: firstly the probable small effect that the time dependent terms will have on the ground state (we note in particular Fig. 1(c) of Ref. \cite{radic} in which a regular vortex lattice is present in both components for a large spin-coupling and a relatively large rotation); secondly, the need to perform meaningful analytical analysis on the ground state profiles requires a `from principles' approach whereby only the fundamental terms of the Hamiltonian are considered, that is the spin-orbit coupling, the rotation and the interaction, as written in the Hamiltonian (\ref{er}). Furthermore to this last point, the experimental infrastructure to create a rotating spin-orbit condensate is relatively new, and there remains the possibility that a new experimental scheme that fully justifies the use of a time-independent Hamiltonian could be proposed. To this end, we believe that our phase diagrams provide interesting and relevant information on the ground states of the rotating spin-orbit coupled condensates.


\section{III. Description of the phase diagrams}

We wish to describe the ground state wave functions of the spin-orbit condensate. In what follows, we assume
  $g_1=g_2\equiv g$ and set $\delta=g_{12}/g$, which measures the effect of interaction between the two components. {{ The experiments of \cite{rashba1,rashba2} have $gN$ large (the Thomas-Fermi limit)}}. Therefore,  our analysis will also be in the case $gN$ large and our system is then described by three parameters: $\Omega$, the rotational velocity; $\kappa$, the spin-orbit interaction strength; and $\delta$. 

We first present numerically obtained phase diagrams for these three parameters, with a Thomas-Fermi analysis following in the next section. These simulations are conducted on the coupled Gross-Pitaevskii equations that result from the energy functional (\ref{er}) through the variation $i\partial\psi_k/\partial t=\delta E/\delta \psi^*_k$ for $k=1,2$:
\begin{equation}
\label{gp}
\begin{split}
i\frac{\partial\psi_k}{\partial t}=&-\frac{1}{2}\nabla^2\psi_k+\frac{1}{2}r^2\psi_k-i\Omega\left(y\frac{\partial\psi_k}{\partial x}-x\frac{\partial\psi_k}{\partial y}\right)\\
&\quad+g|\psi_k|^2\psi_k+g_{12}|\psi_{3-k}|^2\psi_k\\
&\qquad-\kappa\left(i\frac{\partial\psi_{3-k}}{\partial x}+(-1)^{3-k}\frac{\partial\psi_{3-k}}{\partial y}\right).
\end{split}
\end{equation}	
We simulate in imaginary time using the following values of parameters: $g=4$ and $N=200$, together with $\Omega\in[0,1)$, $\kappa\ge0$ and $\delta\ge0$. These parameters place us in the Thomas-Fermi regime. For each parameter set, we classify the ground state according to the densities, $|\psi_k|^2$, and the spin densities, $\bm{S}$. In general, it is difficult to find the true minimizing energy state. But the use of various initial data converging to the same (or similar) final state allows us to determine that the true ground state will be of the same pattern as the one that we exhibit. We break our analysis into three sections; $\Omega=0$, $\Omega$ small and $\Omega$ large.


\subsection*{A. $\Omega=0$}

We begin by considering the non-rotating spin-orbit condensate in which the active parameters are $\kappa$ and $\delta$. In the case when $\kappa=0$, we are  left with a two-component condensate coupled exclusively by the intercomponent interaction strength related by $\delta$. In this case, there are never any topological defects created in the condensate and the ground state density profiles of the condensates are either, for $\delta<0.99$, two co-existing disks (of equal radii), or for $\delta >0.99$, one of the components is a disk while the other is identically zero \cite{ktu}.

\begin{figure}
\begin{center}
\includegraphics[scale=0.3]{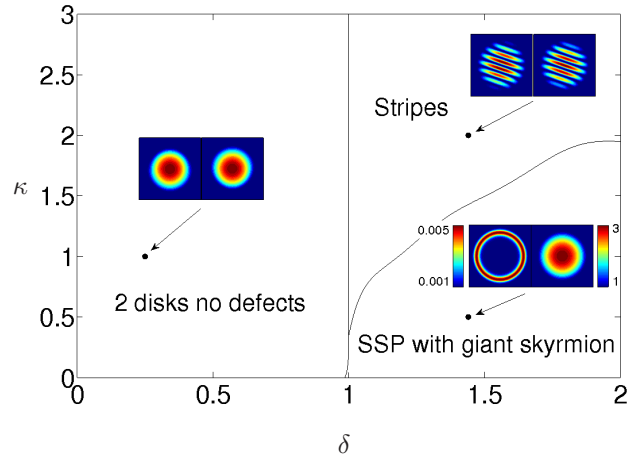}
\end{center}
\begin{picture}(0,0)(10,10)
\put(20,22) {{$\delta$}}
\put(-103,120) {{$\kappa$}}
\end{picture}
\caption{(Color online) $\kappa-\delta$ phase diagram with $\Omega=0$. The numerical parameters are taken as $g=4$ and $N=200$. There are three identified regions: (i) two disks with no defects, (iii) segregated symmetry preserving (SSP) with a giant skyrmion and (iv) stripes. Each region has a typical density plot for each component (left panels, component-1 and right panels, component-2). The numerical values of these simulations correspond to: (i) $(\delta,\kappa)=(0.25,1)$; (iii) $(1.44,0.5)$ and (iv) $(1.44,2)$.}
\label{summ}
\end{figure}

Turning on the spin-orbit coupling term so that $\kappa\neq0$ provides a system which has recently been considered in the literature by a number of authors \cite{galitski,zhai,spin-2,HRPL,KMM,subh,ram,WuMZ,JZ}. A typical example of the phase diagram is shown in Fig. \ref{summ} together with the associated density plots.

When $\delta<1$, the two-components remain co-existing and disk-shaped for all $\kappa$ (Fig. \ref{summ}(i)). We never see  any topological defects in the density profiles of the coexisting disk shaped condensates.
 We can check (as in Fig. \ref{phase_00}) that $S_\perp=(S_x,S_y)$ is almost constant, $S_z$ is almost zero,  and $(\partial\Theta/\partial x, \partial\Theta/\partial y)\sim
  -2\kappa S_\perp$.

  {{ On the other hand, if $\delta\ge1$, the components segregate. For small $\kappa$ (Fig. \ref{summ}(iii)), then one component is a disk, in which most of the particles reside and is surrounded by a thin, low populated annulus for the other component.
The  circulation is $2\pi$  in this annulus which is reminiscent of
the skyrmion computed in \cite{HRPL,ram}. Nevertheless, these authors consider small values of interaction, which leads
to a single Landau level which is populated, and thus a circulation of 1. Here, we fix a large interaction, which leads to a different regime, but find the same type of skyrmion. We will analyze this later in the Thomas-Fermi limit.

As $\kappa$ is increased, the maximum density in the annulus increases, as well as the number of rings (see Fig. \ref{prof2}(a)). We have checked numerically that the circulation is $2\pi$ in each annulus of component 1, as soon as $\delta$ is sufficiently large (leading to segregation of the components). At a critical $\kappa$ (approximately equal to $1.5$ at $\delta=1.5$), symmetry breaking occurs and the ground state becomes a stripe profile as in Fig. \ref{prof2}(b): these stripe density profiles  were studied in \cite{zhai,subh}. The stripes are straight and segregation of
 the components is observed for large $\delta$.

\begin{figure}
\begin{center}
\includegraphics[scale=0.25]{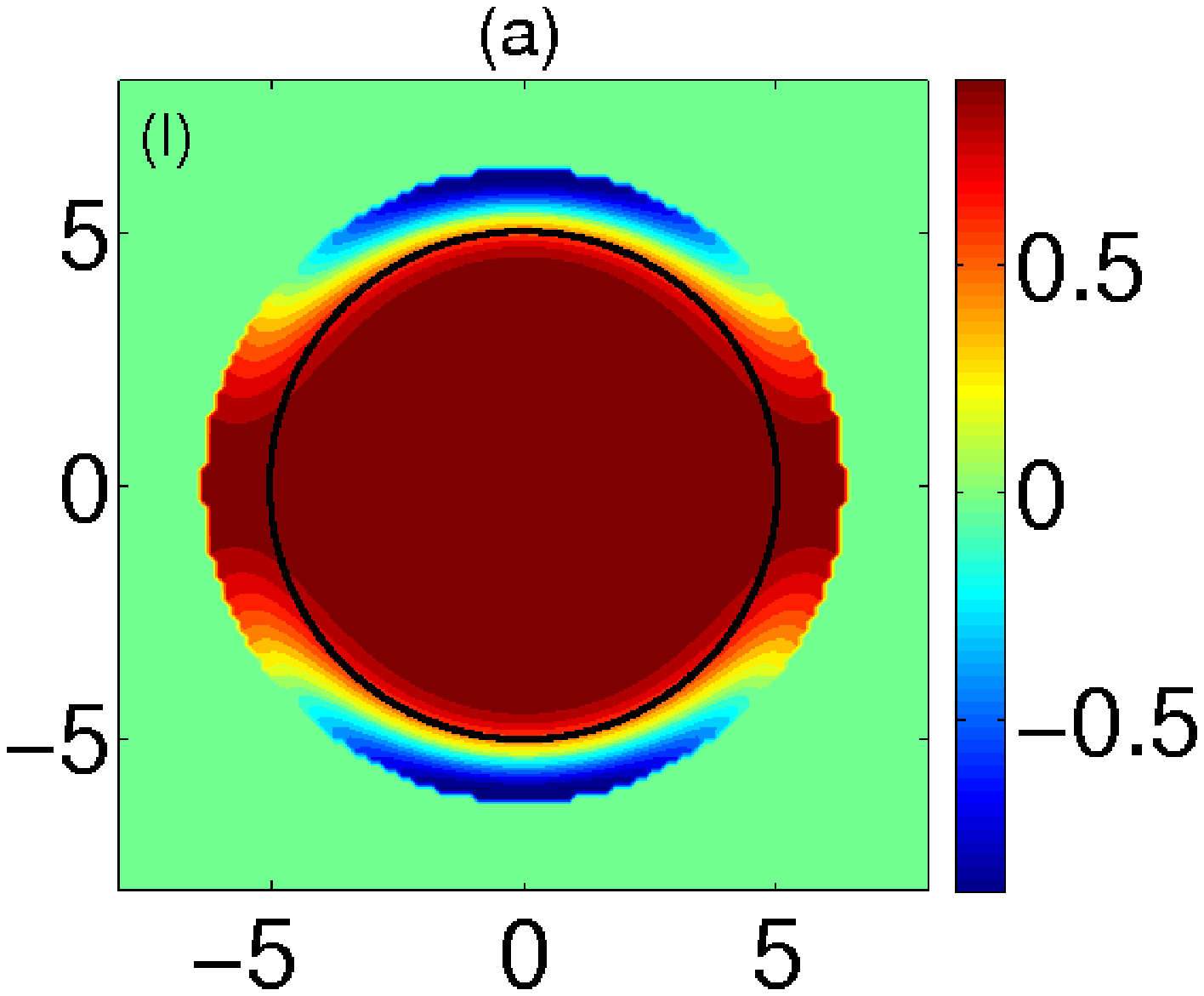}
\includegraphics[scale=0.25]{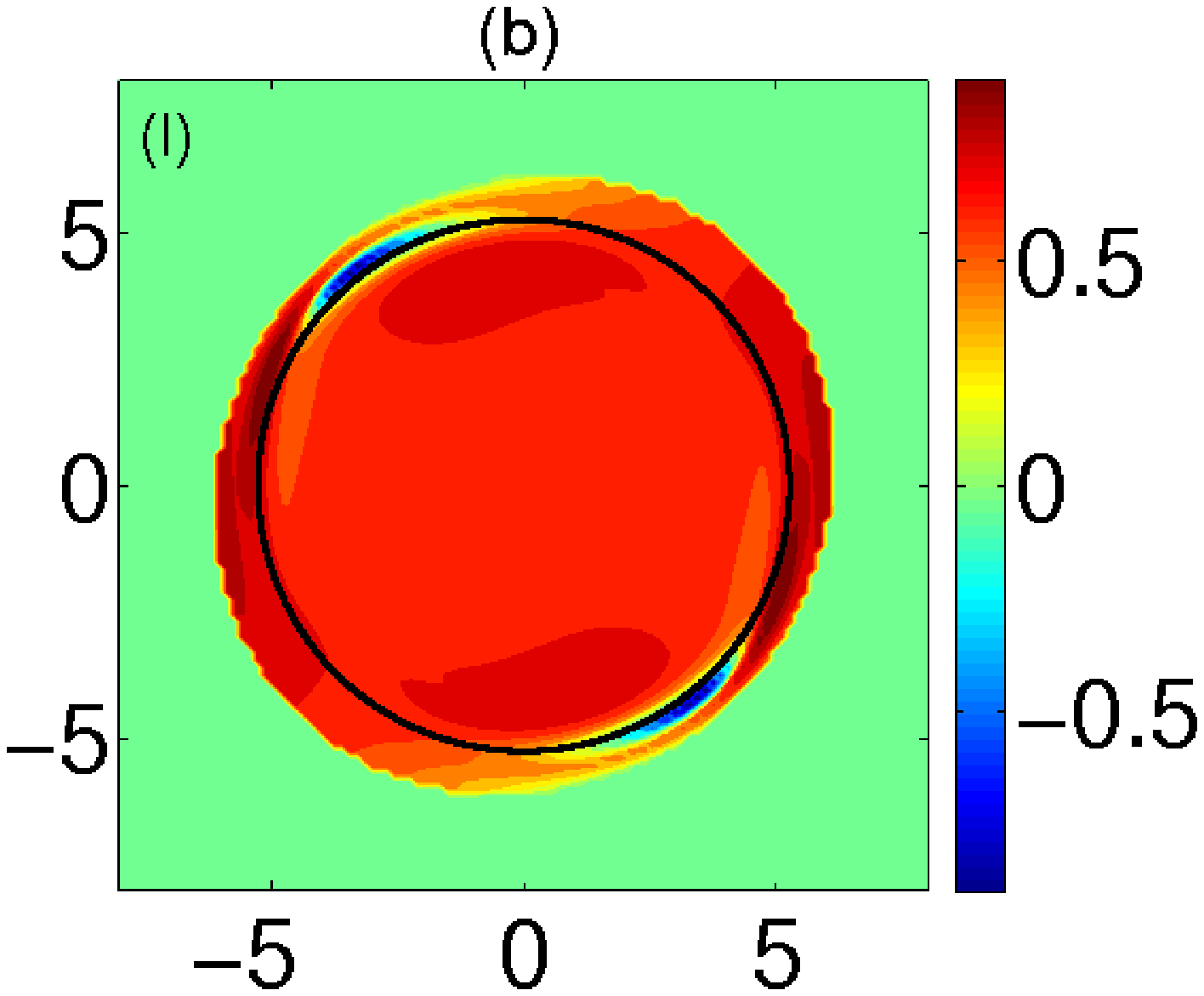}\\
\includegraphics[scale=0.25]{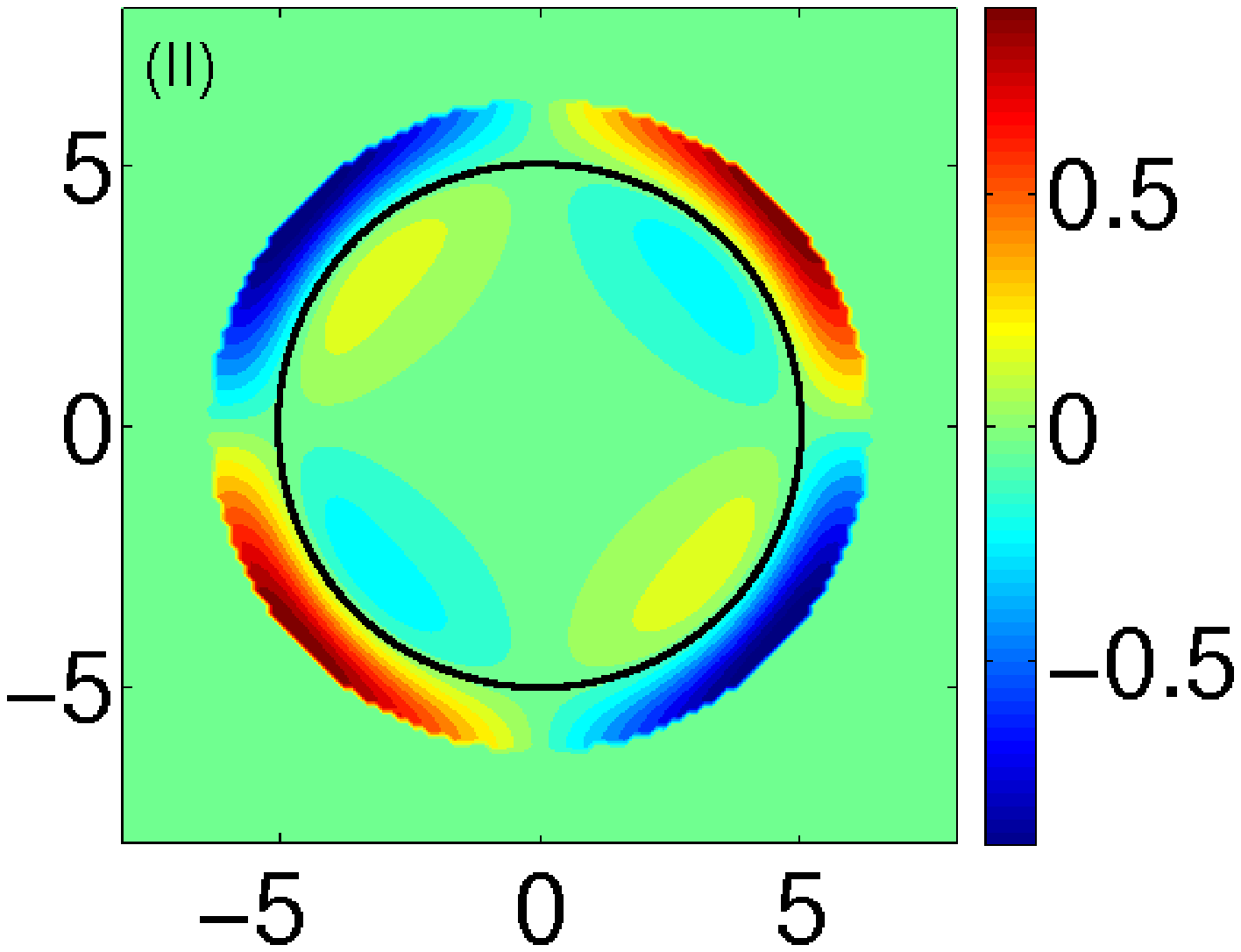}
\includegraphics[scale=0.25]{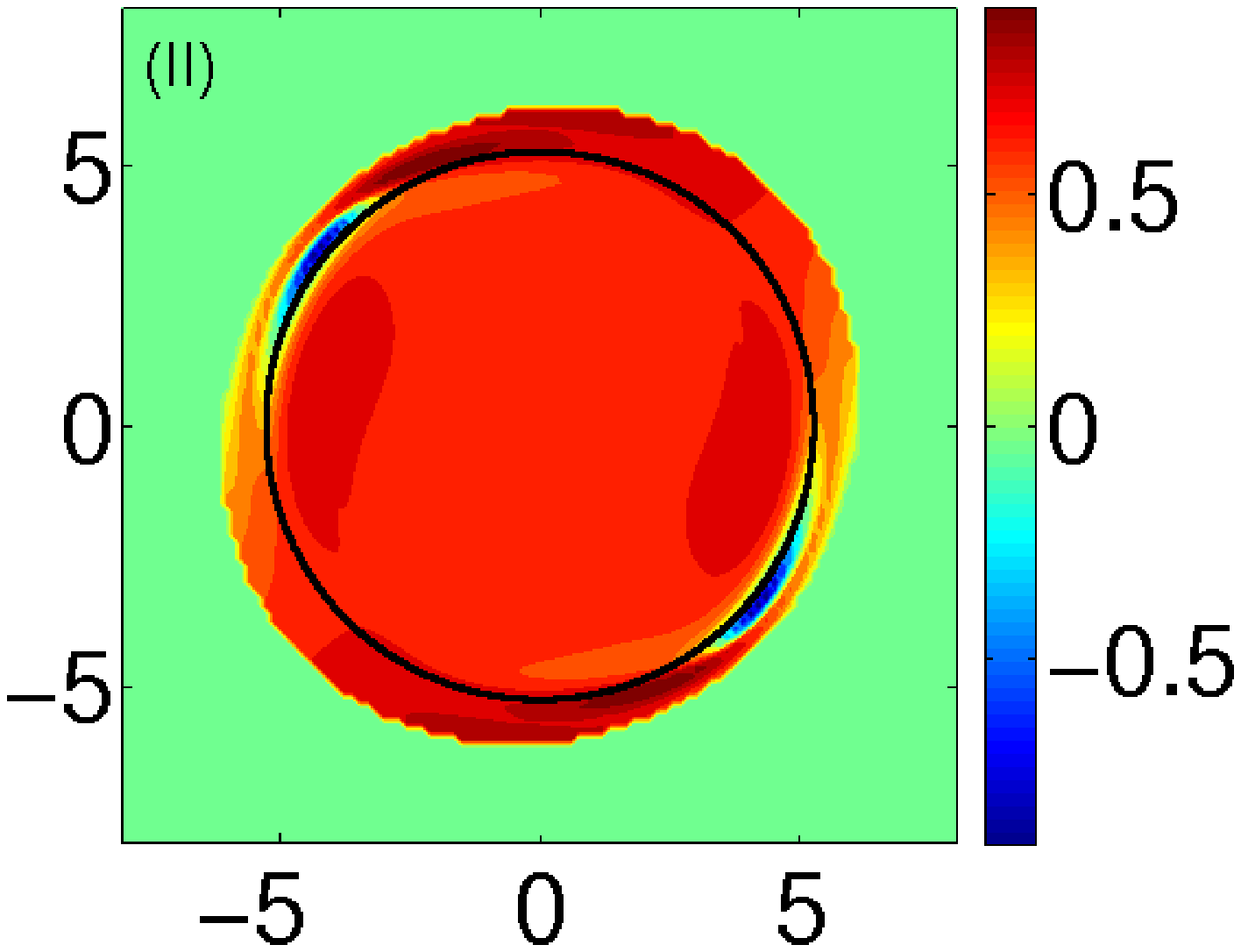}\\
\includegraphics[scale=0.25]{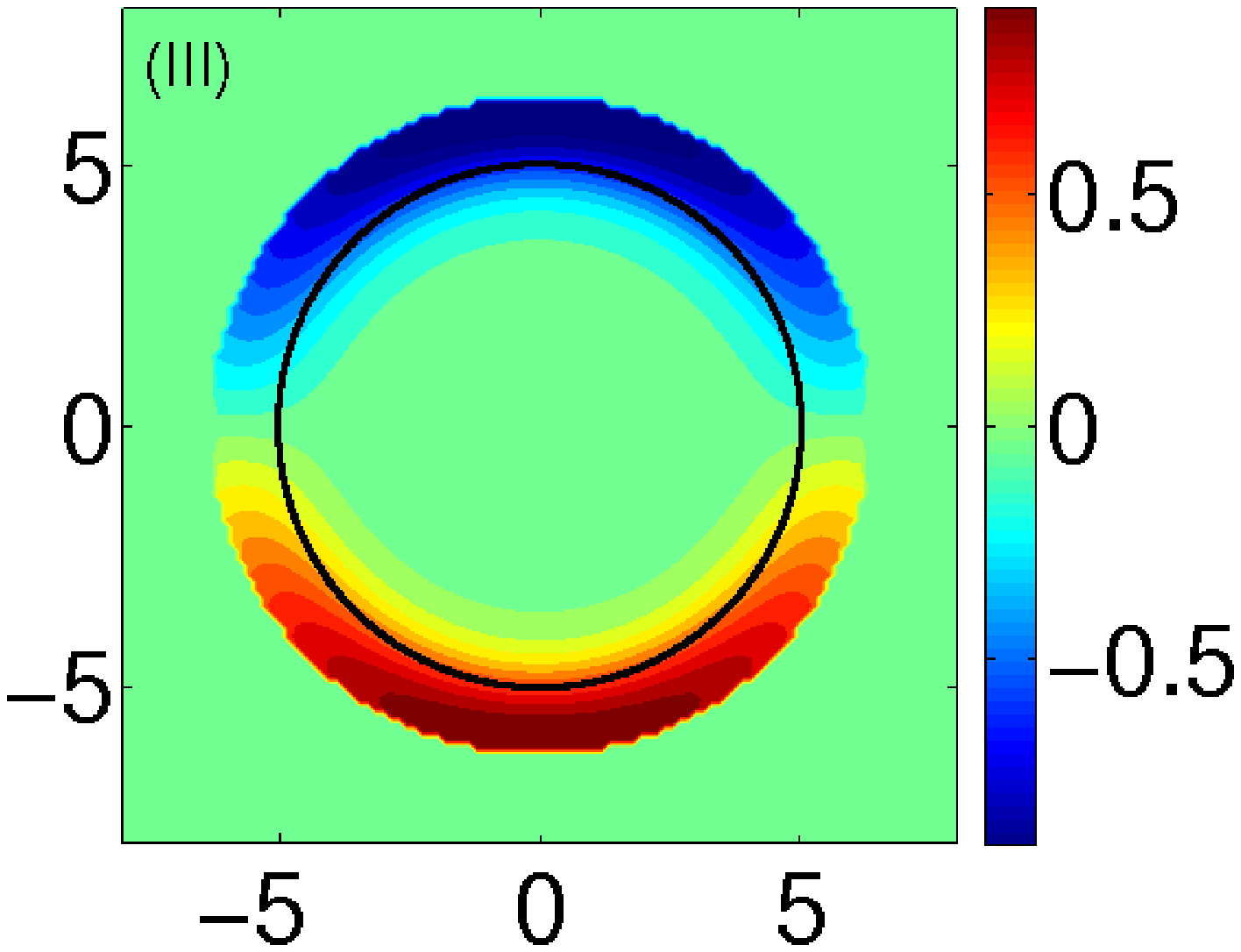}
\includegraphics[scale=0.25]{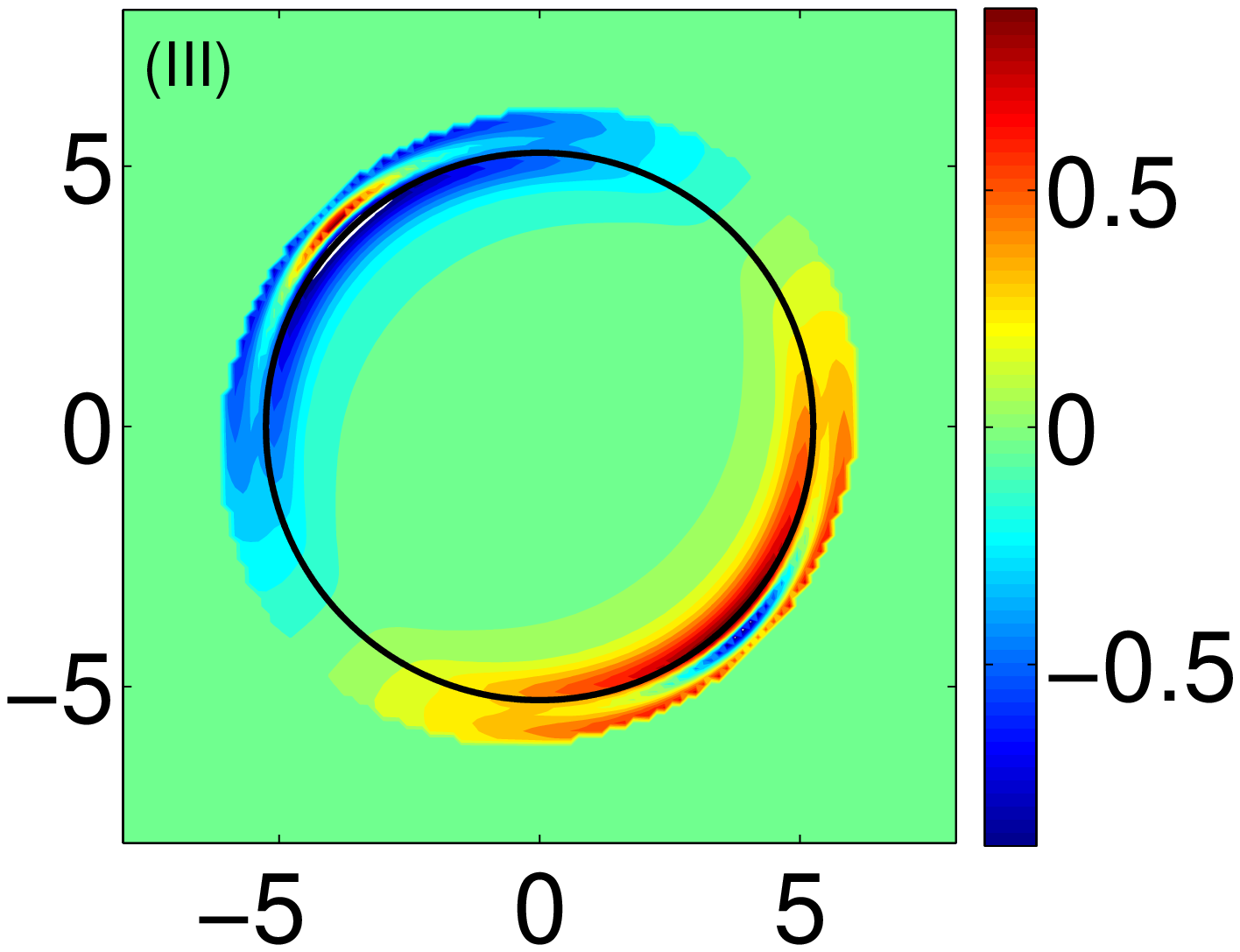}
\end{center}
\begin{picture}(0,0)(10,10)
\put(-53,23) {{$x$}}
\put(63,23) {{$x$}}
\put(-102,245) {{$y$}}
\put(-102,159) {{$y$}}
\put(-102,72) {{$y$}}
\end{picture}
\caption{(Color online) Spin density plots (frame (I) $S_x$, (II) $S_y$ and (III) $S_z$) for numerical simulations carried out when $\Omega=0$ for (a) $(\delta,\kappa)=(0.25,0.5)$ (left column) and (b) $(\delta,\kappa)=(0.25,4.75)$ (right column). The Thomas-Fermi radius is plotted (black circle), calculated in Eq. (\ref{radius_0}).}
\label{phase_00}
\end{figure}

\begin{figure}
\begin{center}
\includegraphics[scale=0.25]{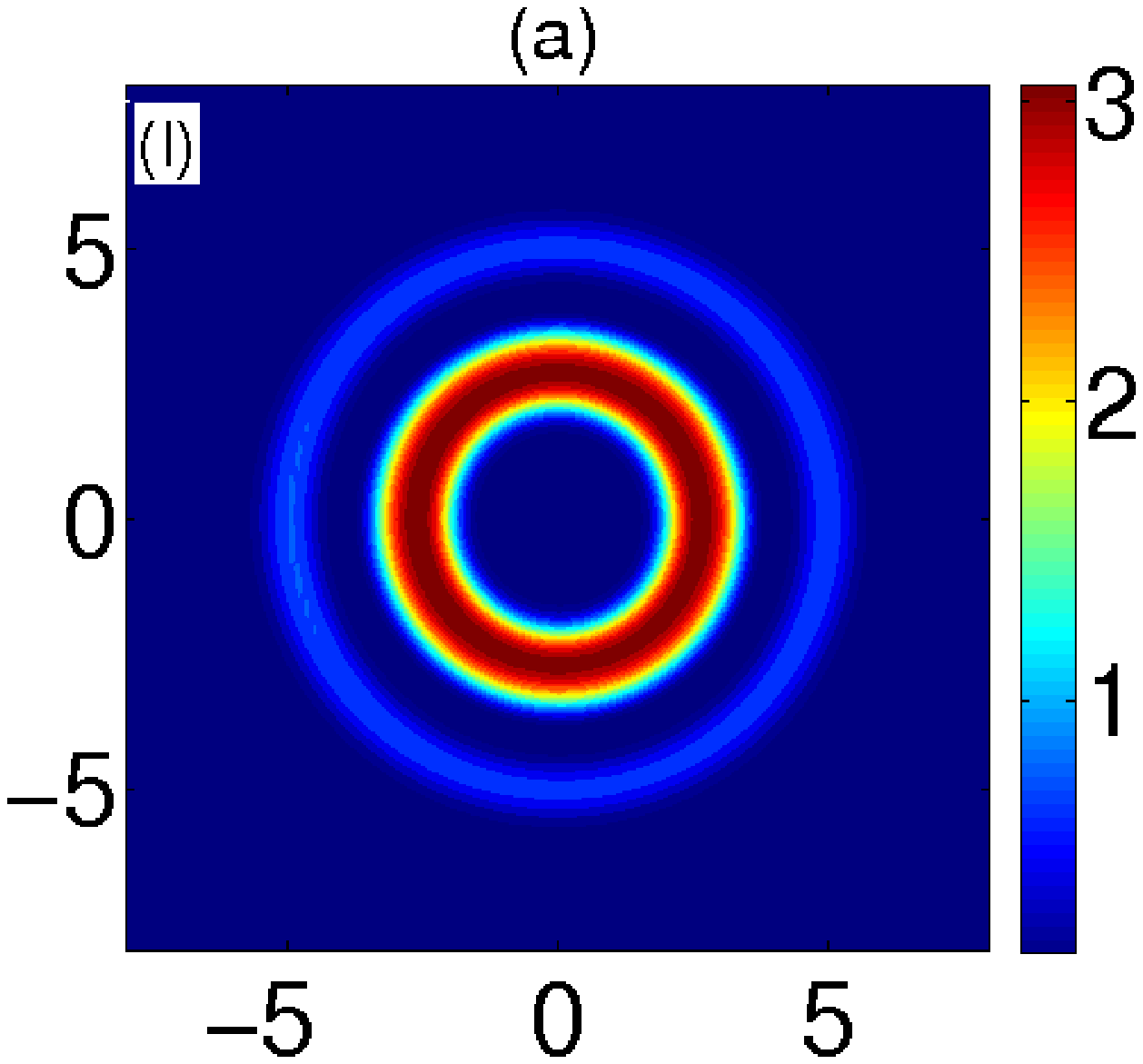}
\includegraphics[scale=0.25]{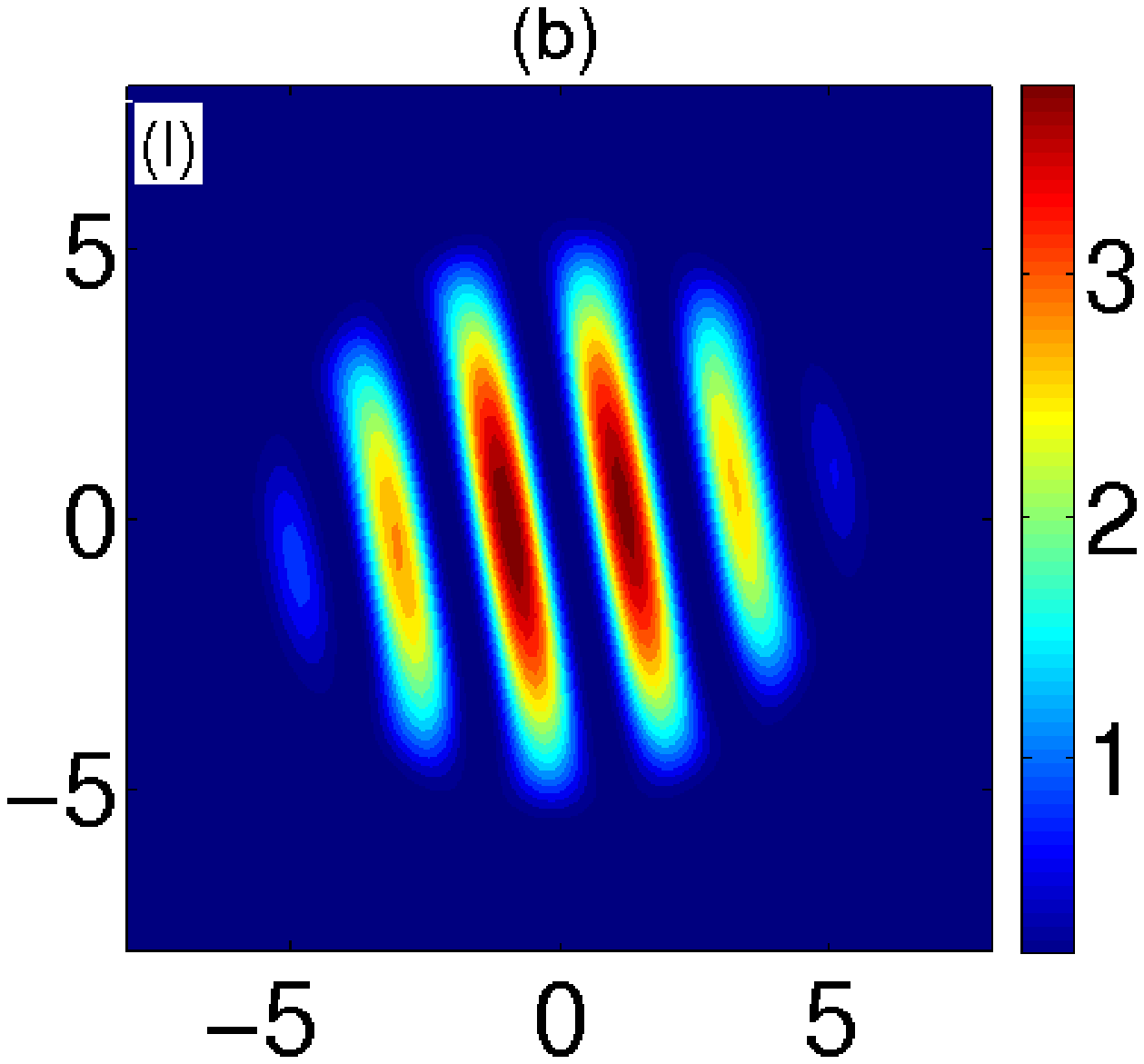}\\
\includegraphics[scale=0.25]{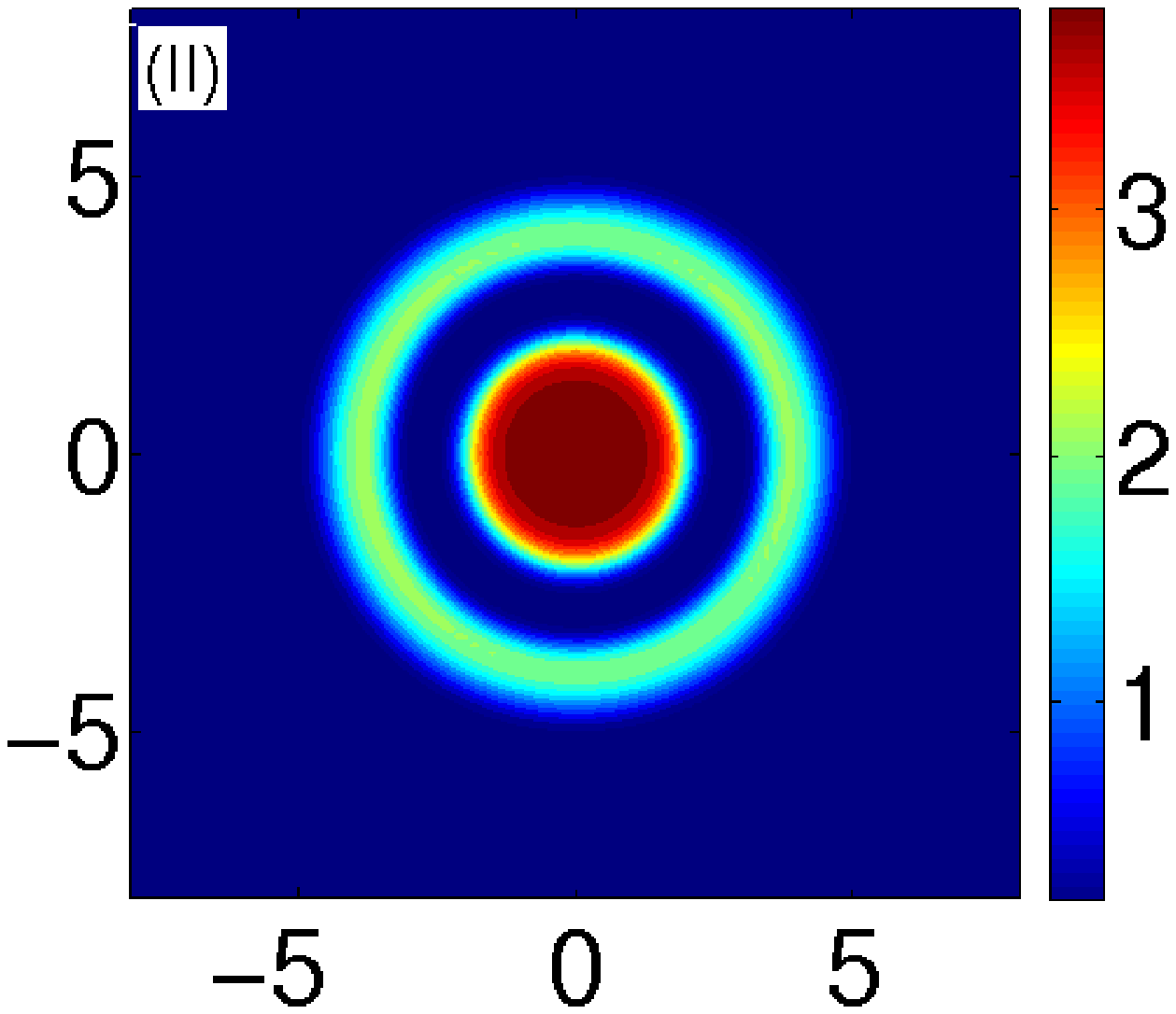}
\includegraphics[scale=0.25]{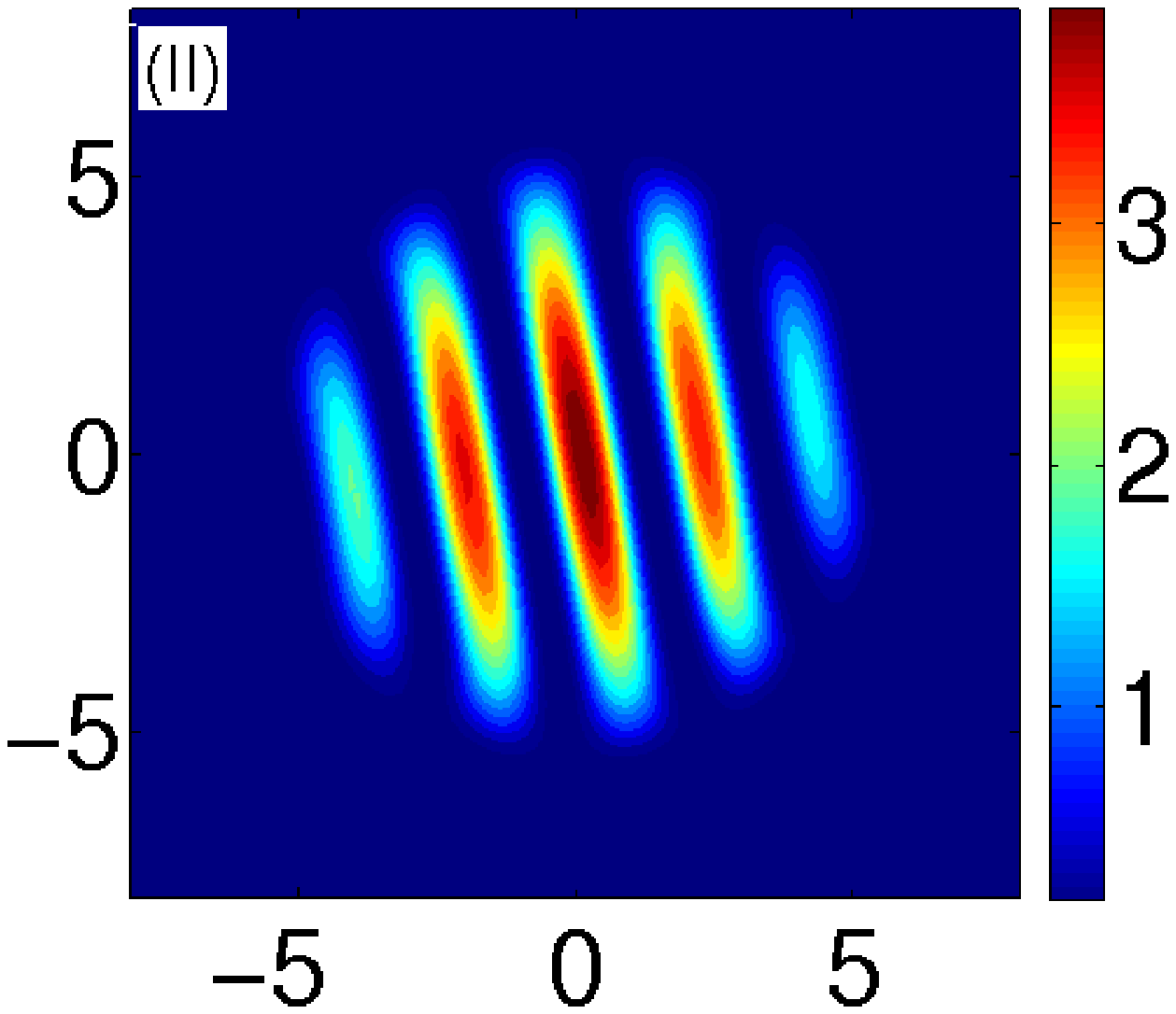}\\
\includegraphics[scale=0.25]{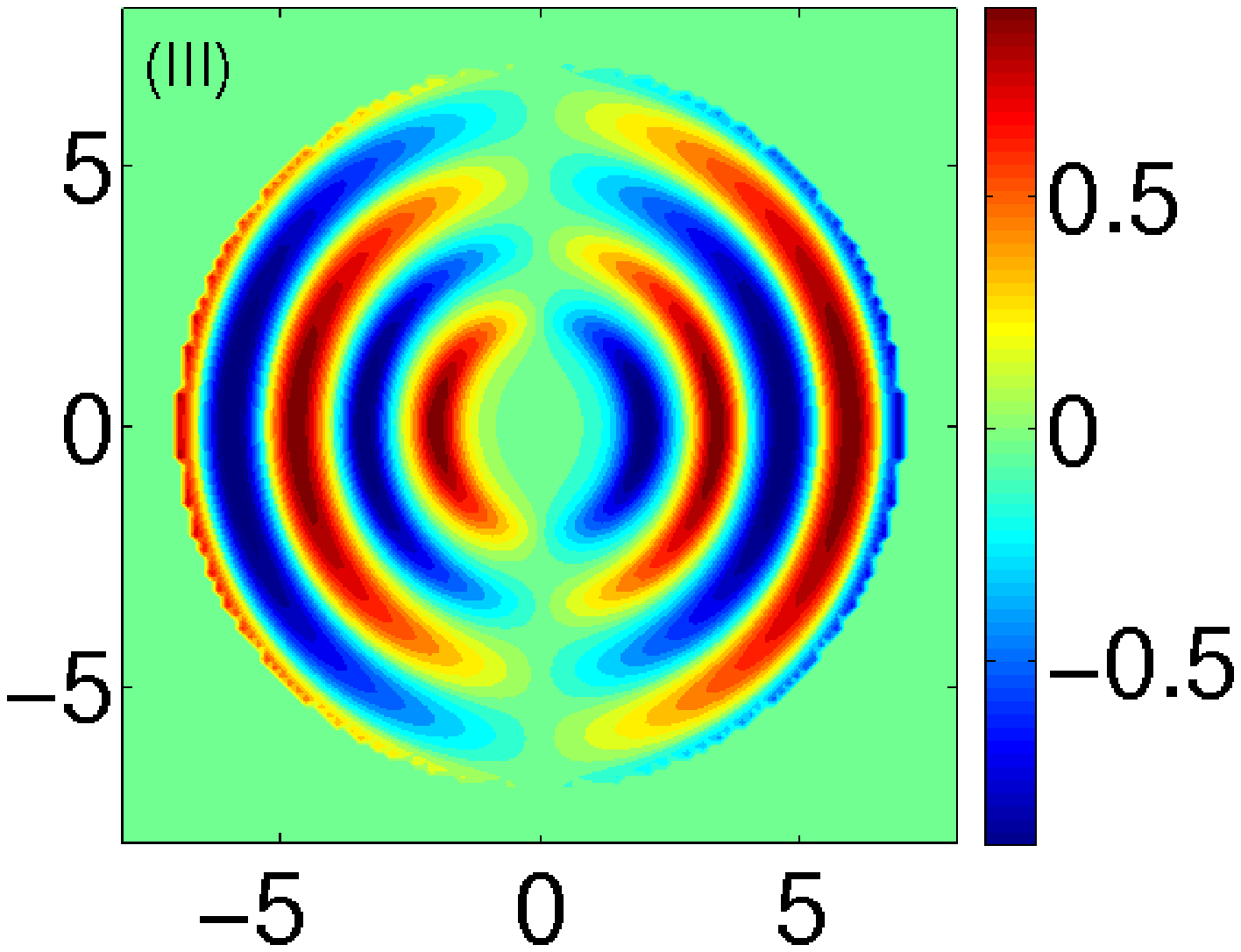}
\includegraphics[scale=0.25]{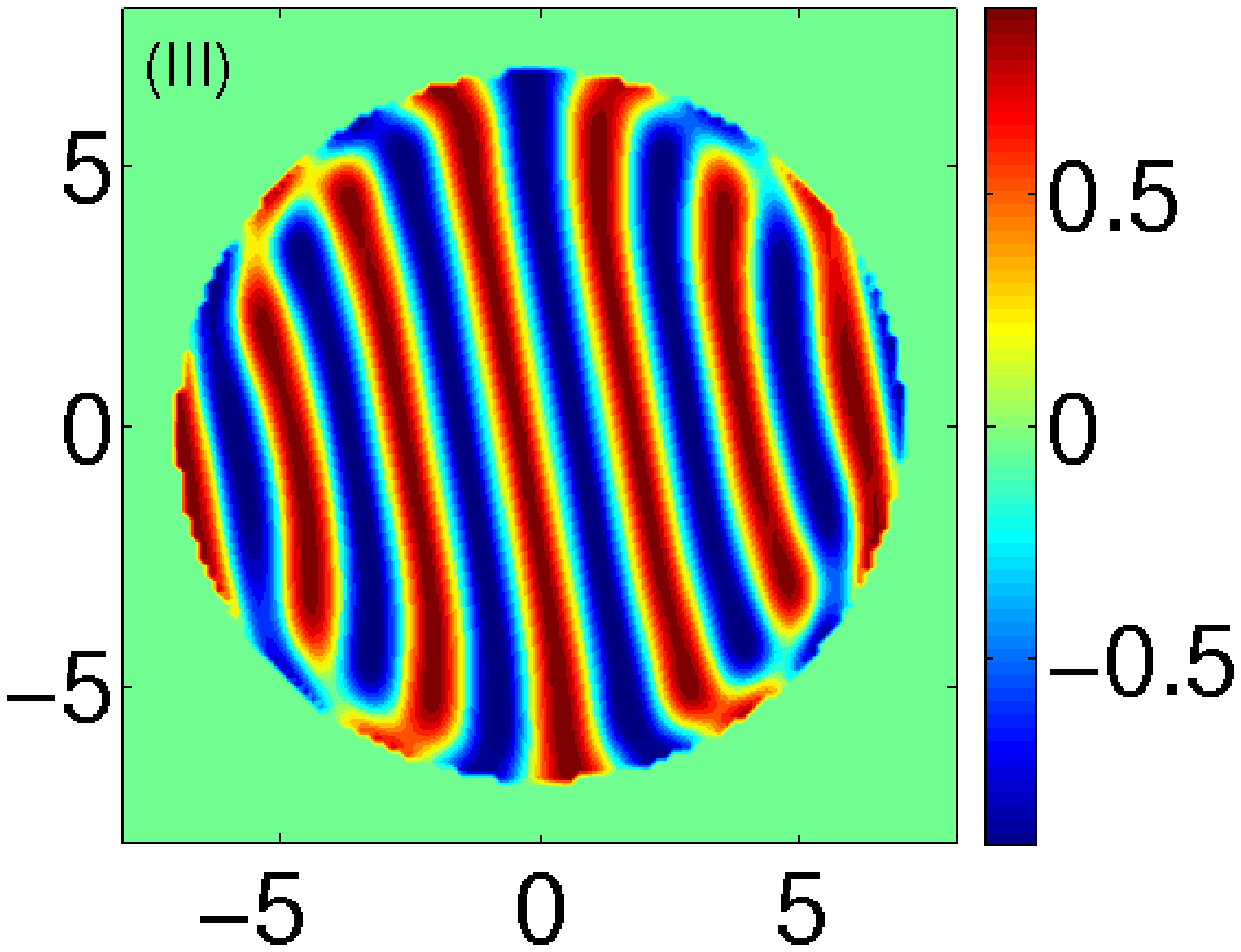}\\
\includegraphics[scale=0.25]{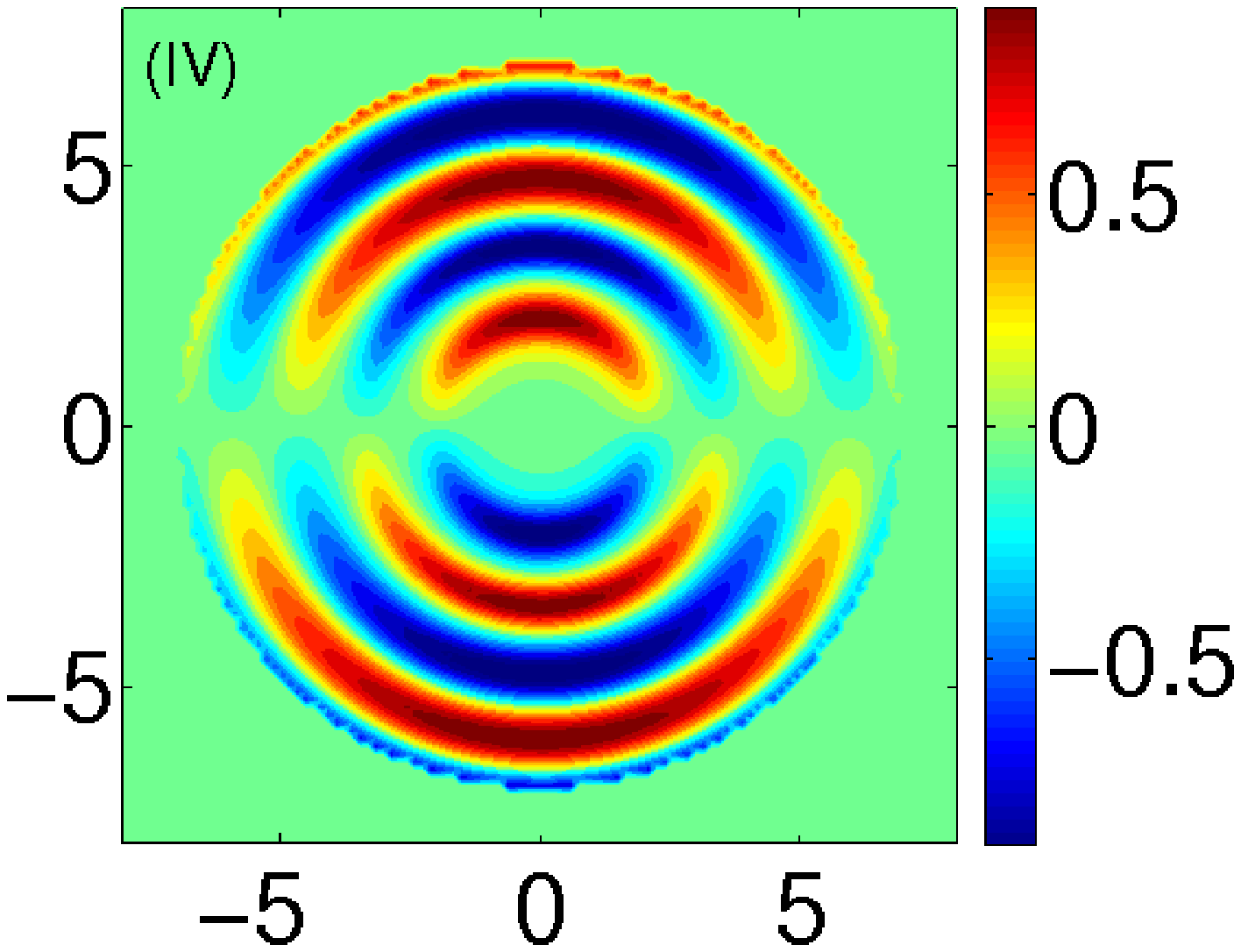}
\includegraphics[scale=0.25]{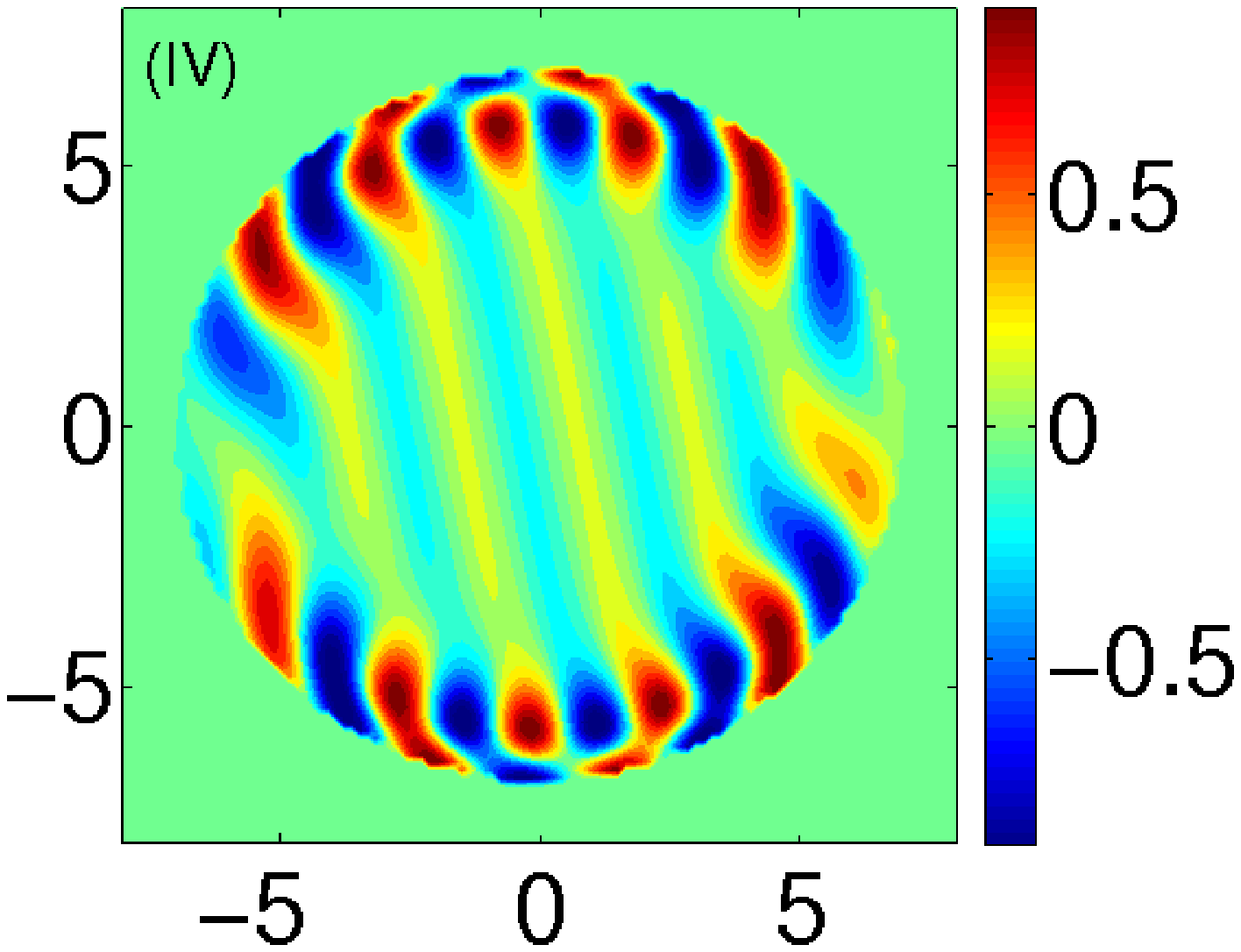}\\
\includegraphics[scale=0.25]{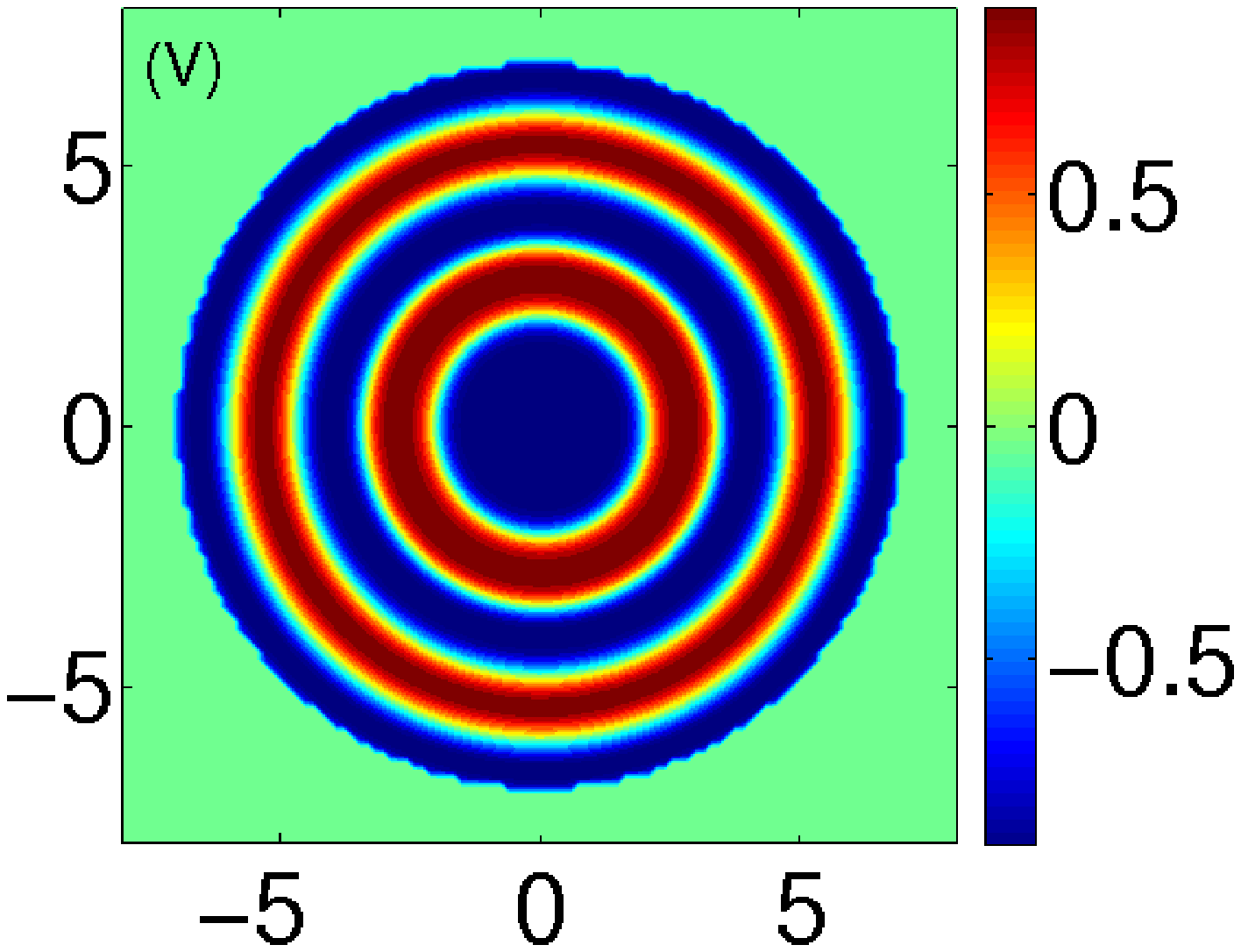}
\includegraphics[scale=0.25]{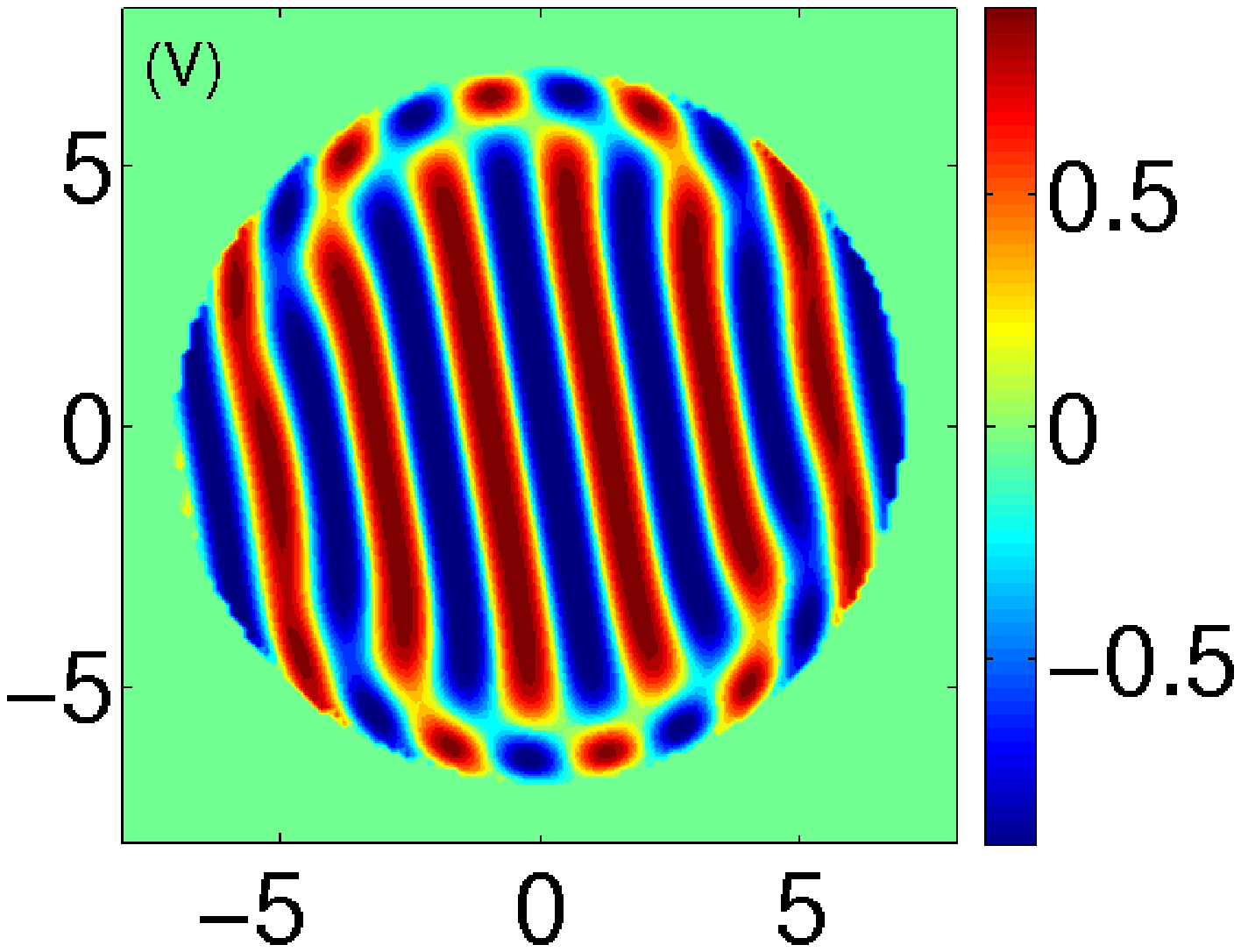}
\end{center}
\begin{picture}(0,0)(10,10)
\put(-53,23) {{$x$}}
\put(63,23) {{$x$}}
\put(-102,416) {{$y$}}
\put(-102,331) {{$y$}}
\put(-102,245) {{$y$}}
\put(-102,159) {{$y$}}
\put(-102,72) {{$y$}}
\end{picture}
\caption{(Color online) Numerical simulations for region (iii) of the $\kappa-\delta$ with $\Omega=0$ phase diagram of Fig. \ref{summ}. Left column (a): $(\delta,\kappa)=(1.5,1.25)$  and right column (b): $(\delta,\kappa)=(1.5,1.5)$. Density plots (frame (I), component-1, and (II), component-2) and spin density plots (frame (III), $S_x$, frame (IV), $S_y$ and frame (V), $S_z$). }
\label{prof2}
\end{figure}


%
%

\subsection*{B. $\Omega$ small}

We now proceed to the case where $\Omega$ and $\kappa$ are in general non-zero. We first present a $\kappa-\delta$ phase diagram for $\Omega=0.1$ (small rotation) in Fig.\ \ref{pd_omega0.1} in which four distinct regions are present. We identify these as (i) two disks with no defects, (ii) two disks with domains, (iii) segregated symmetry preserving (SSP) with a giant skyrmion and (iv) stripes. Each region on the phase diagram of Fig.\ \ref{pd_omega0.1} contains a sample density profile from a simulation carried out within that region (the simulation parameters are noted in the figure). The key difference between a phase diagram with $\Omega=0$ and $\Omega$ small is the development of region (ii), which is not present when $\Omega=0$ (Fig. \ref{summ}).
\begin{figure}
\begin{center}
\includegraphics[scale=0.3]{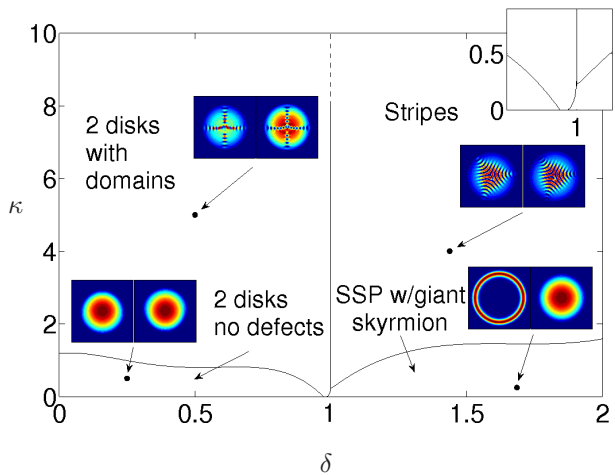}
\end{center}
\begin{picture}(0,0)(10,10)
\put(20,22) {{$\delta$}}
\put(-98,120) {{$\kappa$}}
\end{picture}
\caption{(Color online) $\kappa-\delta$ phase diagram with $\Omega=0.1$. The numerical parameters are taken as $g=4$ and $N=200$. There are four identified regions: (i) two disks with no defects, (ii) two disks with domains, (iii) segregated symmetry preserving (SSP) with a giant vortex and (iv) stripes. Each region has a typical density plot for each component (left panels, component-1 and right panels, component-2). The numerical values of these simulations correspond to: (i) $(\delta,\kappa)=(0.25,0.5)$; (ii) $(0.5,5)$; (iii) $(1.69,0.25)$ and (iv) $(1.44,4)$. The inset shows a zoom around the point $(\delta,\kappa)=(1,0)$. We analyse regions (i) and (ii) in more detail in Fig. \ref{prof_b_0.1}. The transition between regions (iii) and (iv) is shown in Fig. \ref{trans}.}
\label{pd_omega0.1}
\end{figure}

Along the $\kappa=0$ axis, we revert to the case of a rotating two-component condensate for which the ground state profiles are two co-existing disks ($\delta< 1$) or there is spatial separation of the components - one component is a disk and the other has a zero wave function ($\delta >1$) \cite{ktu,am2}. These behaviours are still present for $\kappa$ small (regions (i) and (iii) respectively of Fig. \ref{pd_omega0.1}). The profiles of the spin densities related to these two regions are straightforward: in region (i) we have $\bm{S}=(S_x,S_y,S_z)\approx(1,0,0)$ - much the same as in Fig. \ref{phase_00}(a), whereas in region (iii) we have $\bm{S}\approx(0,0,1)$. As $\kappa$ becomes larger, modulations of the density profiles begin to occur. There is a blurring of the boundary between regions (ii) and (iv). To indicate the uncertainty in the location of this boundary for high $\kappa$, we have used a dashed line in Fig. \ref{pd_omega0.1} above some arbitrary $\kappa$.


For $\delta<1$, the co-existing disk-shaped components each develop vortices that arrange themselves along bands of each component. We classify this region as the region in which both components are `two disks with domains' (region (ii) in Fig. \ref{pd_omega0.1}). The domain we refer to here is related to the profile of $\bm{S}$. We notice - see Fig. \ref{prof_b_0.1} - that across some bands, the behaviours of the $S_x$ and $S_y$ components of the spin density change sign. For example, Fig. \ref{prof_b_0.1}(III,IV) plots the $S_x$ component and the $S_y$ component for the parameters $(\delta,\kappa)=(0.5,1.25)$ and $(\delta,\kappa)=(0.5,5)$ [$\Omega=0.1$]. In the simulation with $(\delta,\kappa)=(0.5,1.25)$, two bands of vortices have been created along the $x$ axis, while for $(\delta,\kappa)=(0.5,5)$ there are four bands of vortices, each along one of the principal axes. In both cases we see that for $y>0$ ($<0$), $S_x>0$ ($<0$) and for $x>0$ ($<0$), $S_y<0$ ($>0$). This creates domains within the $S_x$ and $S_y$ component profiles (we note that $S_z\sim0$ away from the vortex lines). For this particular example, we say that there are two domains. A particular feature of the domain structure of the $S_x$ and $S_y$ is that, away from the vortex lines, they become (approximately) constant. For example, in Fig. \ref{prof_b_0.1}(III,IV), $S_x\approx1\sqrt{2}$ ($\approx-1\sqrt{2}$) for $y>0$ ($<0$) and $S_y\approx-1\sqrt{2}$ ($\approx1\sqrt{2}$) for $x>0$ ($<0$) [note that $S_x^2+S_y^2\approx1$ as we have $S_z\approx0$ everywhere]. As $\kappa$ is increased to higher values, we see examples with more domains. As for the total phase, we still have numerically the relation  $(d\Theta/dx,d\Theta/dy)=-2\kappa(S_x,S_y)$.

\begin{figure}
\begin{center}
\includegraphics[scale=0.25]{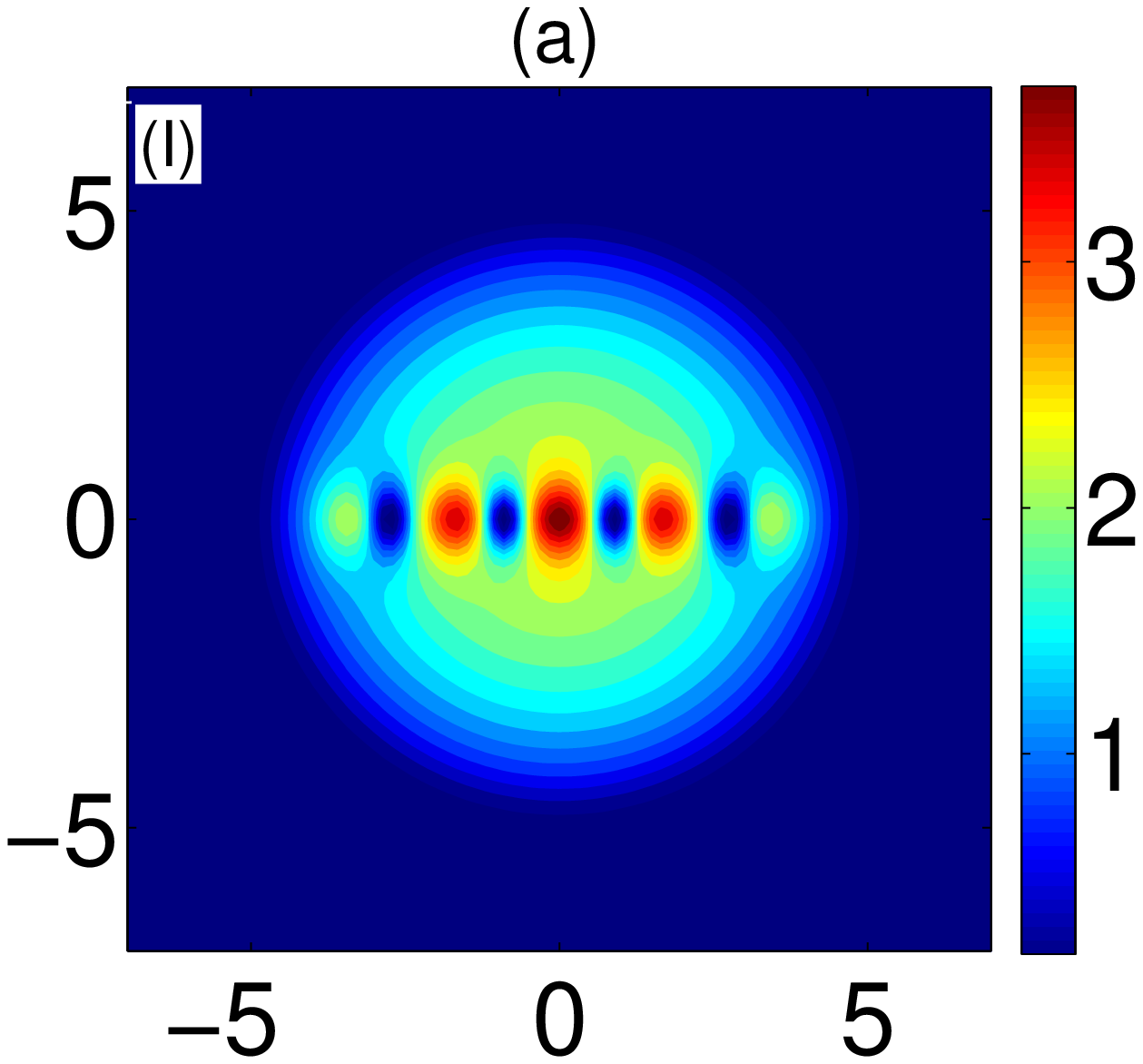}
\includegraphics[scale=0.25]{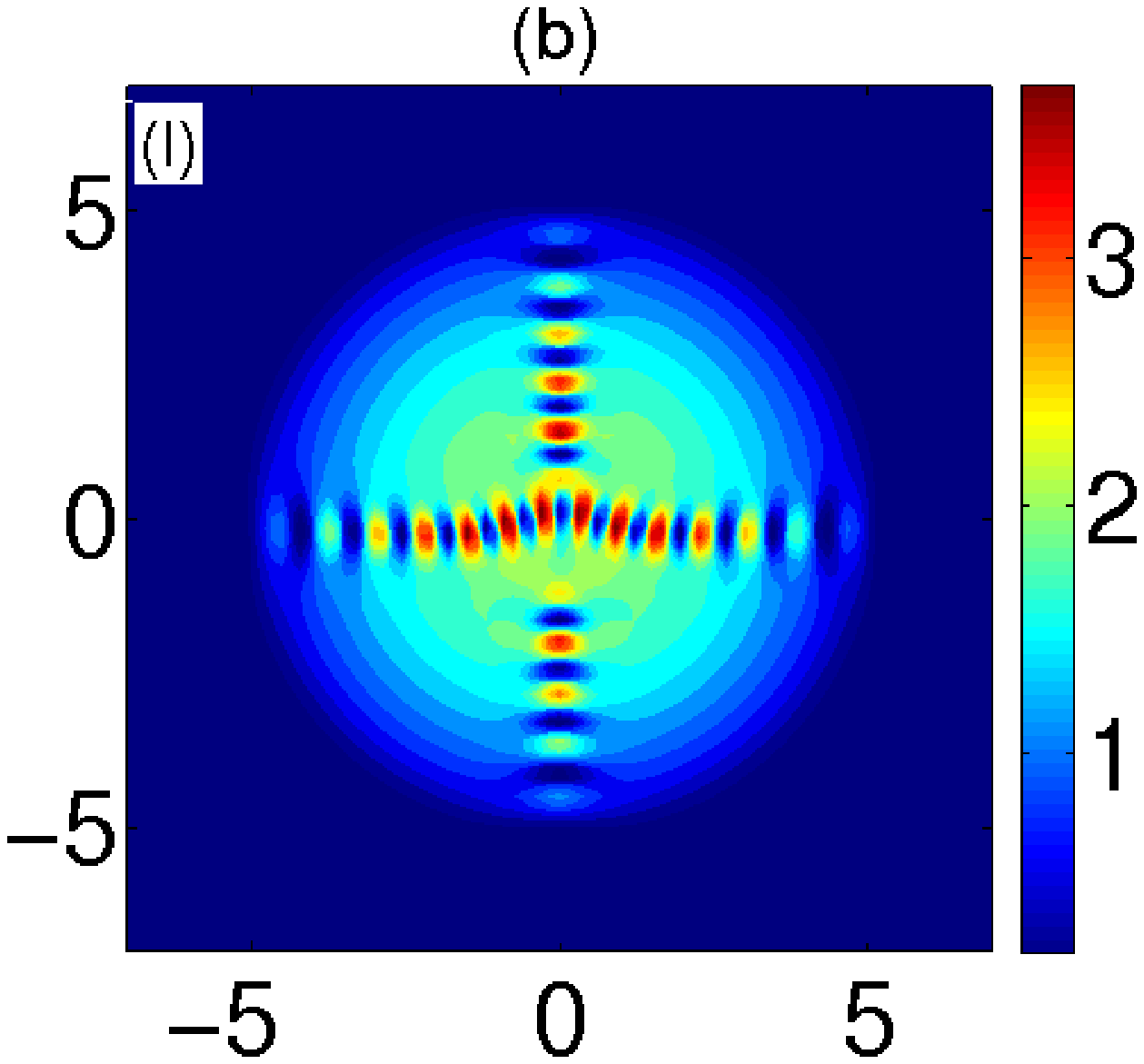}\\
\includegraphics[scale=0.25]{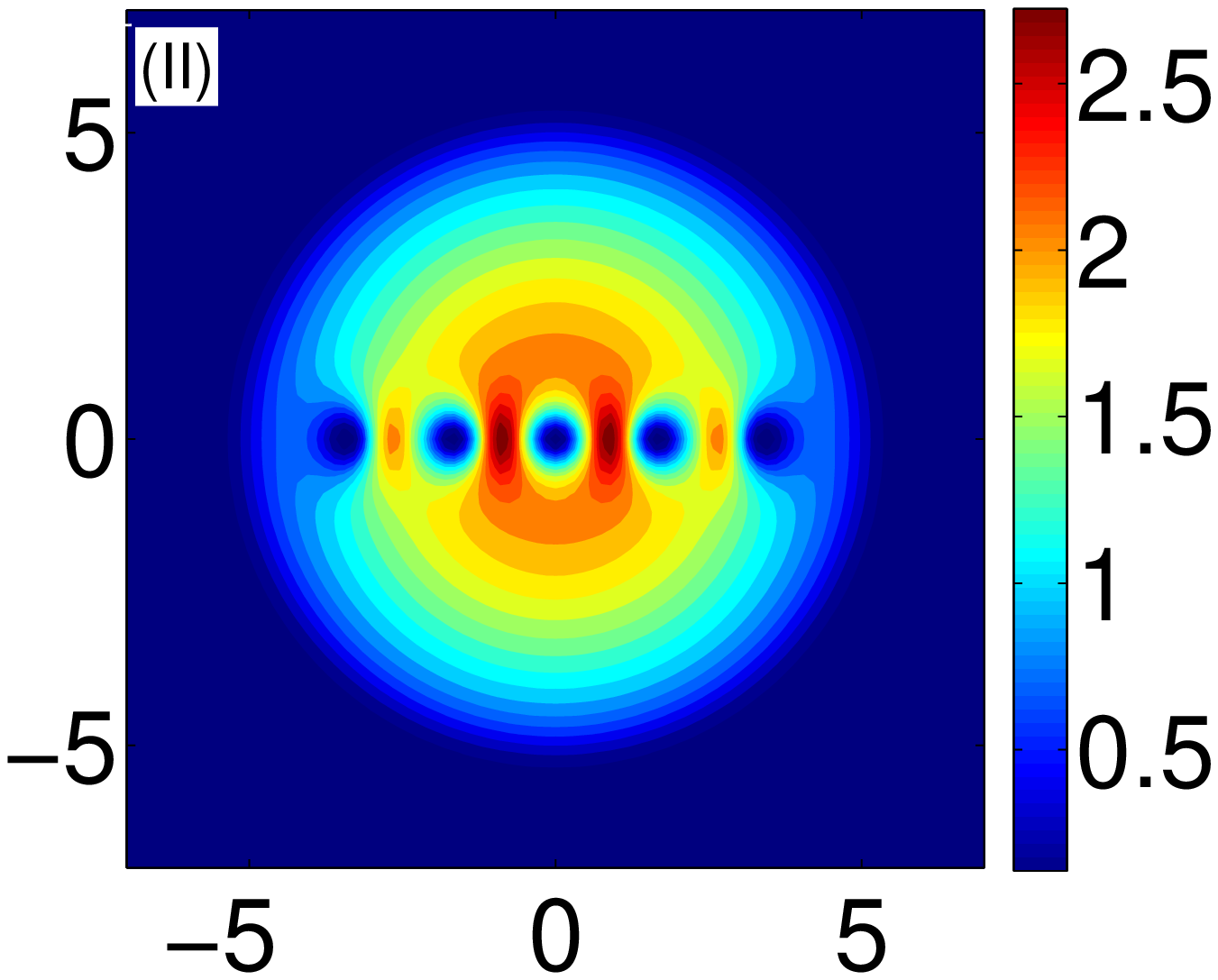}
\includegraphics[scale=0.25]{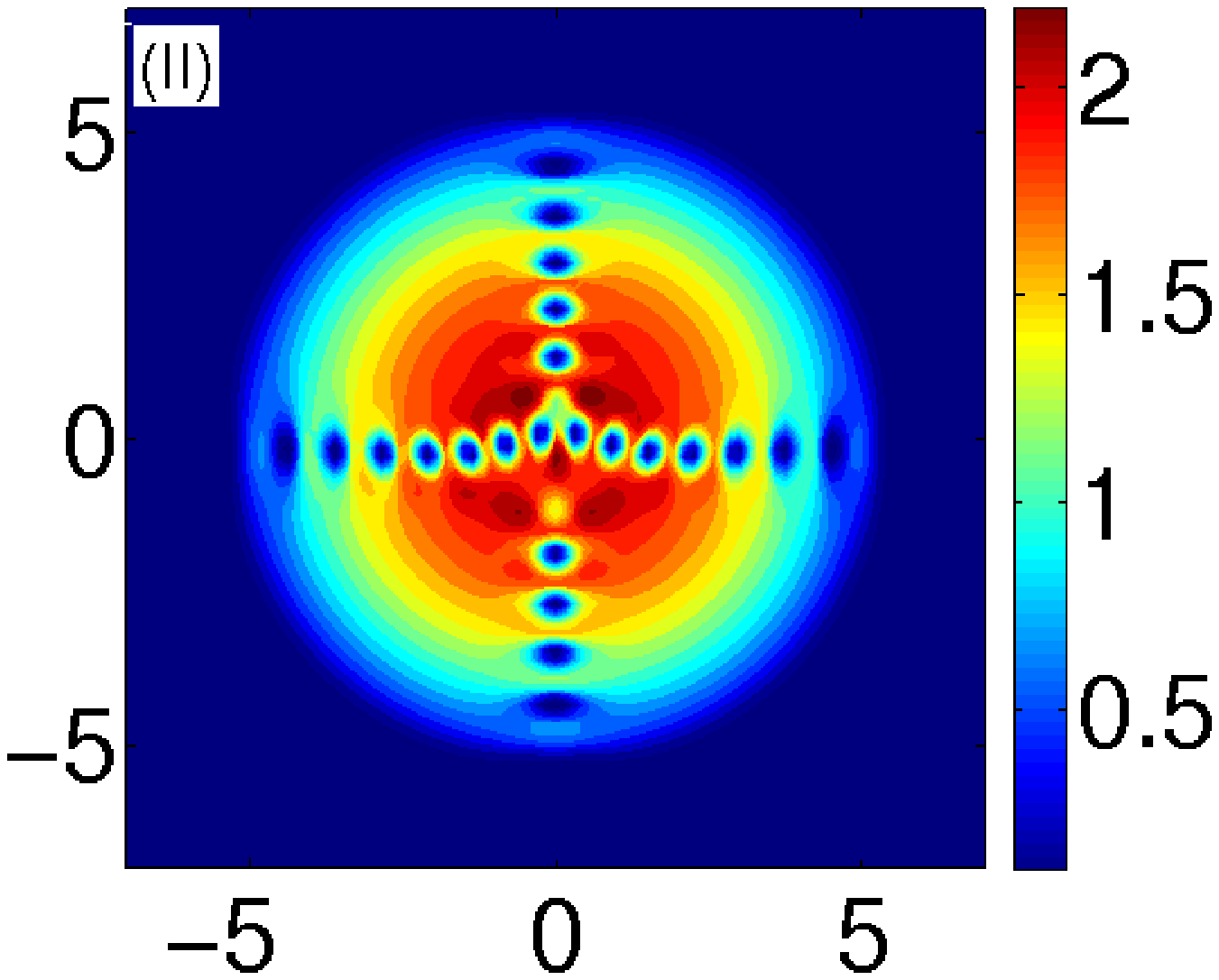}\\
\includegraphics[scale=0.25]{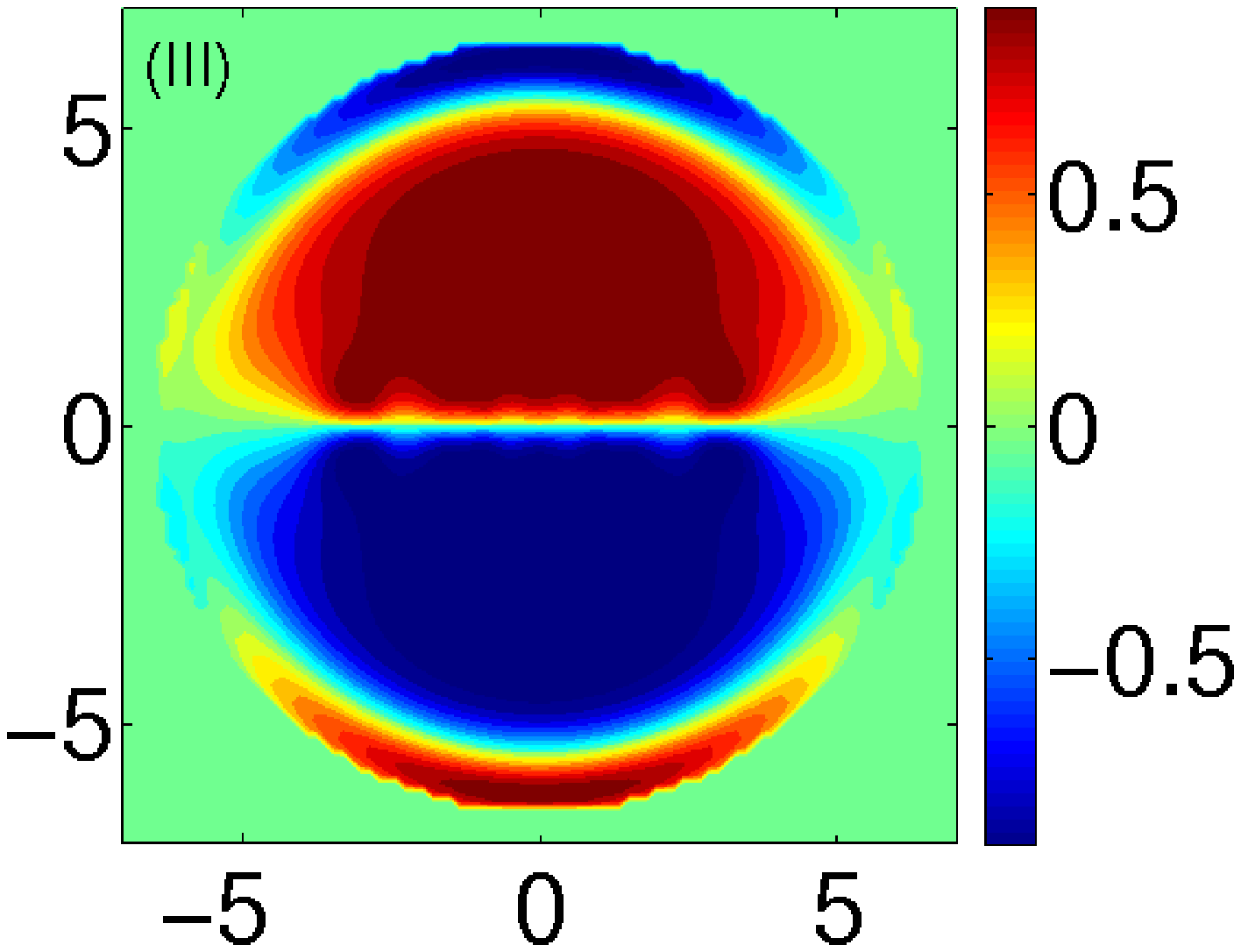}
\includegraphics[scale=0.25]{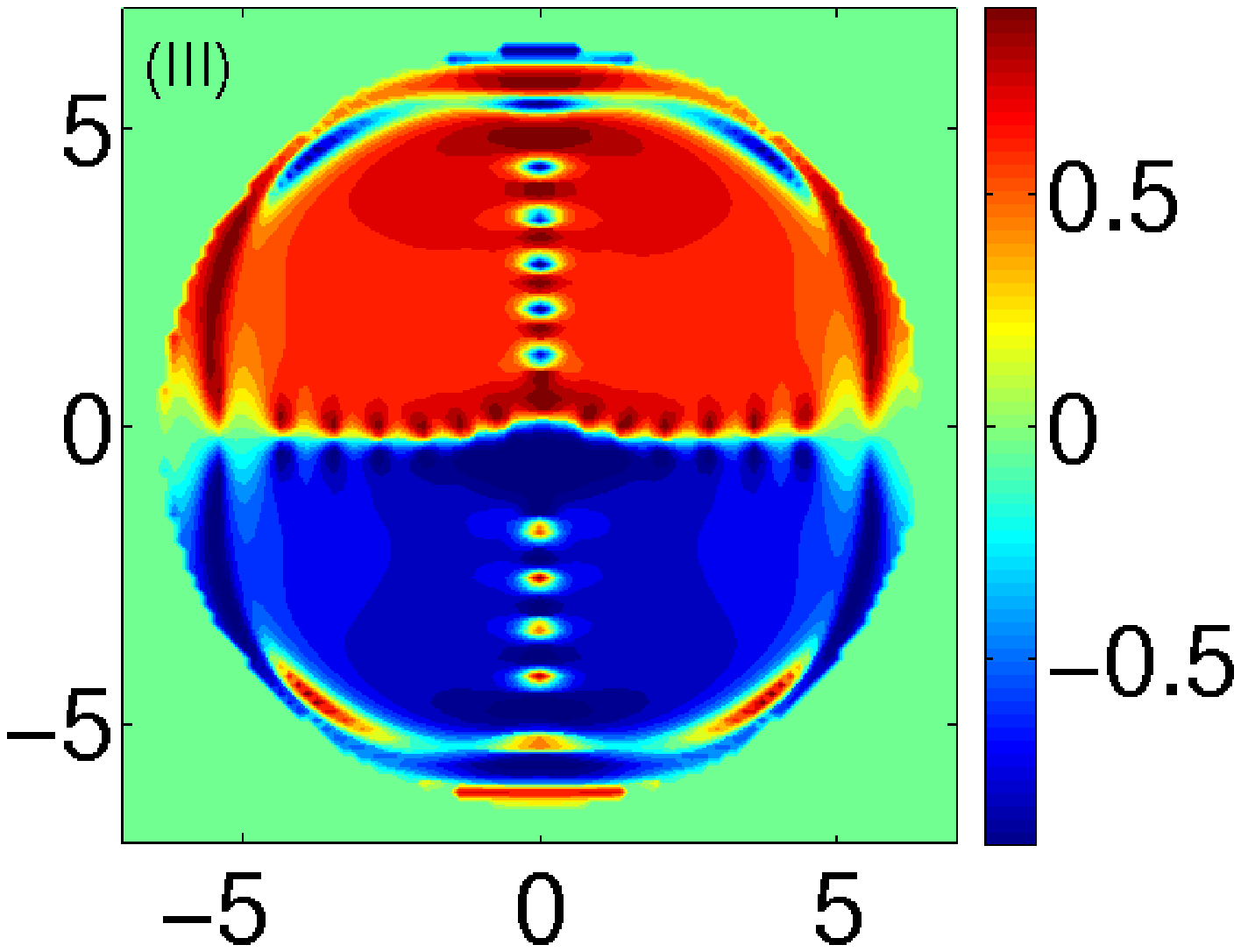}\\
\includegraphics[scale=0.25]{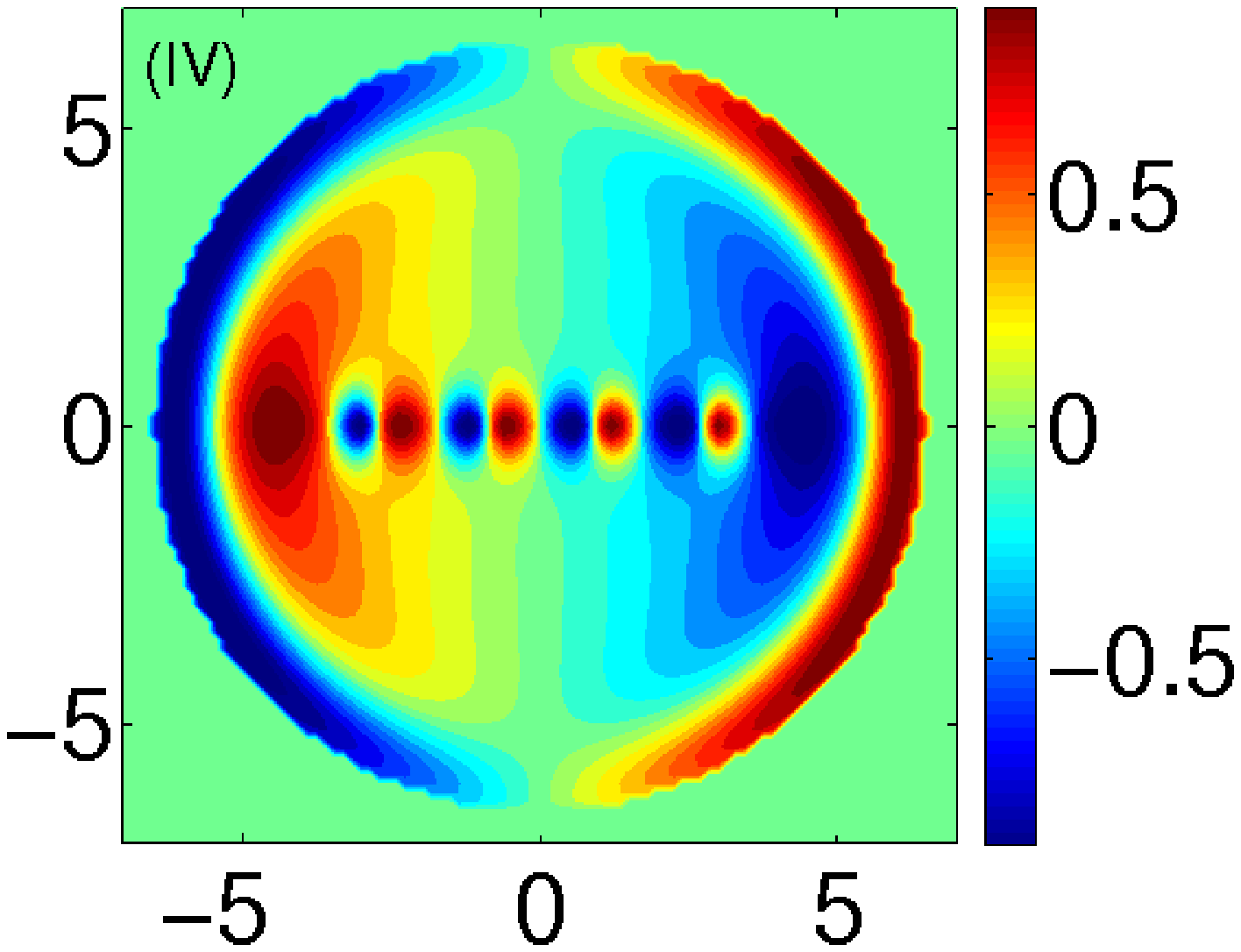}
\includegraphics[scale=0.25]{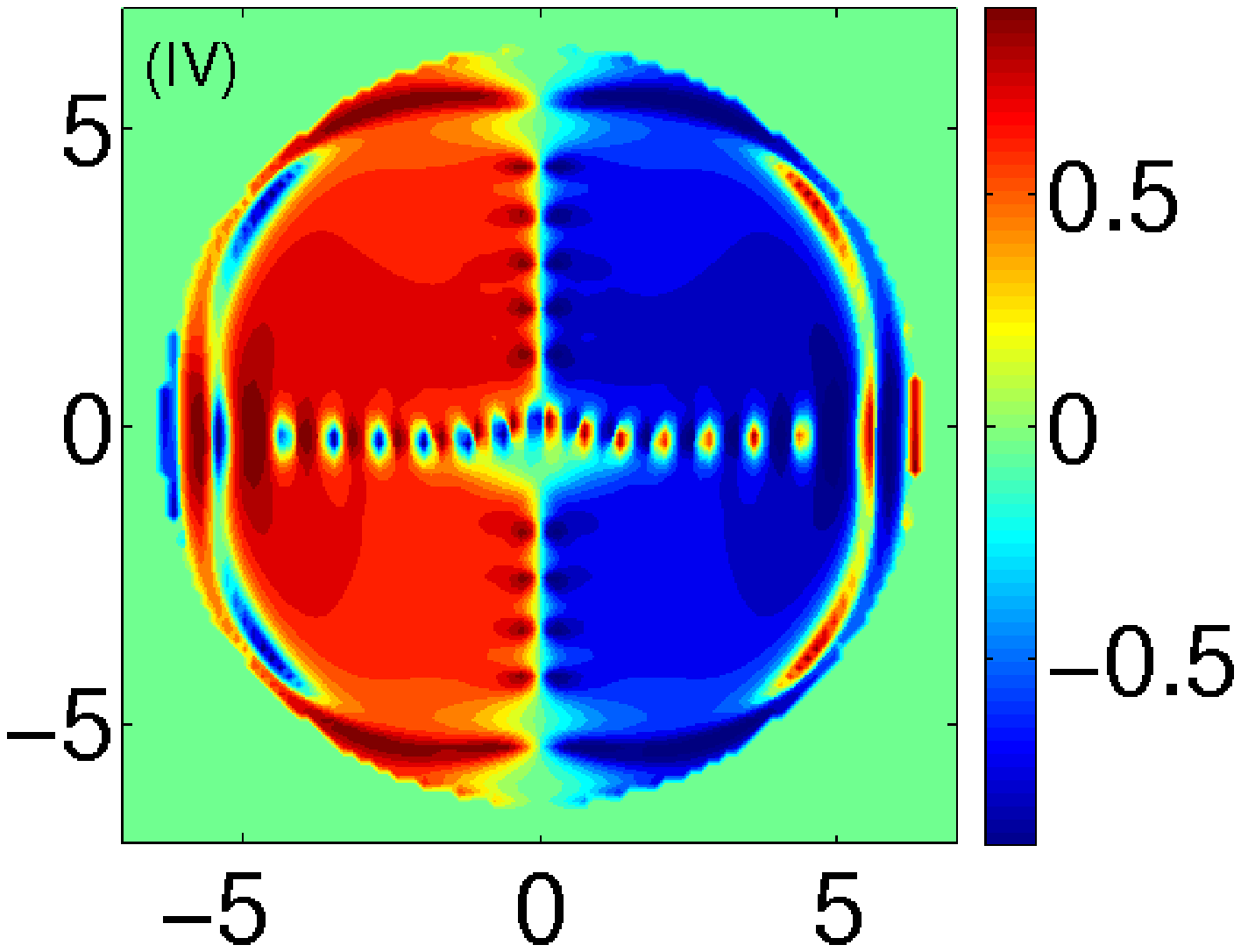}\\
\includegraphics[scale=0.25]{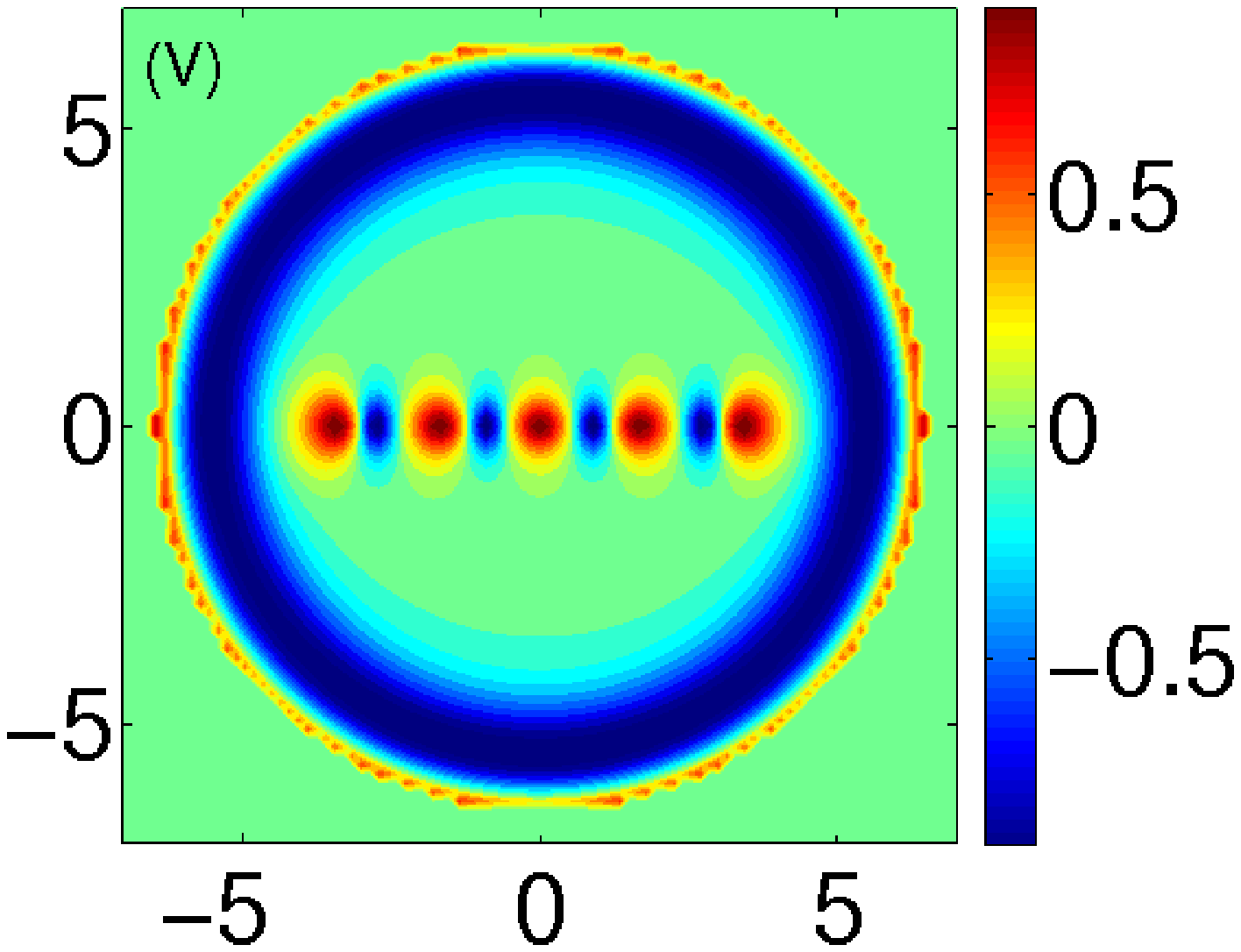}
\includegraphics[scale=0.25]{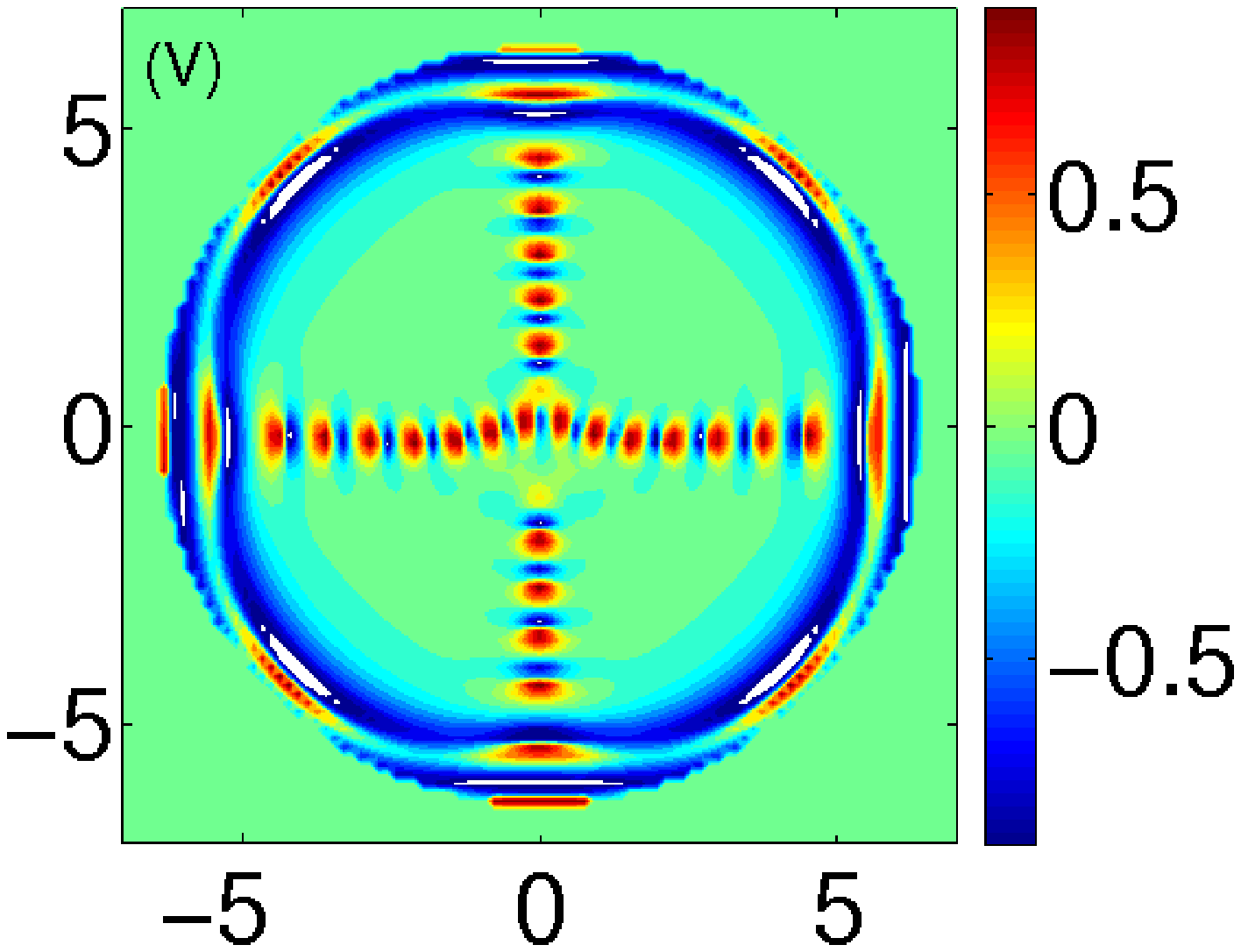}
\end{center}
\begin{picture}(0,0)(10,10)
\put(-53,23) {{$x$}}
\put(63,23) {{$x$}}
\put(-102,416) {{$y$}}
\put(-102,331) {{$y$}}
\put(-102,245) {{$y$}}
\put(-102,159) {{$y$}}
\put(-102,72) {{$y$}}
\end{picture}
\caption{(Color online) Numerical simulations for region (ii) of the $\kappa-\delta$ with $\Omega=0.1$ phase diagram of Fig. \ref{pd_omega0.1}. Left column (a): $(\delta,\kappa)=(0.5,1.25)$  and right column (b): $(\delta,\kappa)=(0.5,5)$. Density plots (frame (I), component-1, and (II), component-2) and spin density plots (frame (III), $S_x$, frame (IV), $S_y$ and frame (V), $S_z$). }
\label{prof_b_0.1}
\end{figure}

 For $\delta>1$, when $\kappa$ is small, the density profiles are a disk for one component and an annulus for the other with a circulation 1 around the annulus. As $\kappa$ increases, the rotation forces more circulation, while $\kappa$ is not large enough
  to have the transition to the stripe.  In Fig. \ref{trans}, we show some density profiles that correspond to values of $\kappa$ taken around this transition. Fig. \ref{trans}.III illustrates the combined effect of rotation and spin orbit. In the phase diagram of Fig. \ref{pd_omega0.1}, we have drawn the transition between the two regimes [from regime SSP with giant skyrmion to stripes] as being instantaneous, but in reality there is a smooth transition from one profile to the other. For large $\kappa$, the ground state corresponds to stripes, that are no longer straight.

\begin{figure}
\begin{center}
\includegraphics[scale=0.25]{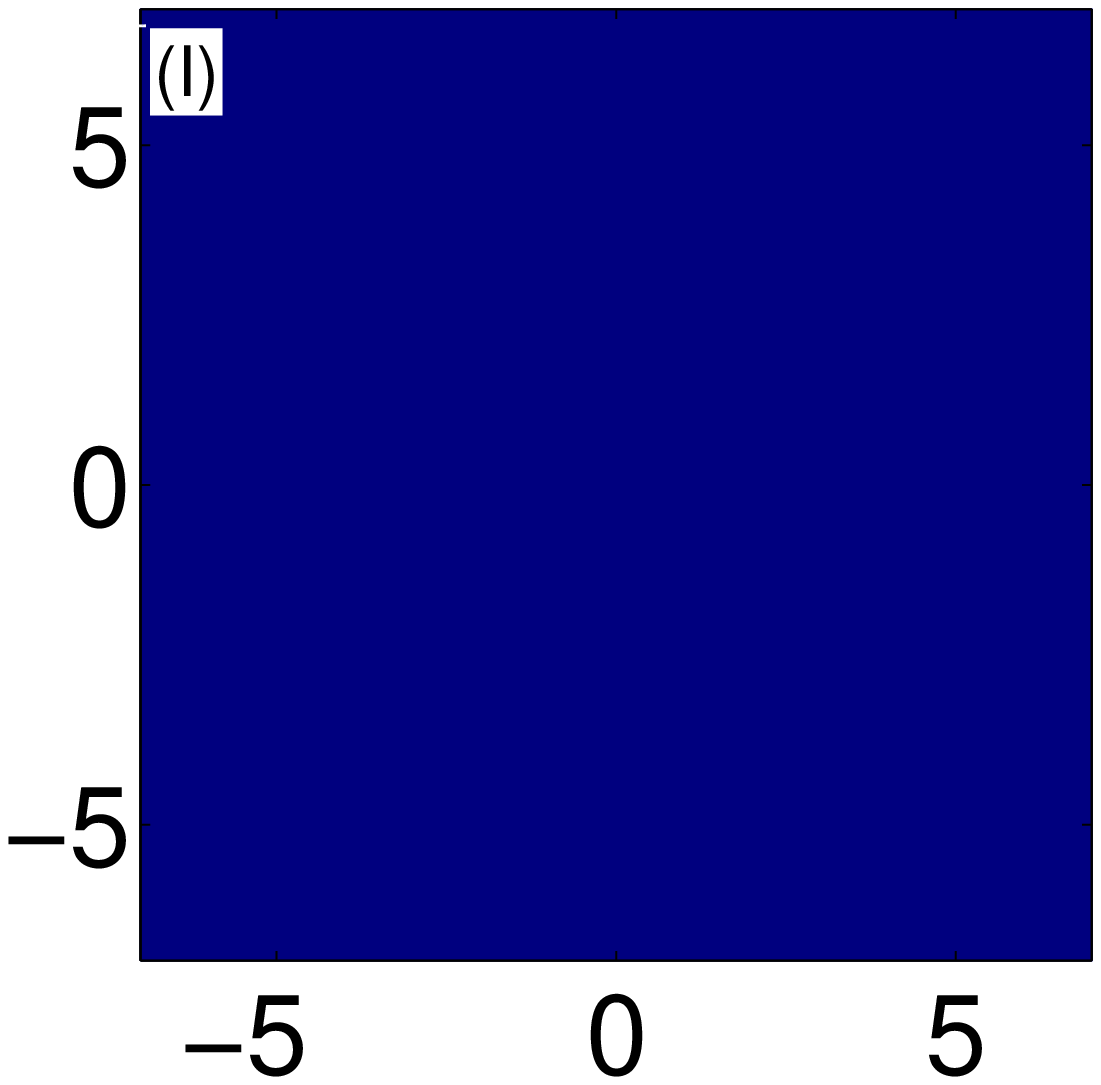}
\includegraphics[scale=0.25]{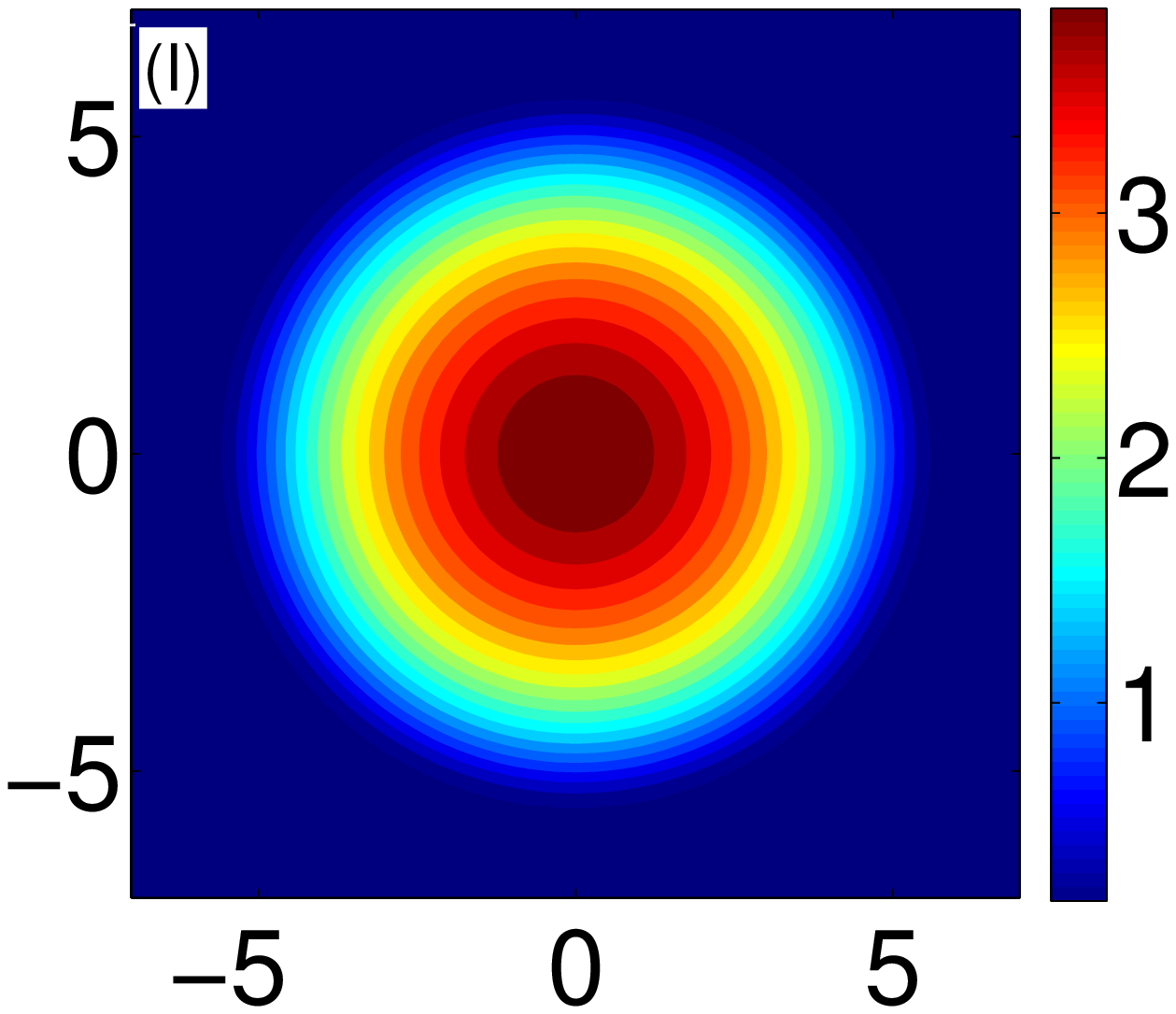}\\
\includegraphics[scale=0.25]{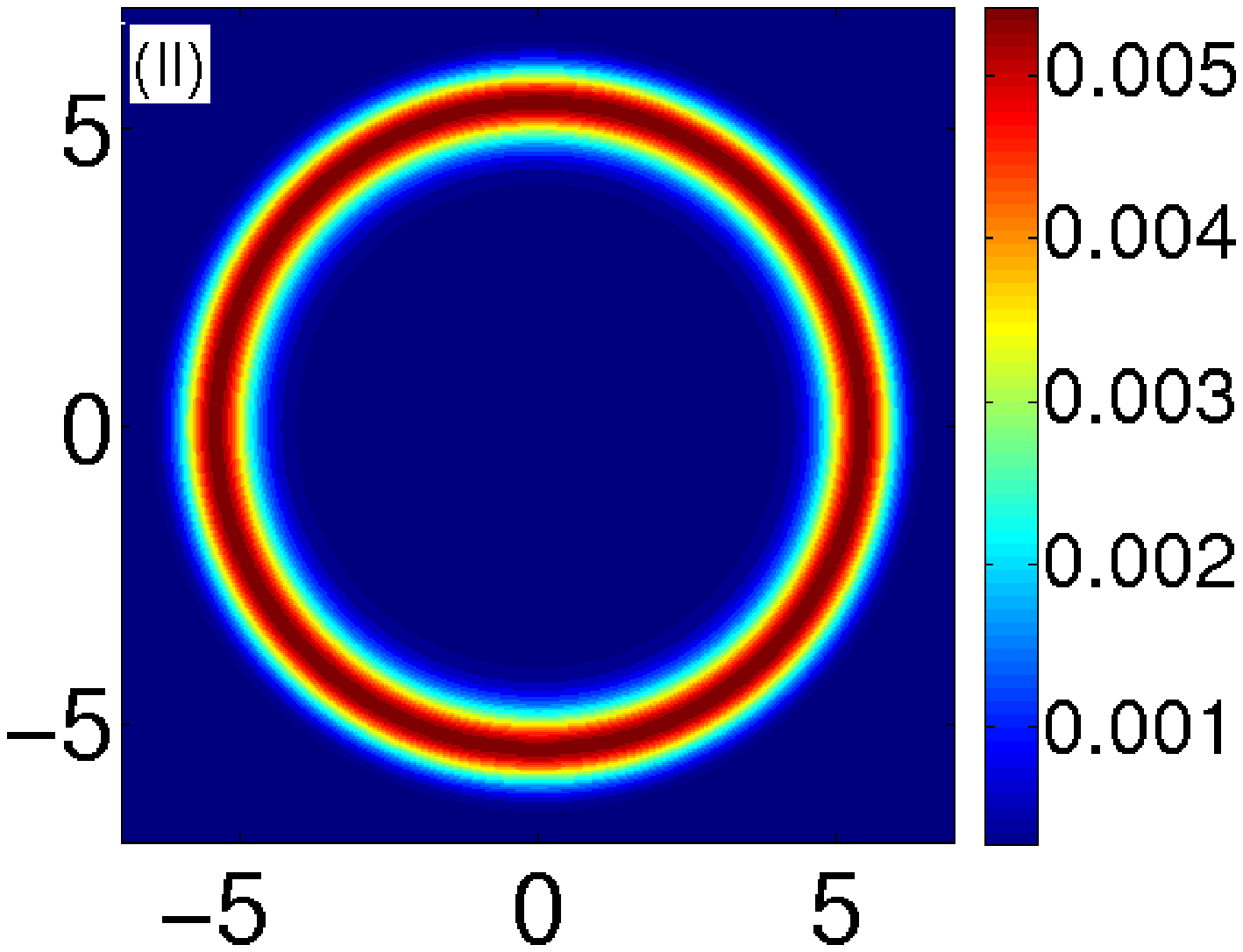}
\includegraphics[scale=0.25]{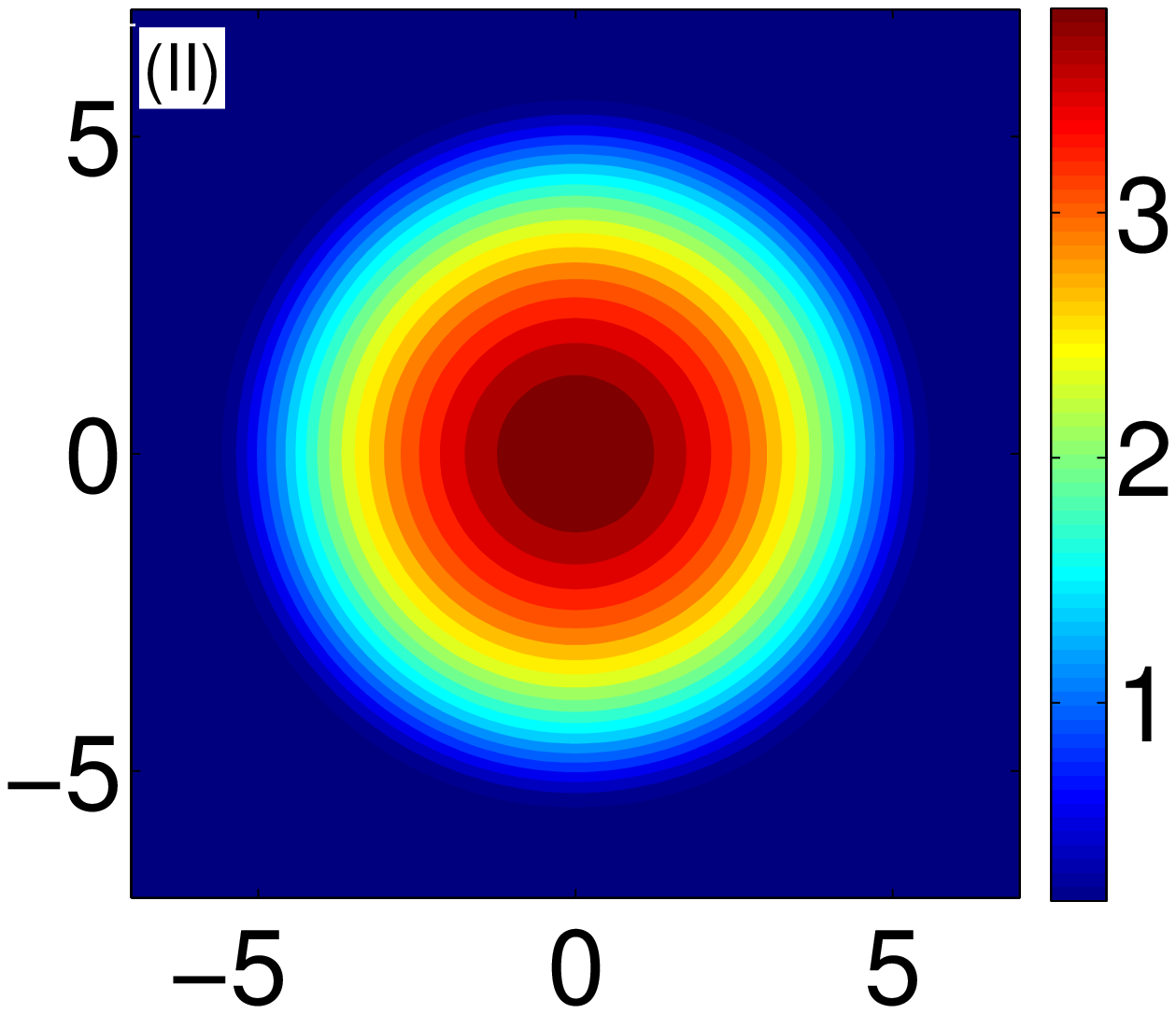}\\
\includegraphics[scale=0.25]{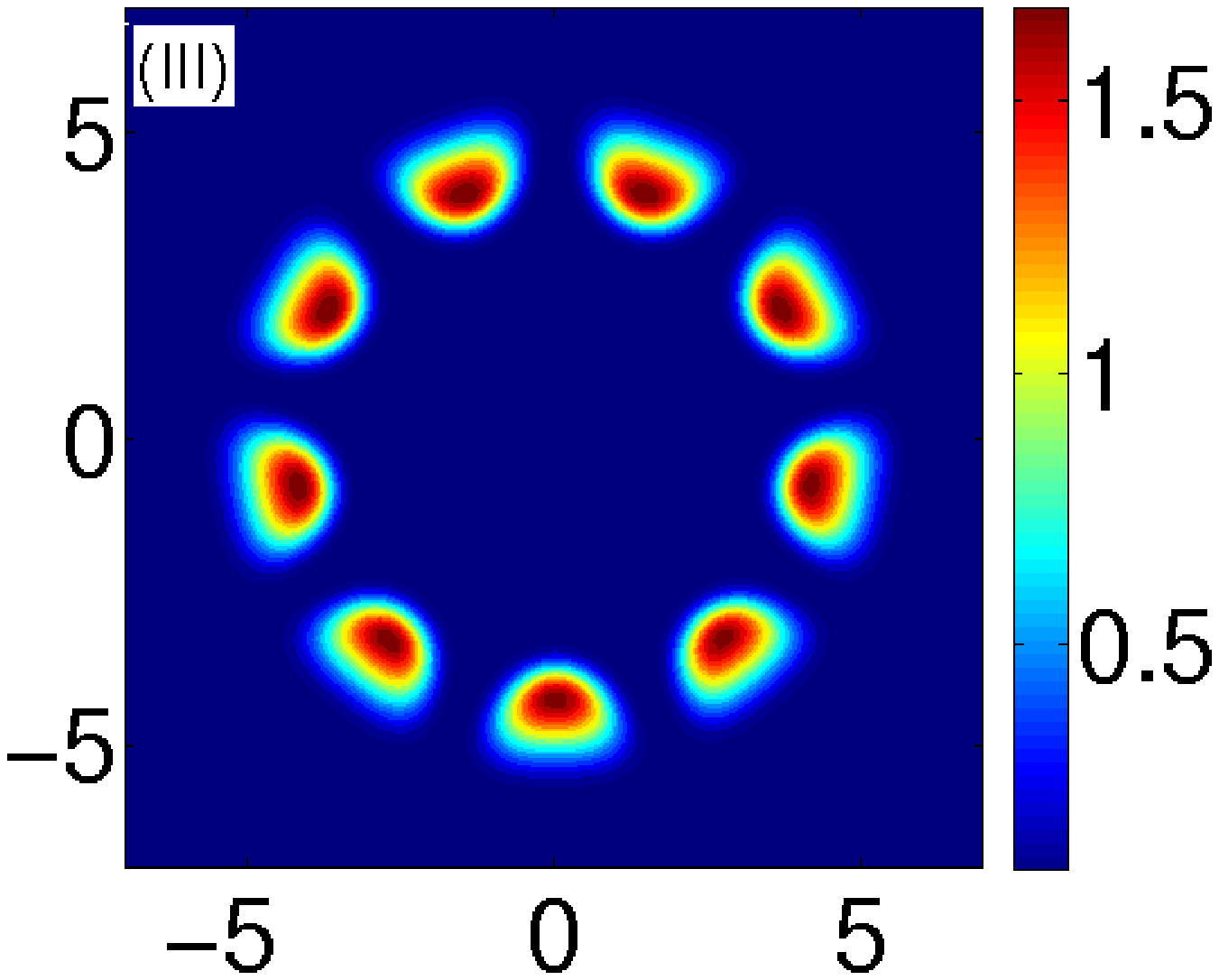}
\includegraphics[scale=0.25]{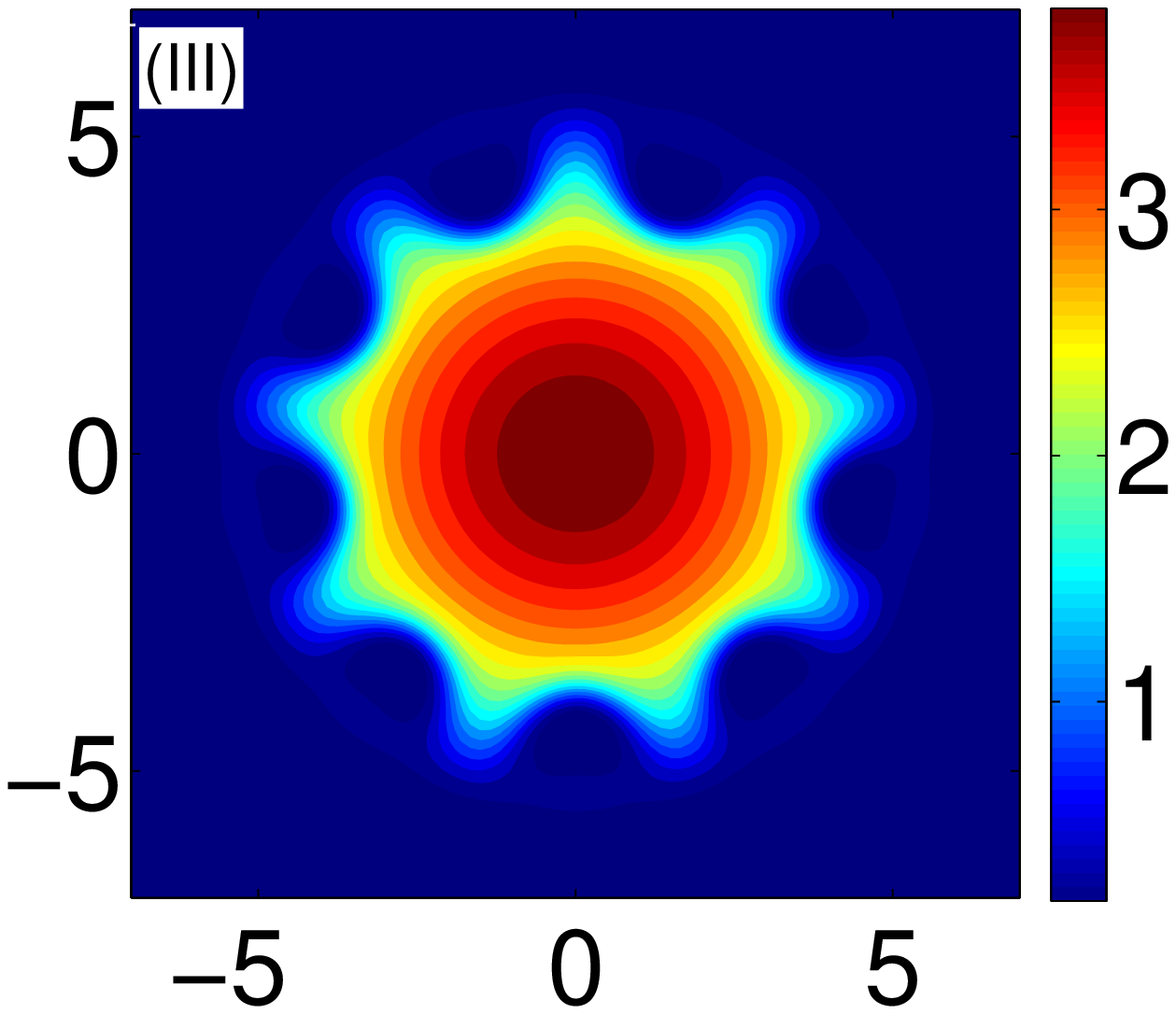}\\
\includegraphics[scale=0.25]{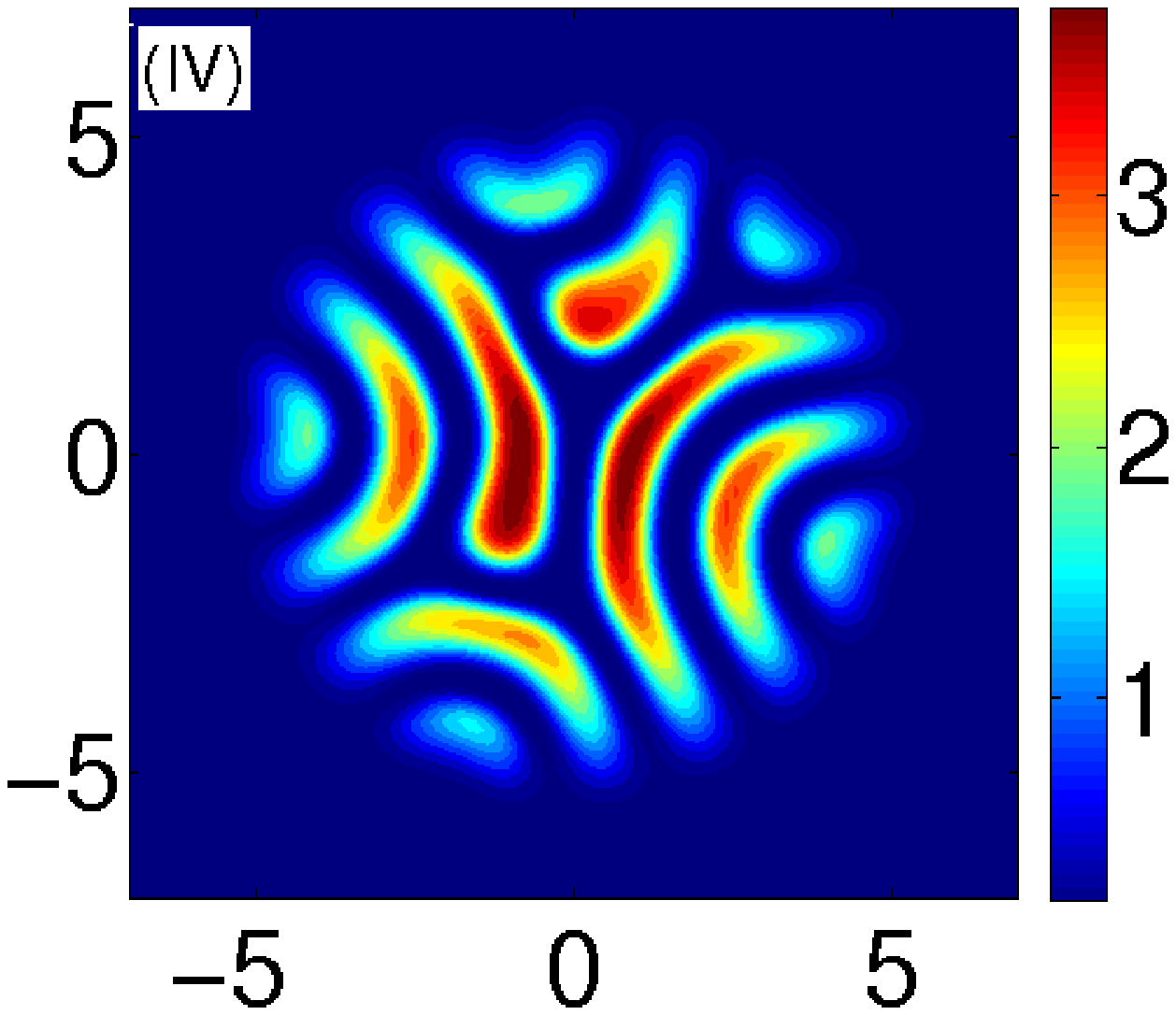}
\includegraphics[scale=0.25]{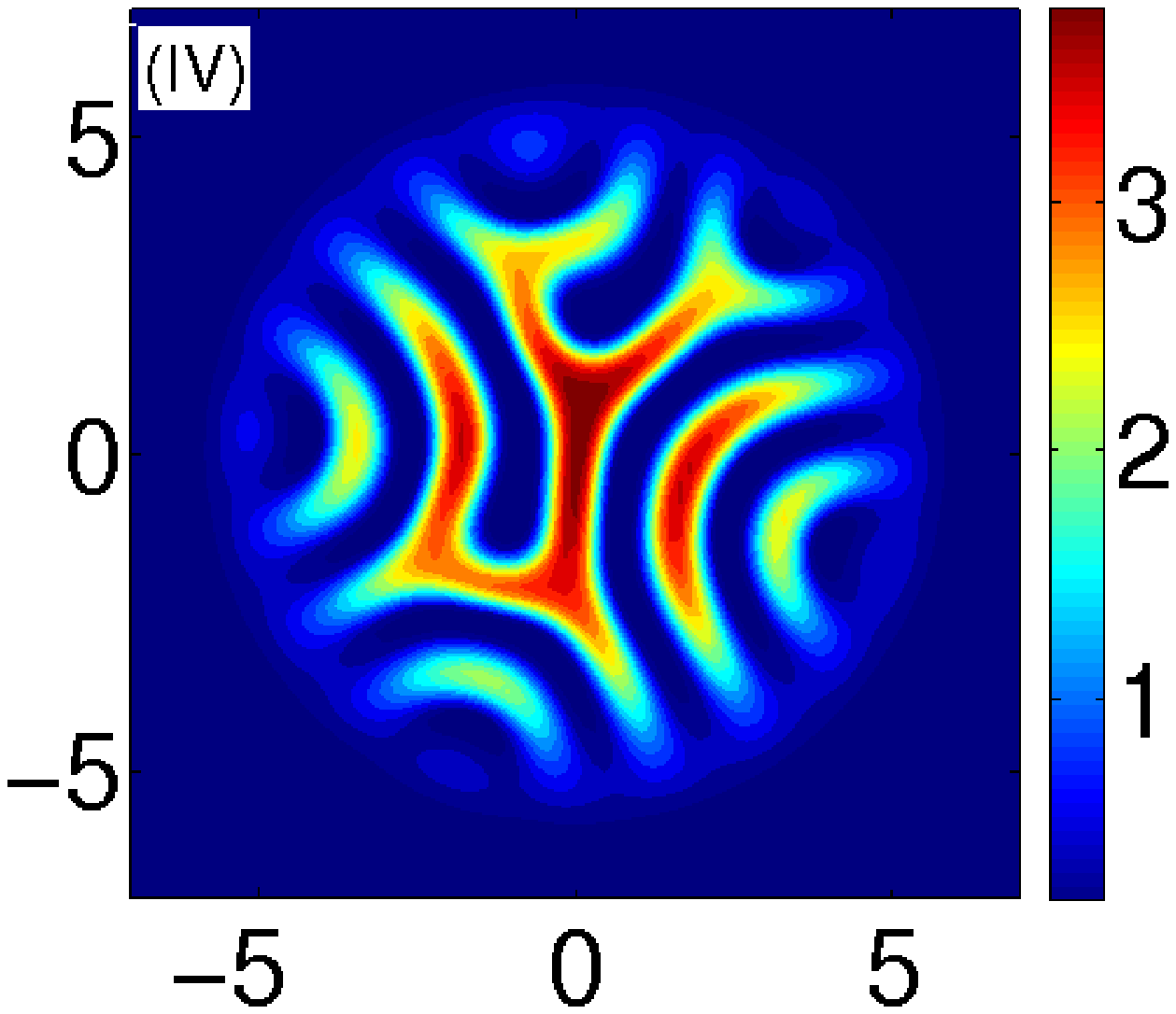}
\end{center}
\begin{picture}(0,0)(10,10)
\put(-53,23) {{$x$}}
\put(63,23) {{$x$}}
\put(-102,331) {{$y$}}
\put(-102,245) {{$y$}}
\put(-102,159) {{$y$}}
\put(-102,72) {{$y$}}
\end{picture}
\caption{(Color online) Density plots (left column, component-1 and right column, component-2) with $\Omega=0.1$ and $\delta=2$. The numerical simulations are carried out for (I) $\kappa=0$, (II) $\kappa=0.25$, (III) $\kappa=1.25$ and (IV) $\kappa=1.75$. Note that the density of component-1 for (I) is identically zero.}
\label{trans}
\end{figure}

\subsection*{C. Large $\Omega$}

If instead we look to the large $\Omega$ limit, then we see the rotational effect dominating. Figure \ref{pd_omega0.9} shows a $\kappa-\delta$ phase diagram for $\Omega=0.9$ (large rotation: note that $\Omega$ must stay below $1$) in which three distinct regions are present. We identify these as (v) two disks with vortex lattices and peaks, (vi) one component is a disk with a vortex lattice and the other contains peaks of density and (vii) two annuli with vortex lattices. Each region on the phase diagram of Fig.\ \ref{pd_omega0.9} contains a sample density profile from a simulation carried out within that region (the simulation parameters are noted in the figure).
\begin{figure}
\begin{center}
\includegraphics[scale=0.3]{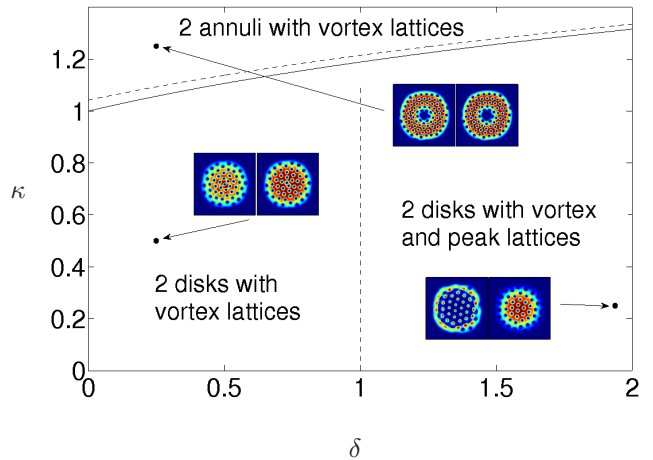}
\end{center}
\begin{picture}(0,0)(10,10)
\put(20,22) {{$\delta$}}
\put(-108,120) {{$\kappa$}}
\end{picture}
\caption{(Color online) $\kappa-\delta$ phase diagram with $\Omega=0.9$. The numerical parameters are taken as $g=4$ and $N=200$. There are three identified regions: (v) two disks with vortex lattices, (vi) a disk with vortex lattice and peaks and (vii) two annuli with vortex lattices. Each region has a typical density plot for each component (left panels, component-1 and right panels, component-2). The numerical values of these simulations correspond to: (v) $(\delta,\kappa)=(0.25,0.5)$; (vi) $(1.94,0.25)$ and (vii) $(0.25,1.25)$. The boundary between regions (v) and (vii) is plotted according to the numerical simulations (solid line) and analytically (dashed line), calculated according to Eq. (\ref{crit_val}). We analyse regions (v) and (vii) in more detail in Fig. \ref{prof_d_0.9}.}
\label{pd_omega0.9}
\end{figure}

Again if we consider the $\kappa=0$ axis then we revert to the two-component condensate rotating at high angular velocities \cite{am2}. In these cases, the large rotational effect leads to angular momentum being imparted onto the condensate and therefore to the existence of vortices. For $\delta<1$, the condensate is made of two co-existing disk-shaped components both with a triangular coreless vortex lattice. For $\delta\ge1$,  it is a single component with a triangular vortex lattice:  the other component has zero wave function. As $\kappa$ becomes non-zero, then for $\delta<1$, each component has a lattice
 of vortices (no peaks), while if $\delta>1$, spatial separation of the component occurs and vortices in the dominating component
  correspond to isolated peaks in the other. As $\kappa$ increases further, the less populated component starts to grow and the peaks get localized only in the center until they disappear, leading eventually to the formation of an annulus in one component. This is illustrated in Fig. \ref{trans3}. In Fig. \ref{trans2} we also show some density profiles that correspond to this transition for $\Omega=0.5$.

\begin{figure}
\begin{center}
\includegraphics[scale=0.25]{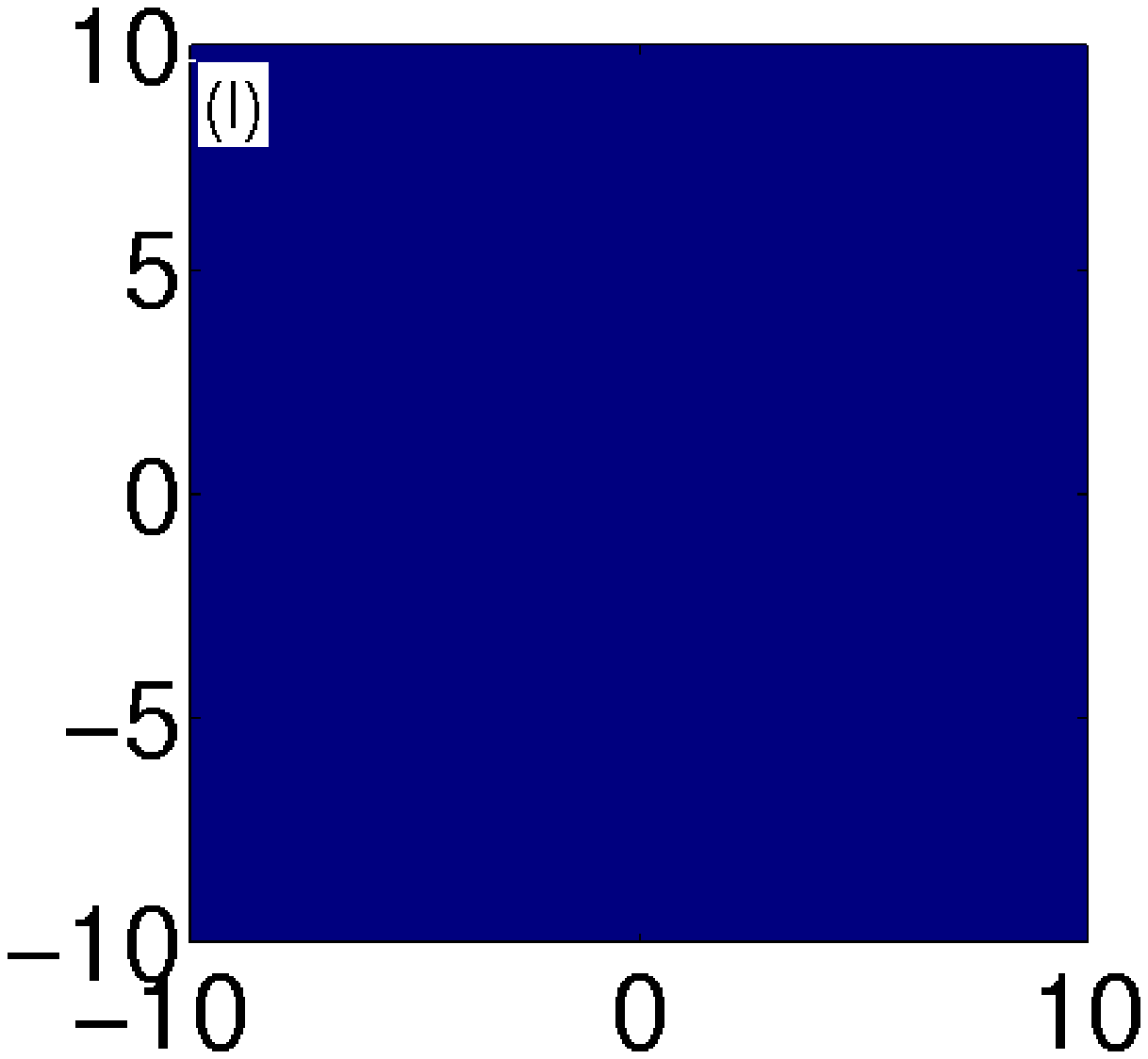}
\includegraphics[scale=0.25]{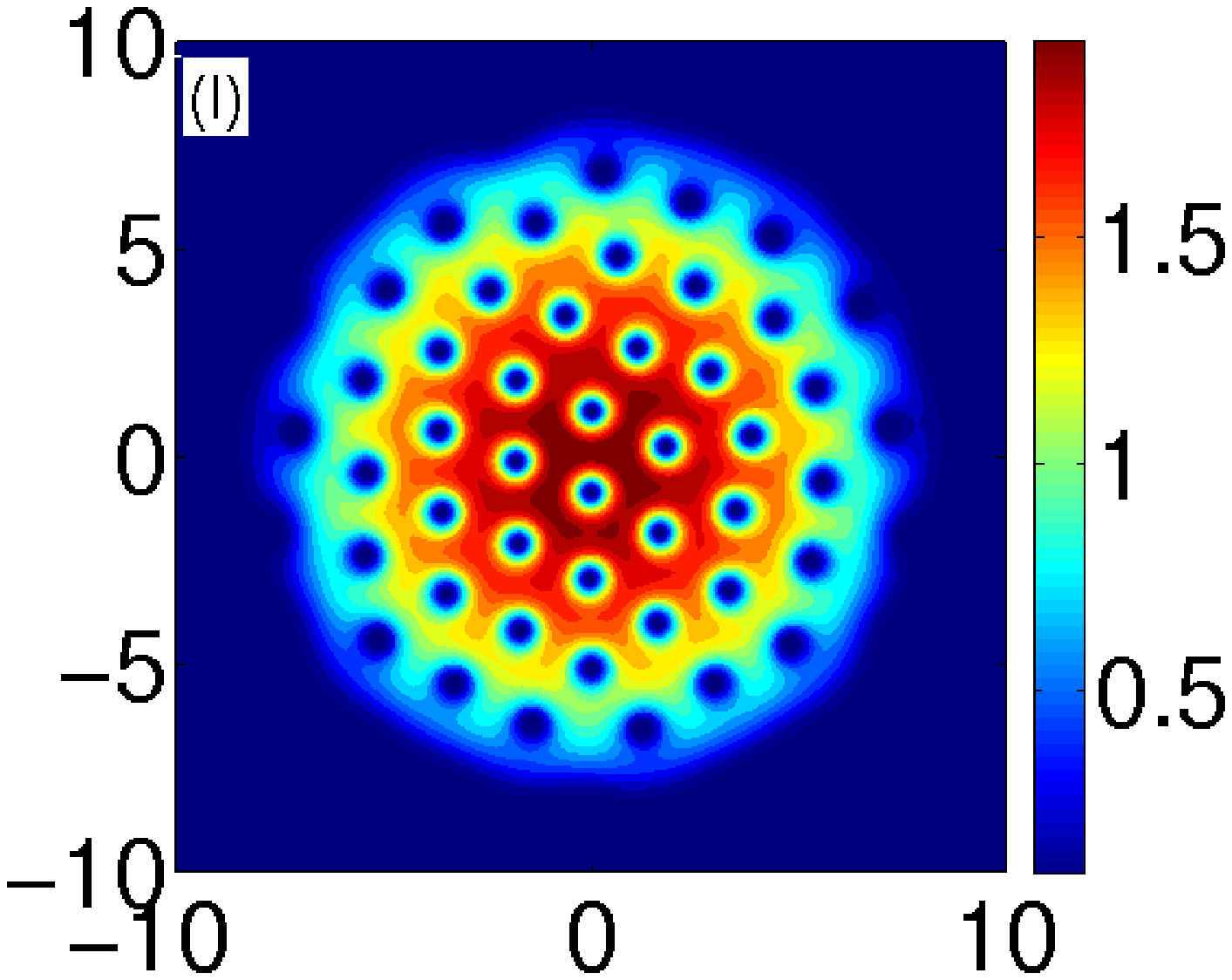}\\
\includegraphics[scale=0.25]{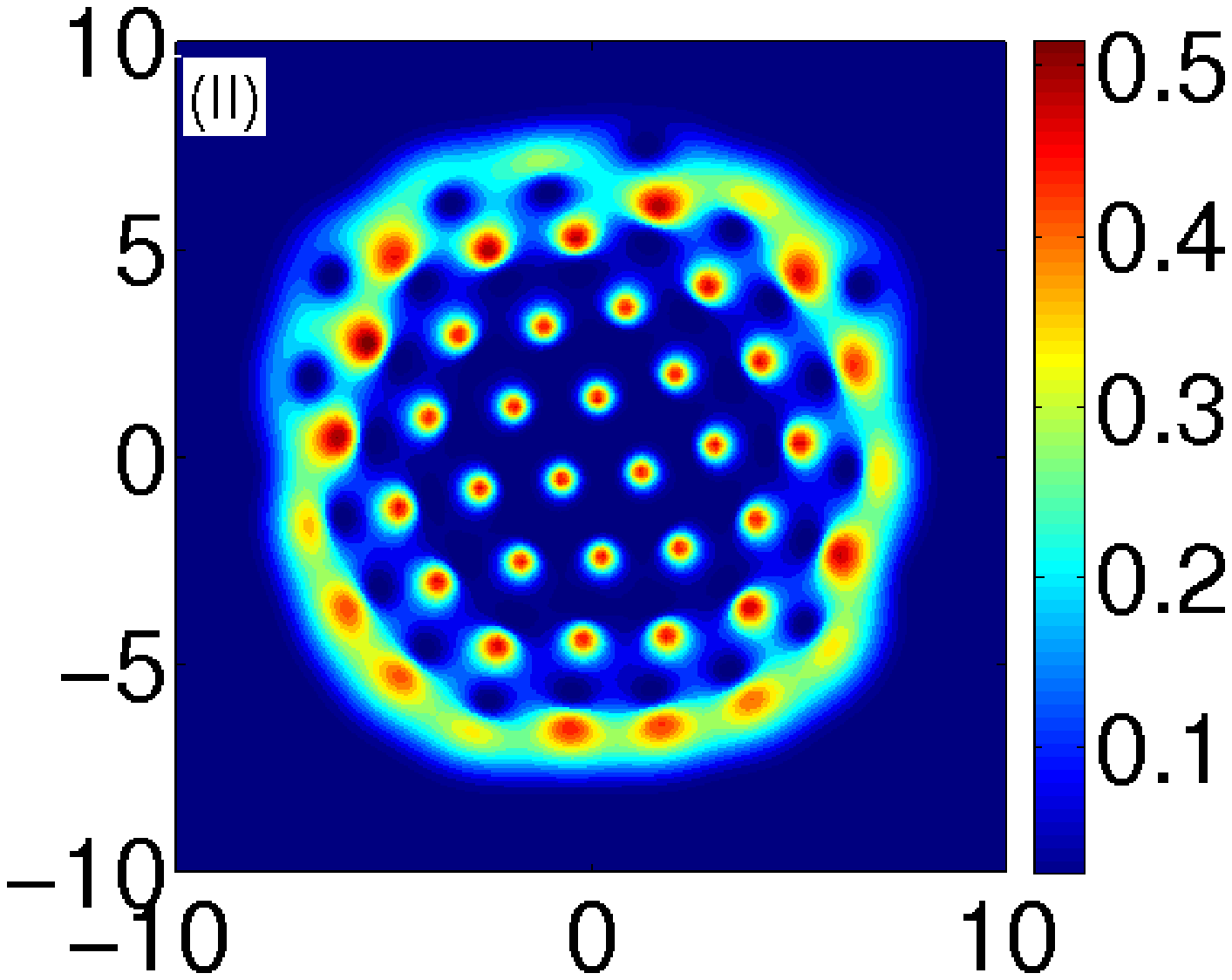}
\includegraphics[scale=0.25]{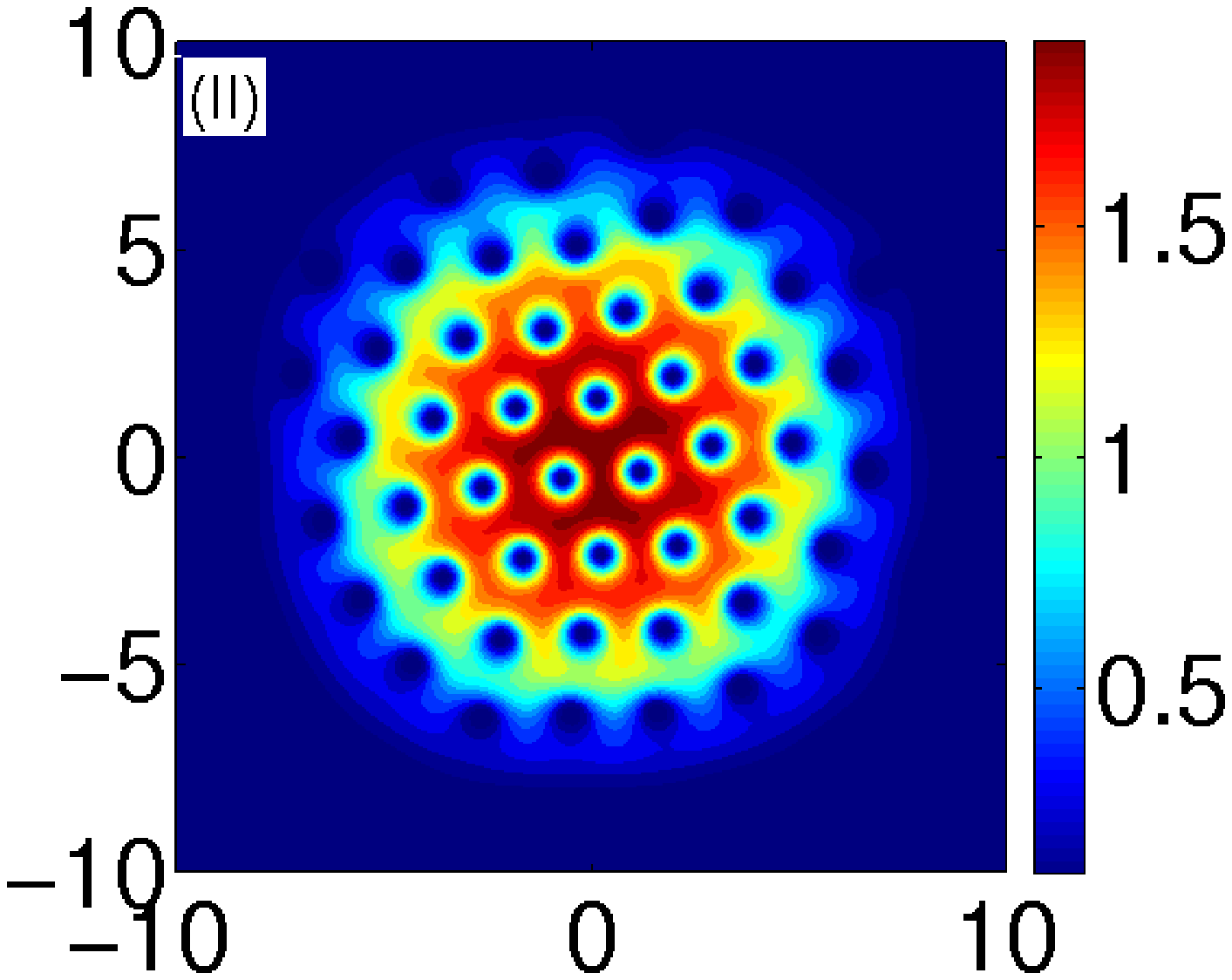}\\
\includegraphics[scale=0.25]{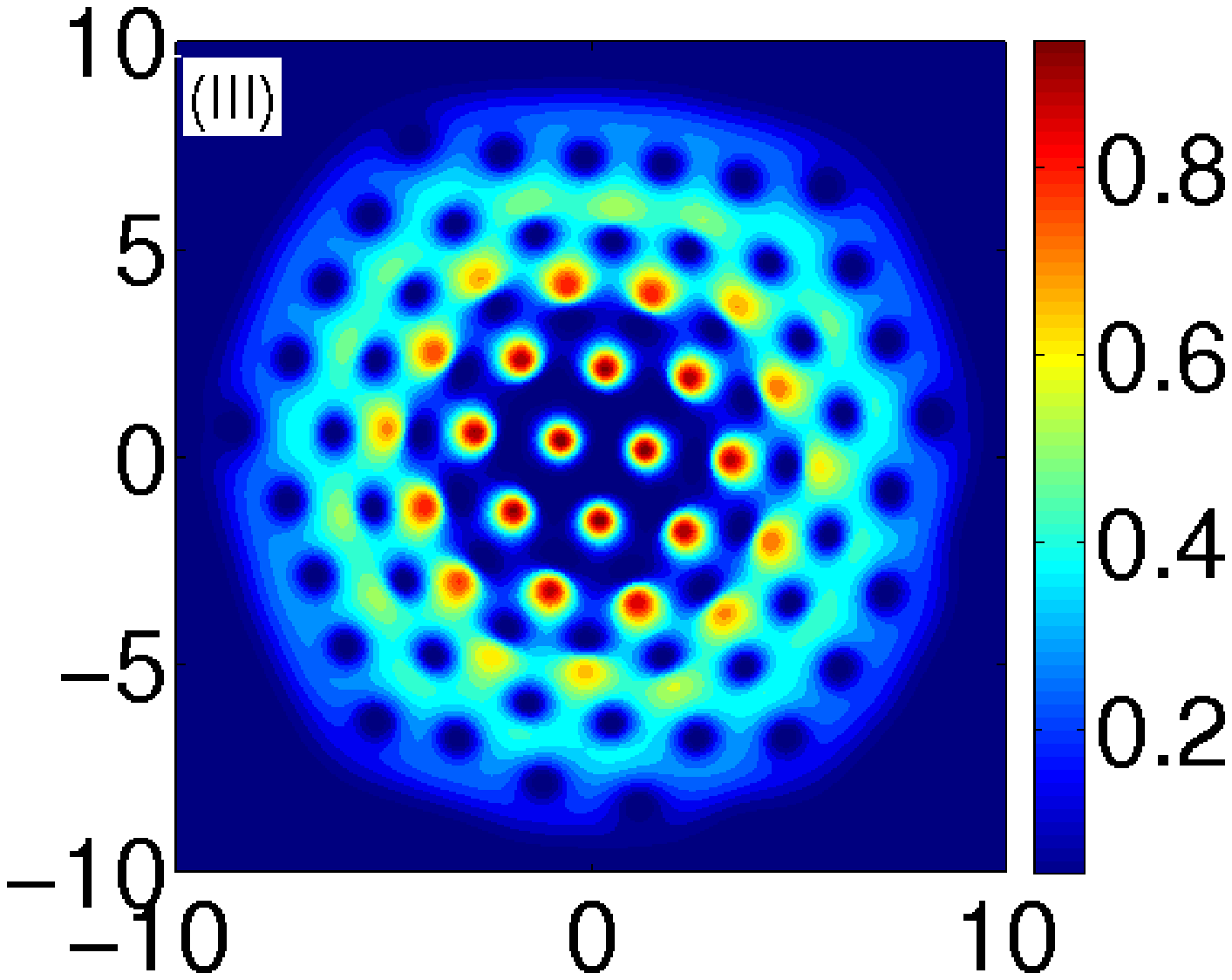}
\includegraphics[scale=0.25]{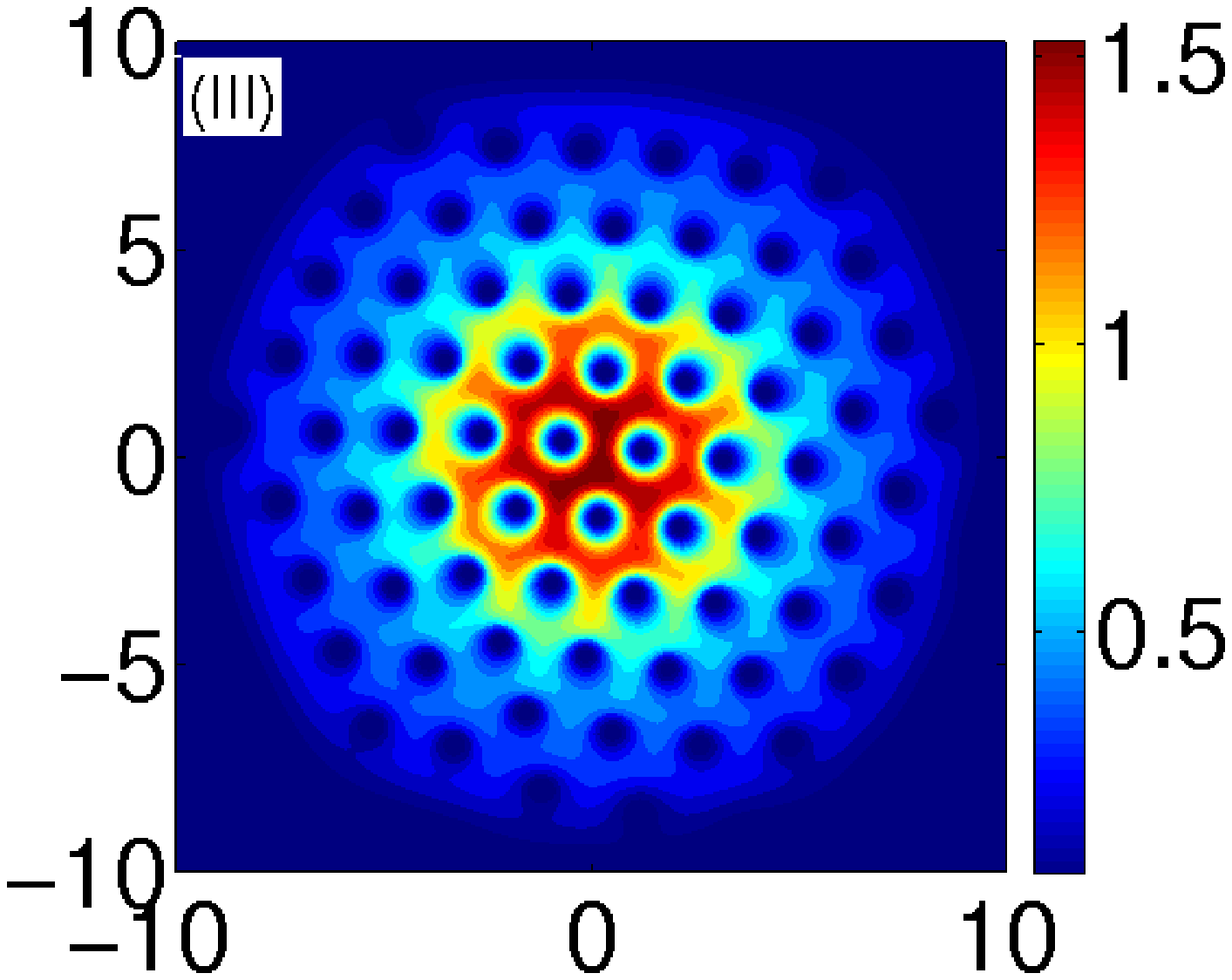}\\
\includegraphics[scale=0.25]{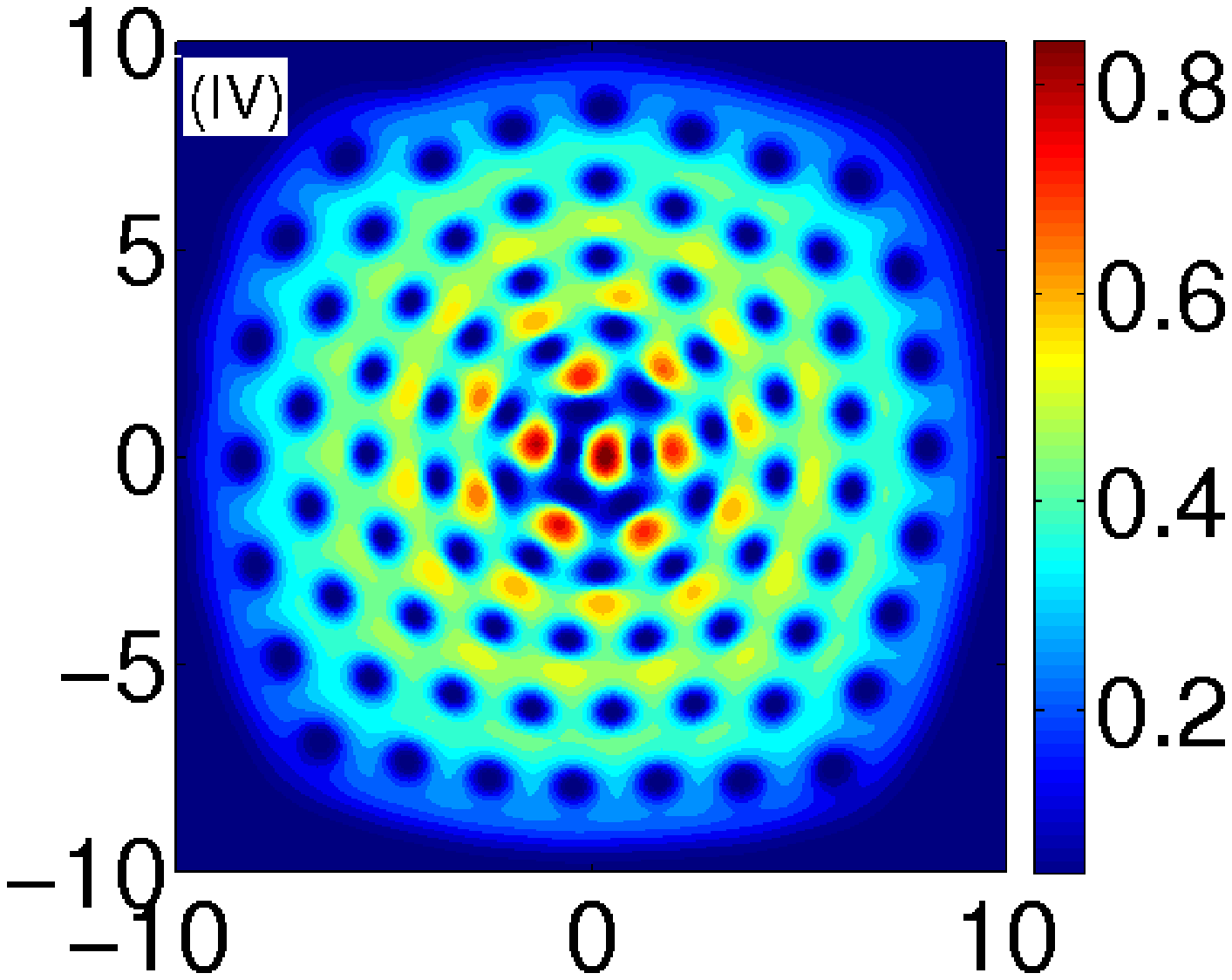}
\includegraphics[scale=0.25]{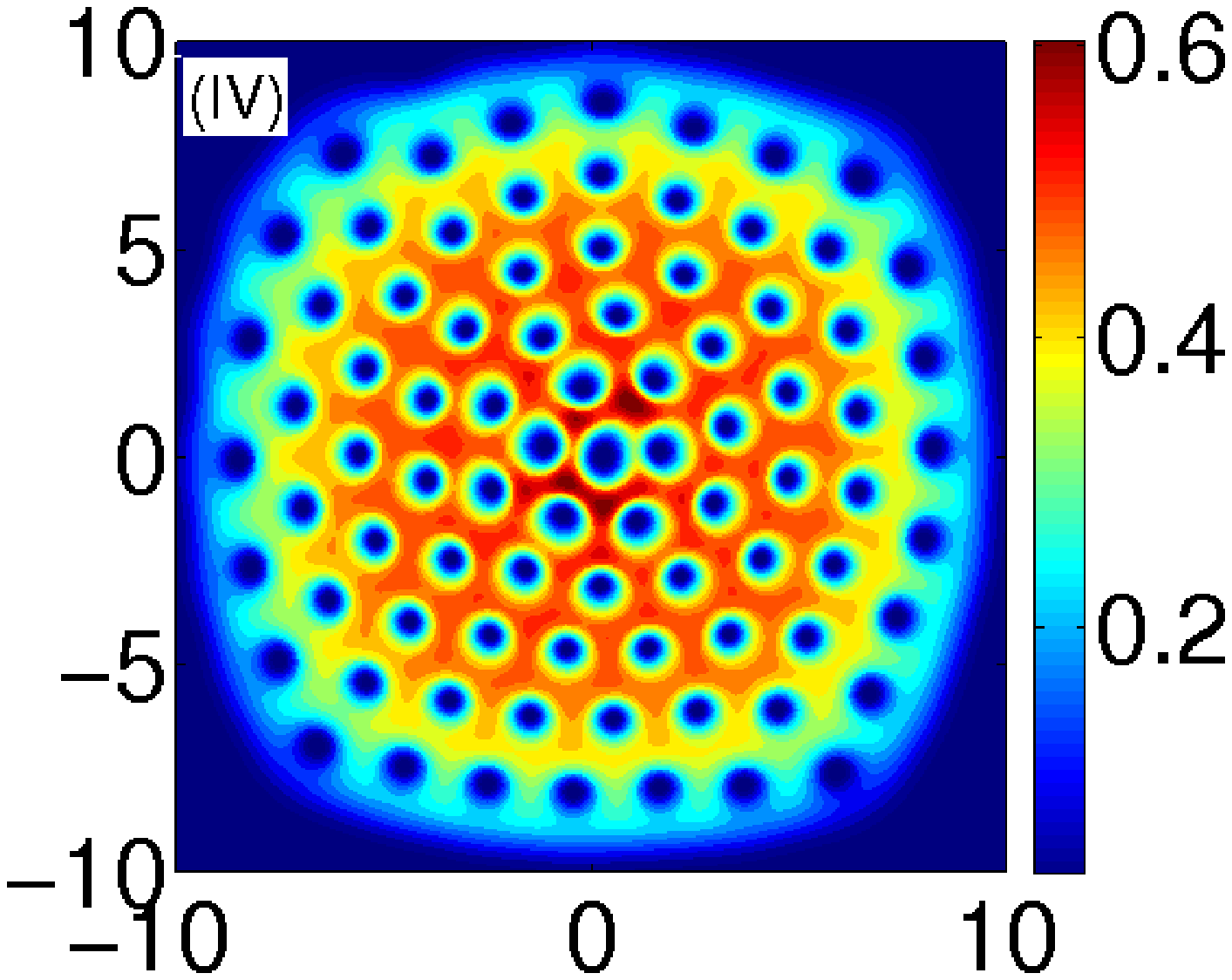}
\end{center}
\begin{picture}(0,0)(10,10)
\put(-53,23) {{$x$}}
\put(63,23) {{$x$}}
\put(-102,331) {{$y$}}
\put(-102,245) {{$y$}}
\put(-102,159) {{$y$}}
\put(-102,72) {{$y$}}
\end{picture}
\caption{(Color online) Density plots (left column, component-1 and right column, component-2) with $\Omega=0.9$ and $\delta=1.94$. The numerical simulations are carried out for (I) $\kappa=0$, (II) $\kappa=0.25$, (III) $\kappa=0.5$ and (IV) $\kappa=1$. Note that the density of component-1 for (I) is identically zero.}
\label{trans3}
\end{figure}

\begin{figure}
\begin{center}
\includegraphics[scale=0.25]{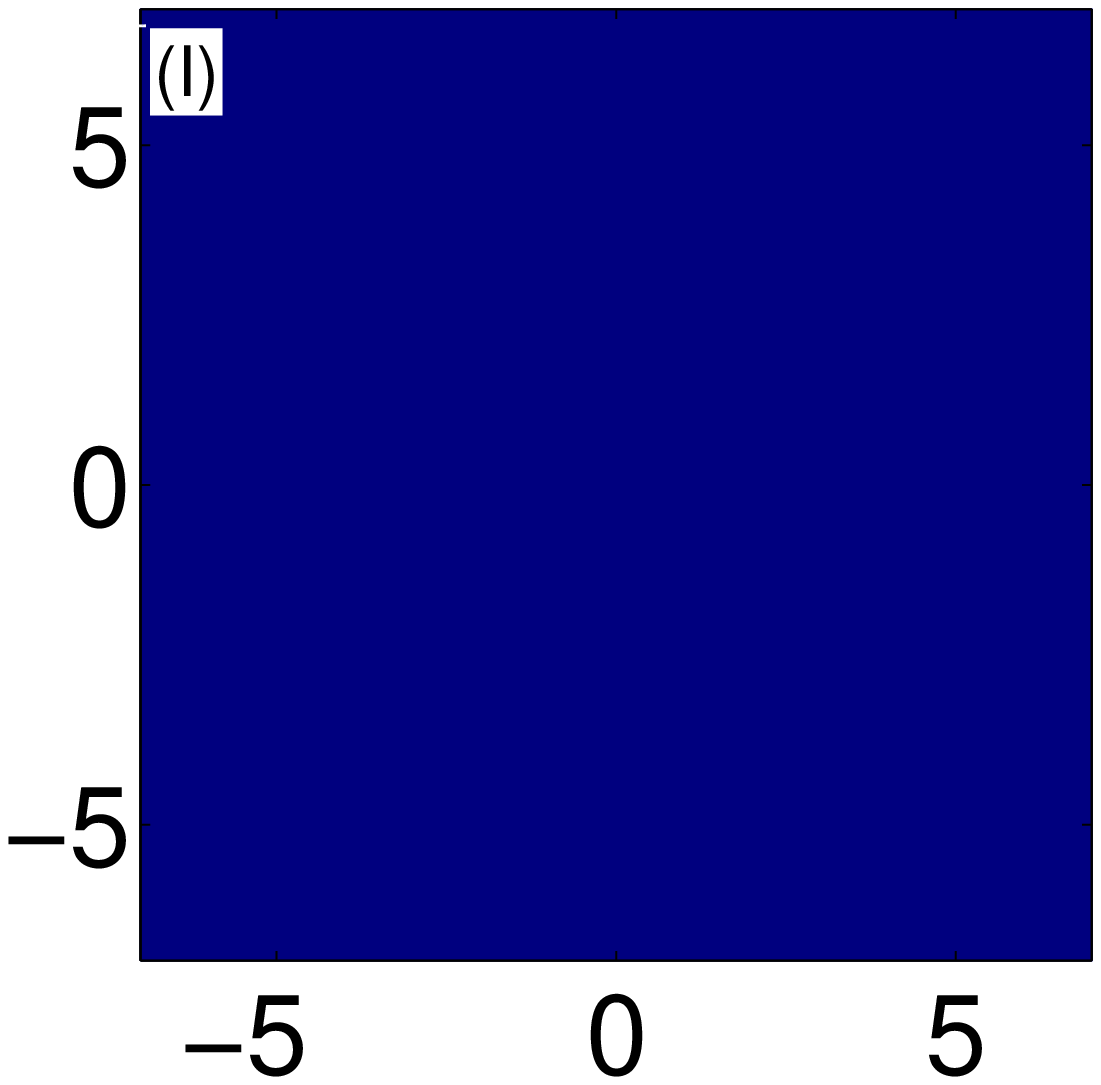}
\includegraphics[scale=0.25]{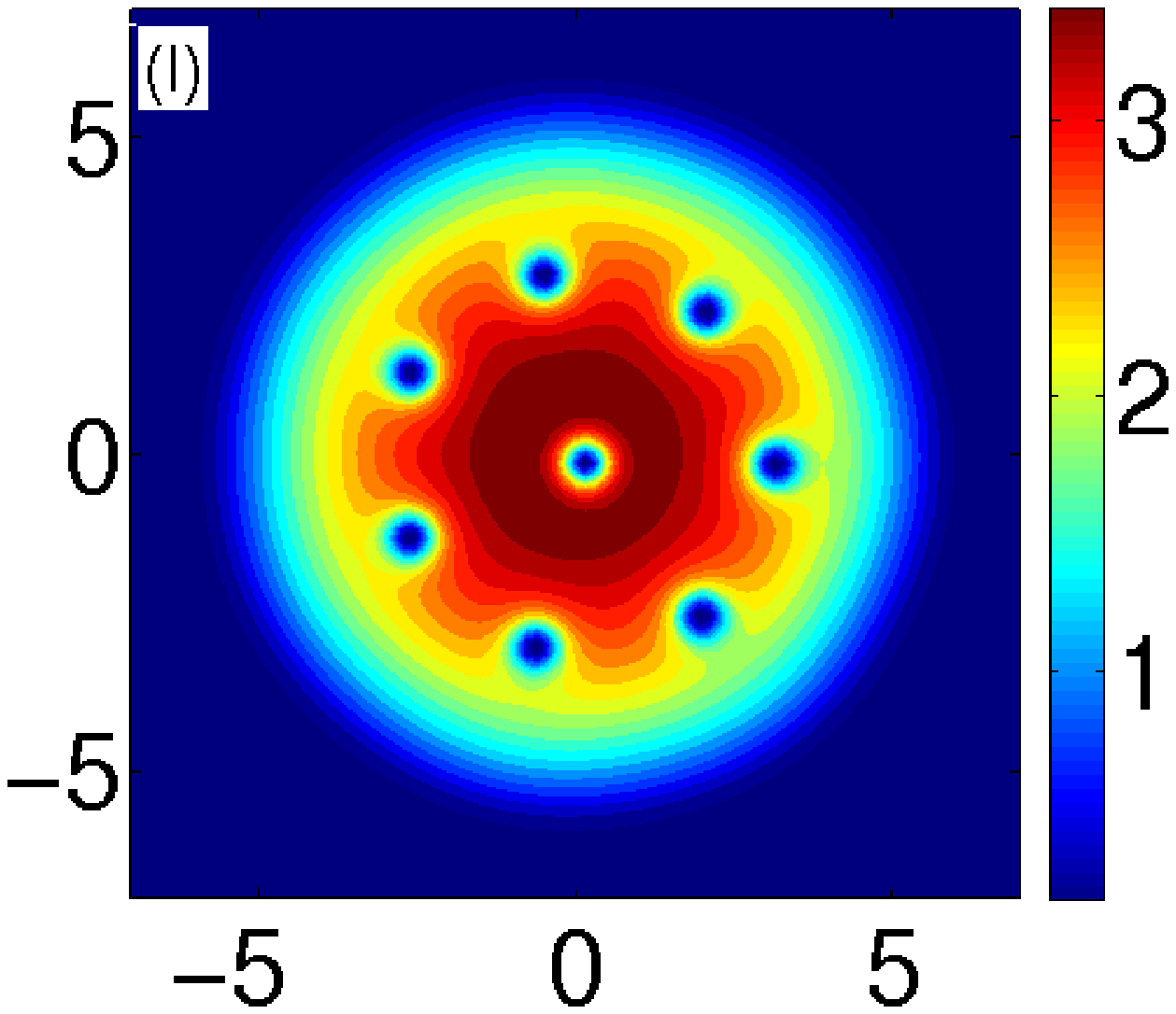}\\
\includegraphics[scale=0.25]{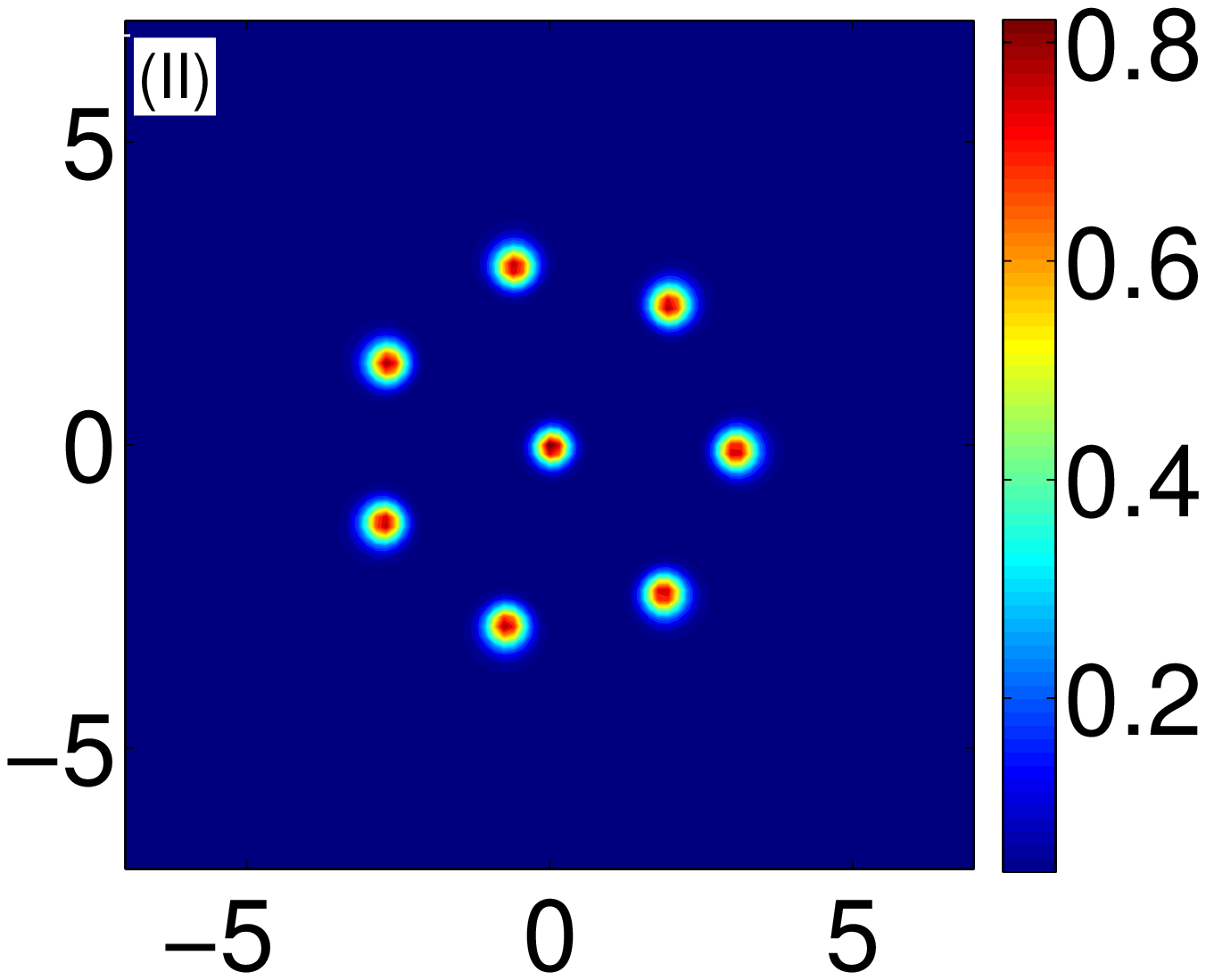}
\includegraphics[scale=0.25]{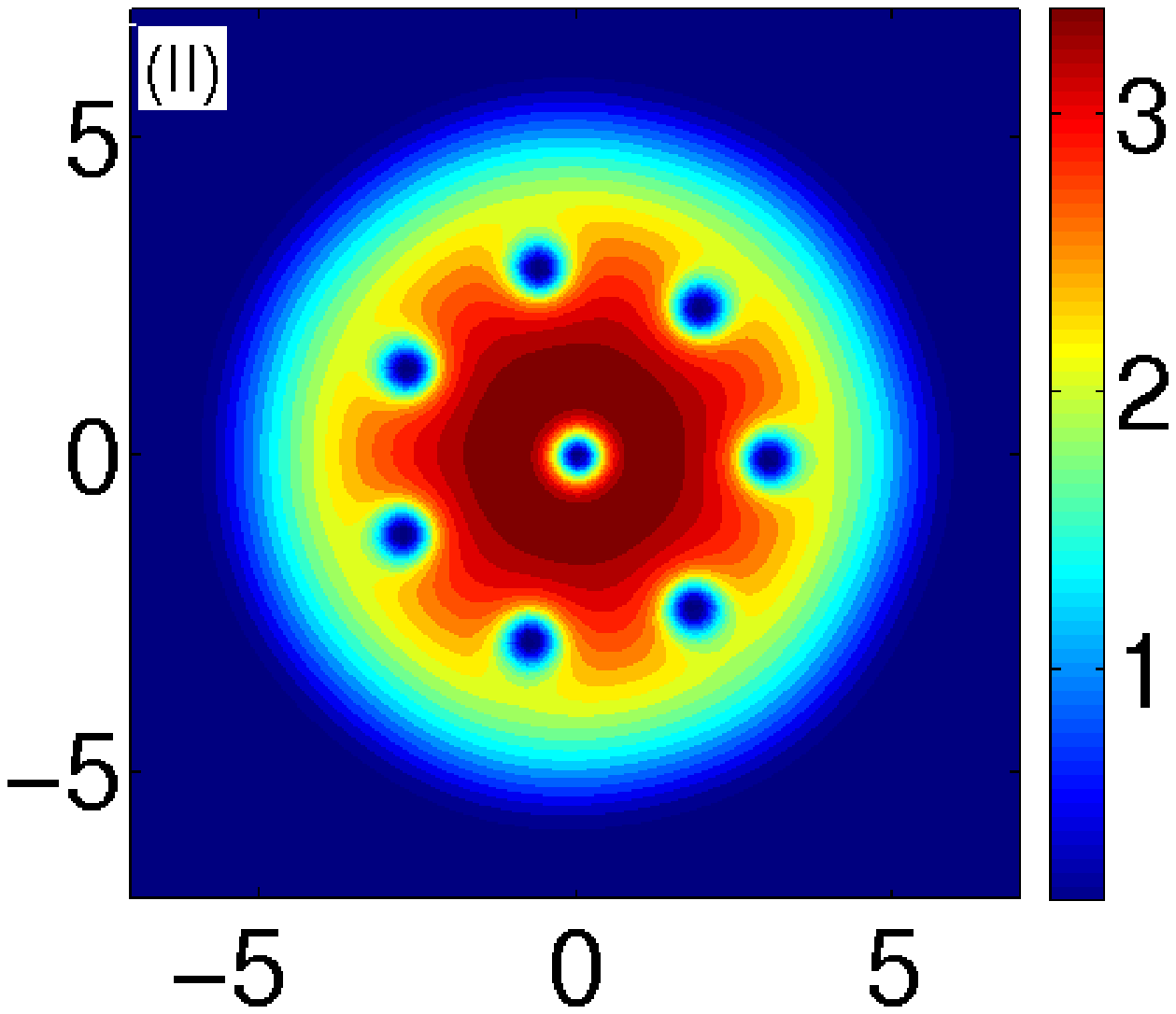}\\
\includegraphics[scale=0.25]{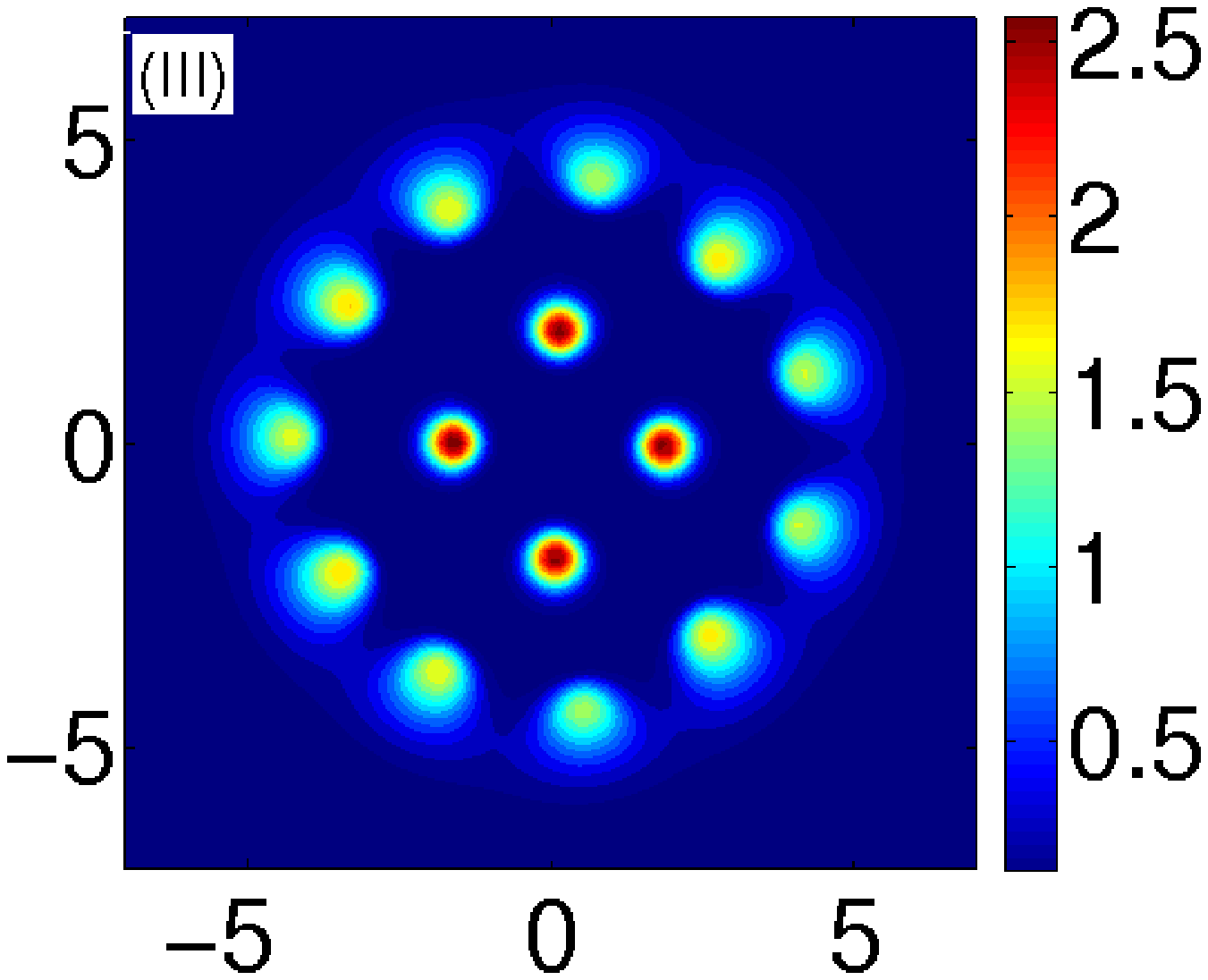}
\includegraphics[scale=0.25]{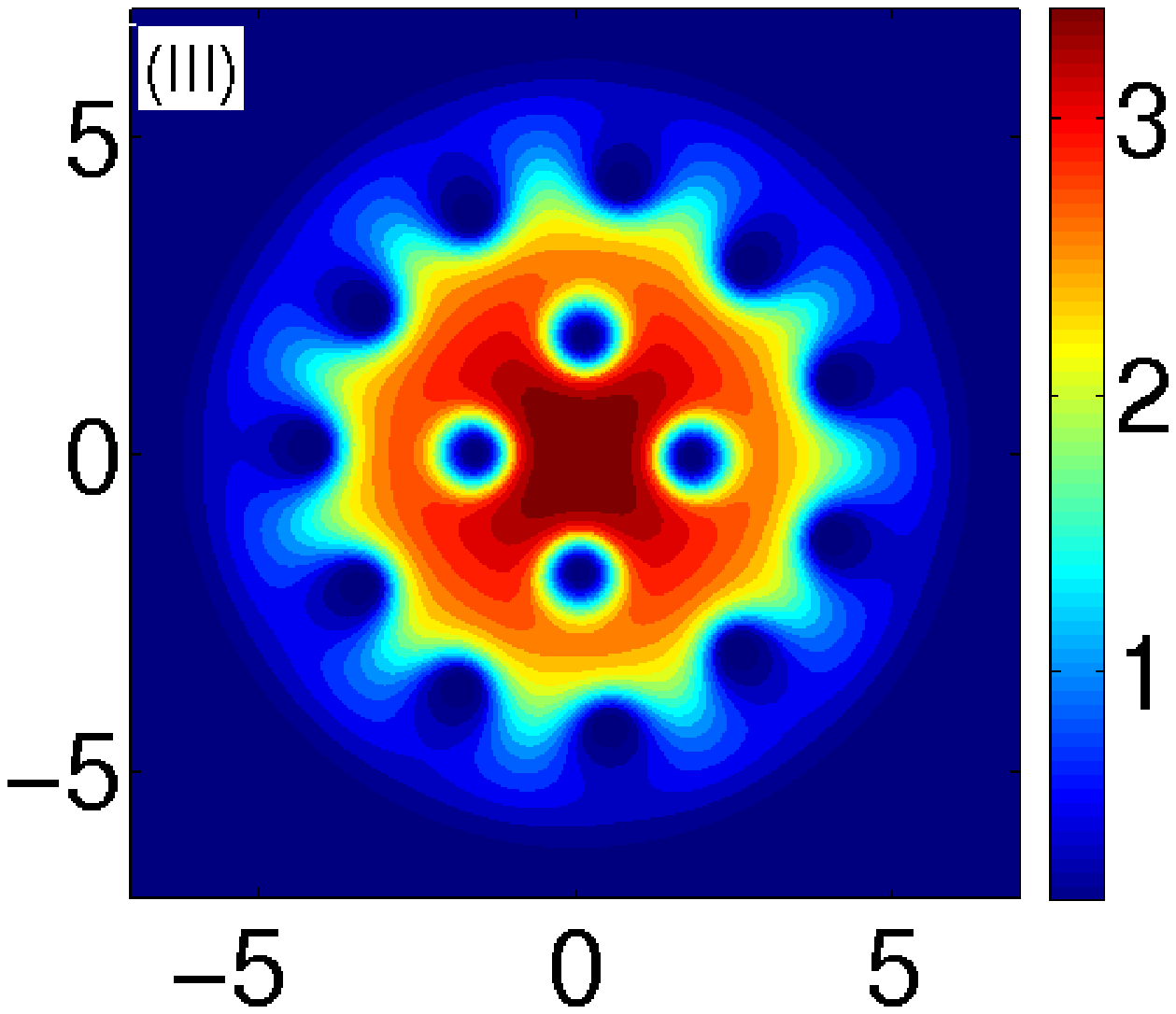}\\
\includegraphics[scale=0.25]{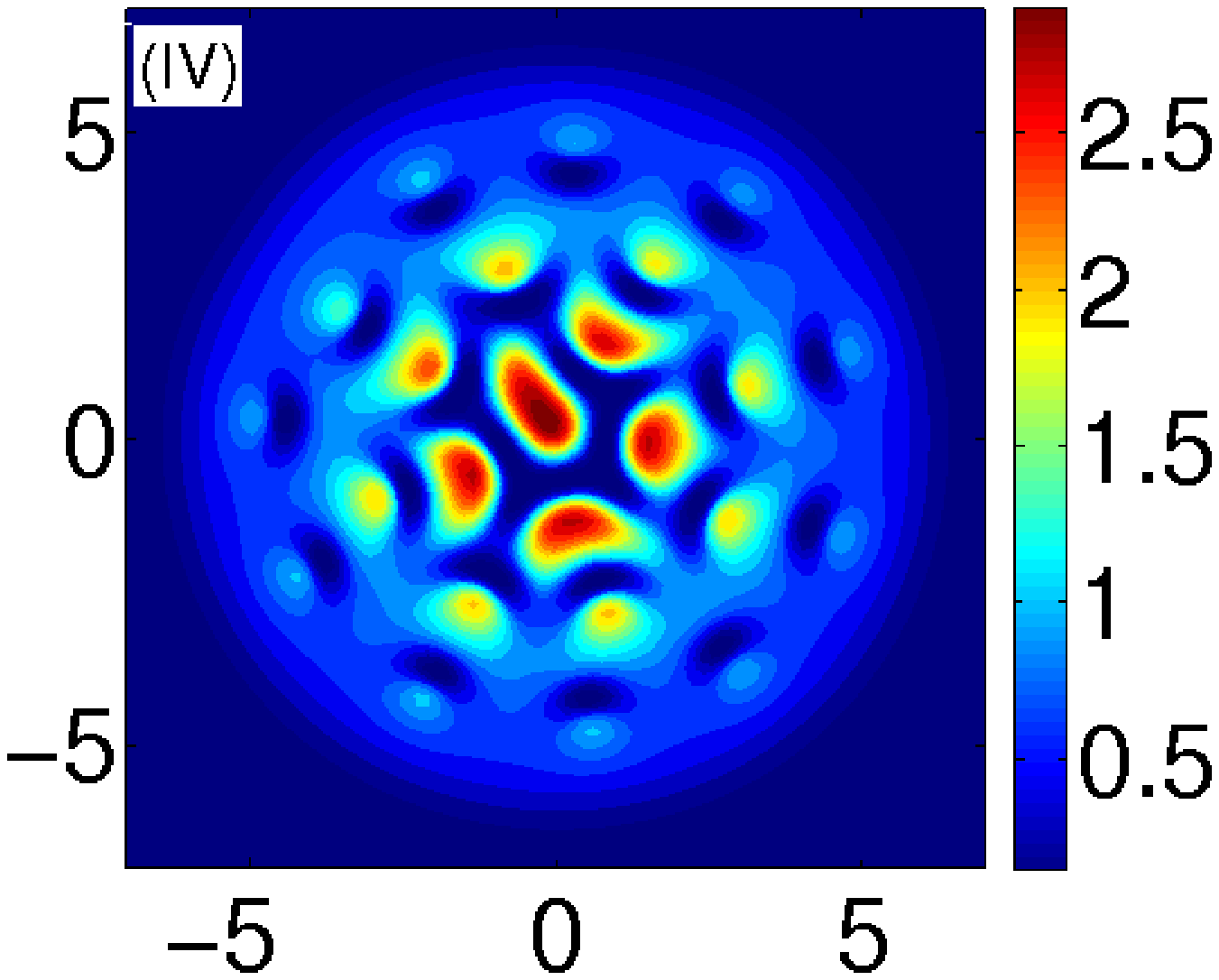}
\includegraphics[scale=0.25]{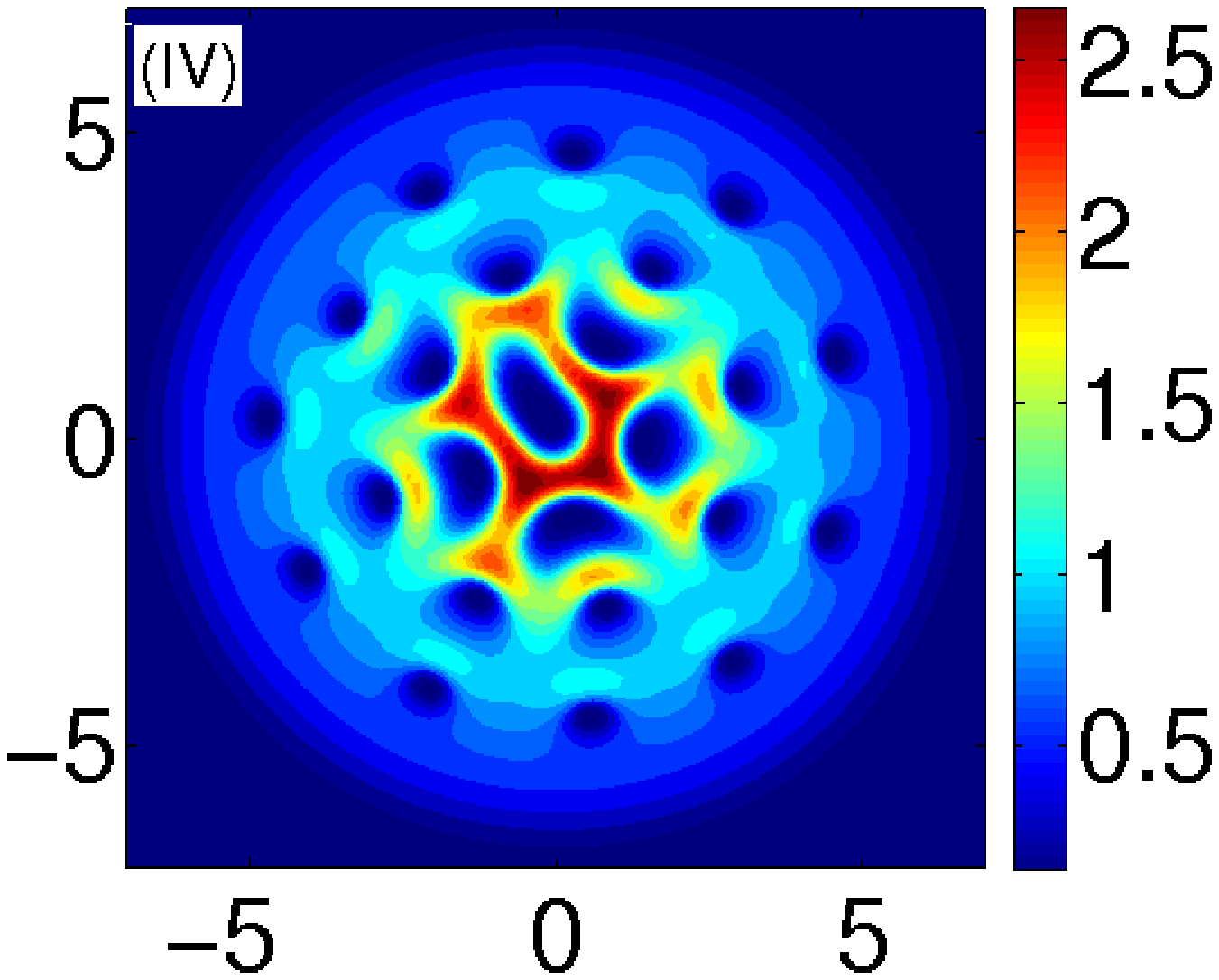}
\end{center}
\begin{picture}(0,0)(10,10)
\put(-53,23) {{$x$}}
\put(63,23) {{$x$}}
\put(-102,331) {{$y$}}
\put(-102,245) {{$y$}}
\put(-102,159) {{$y$}}
\put(-102,72) {{$y$}}
\end{picture}
\caption{(Color online) Density plots (left column, component-1 and right column, component-2) with $\Omega=0.5$ and $\delta=2$. The numerical simulations are carried out for (I) $\kappa=0$, (II) $\kappa=0.25$, (III) $\kappa=0.75$ and (IV) $\kappa=1.75$. Note that the density of component-1 for (I) is identically zero.}
\label{trans2}
\end{figure}

As $\delta$ crosses $1$ for small $\kappa$, there is a transition from $\bm{S}\approx(0,0,1)$ to $\bm{S}=(S_x,S_y,0)$ where the $S_x$ and $S_y$ components of the spin density are in general non-zero. Figure \ref{prof_d_0.9}(a) plots the component densities and spin densities for region (v) of the $\Omega=0.9$ phase diagram. But if $\delta>1$, and $\kappa$ increases, the region where $S_z=1$ gets smaller and eventually disappears, at which point the annulus develops.
  While in Fig. \ref{prof_b_0.1} the $S_x$ and $S_y$ are almost constant, in Fig. \ref{prof_d_0.9} the $S_x$ and $S_y$ are sine/cosine-like functions. We will show this to be the case later, but we note for now that, in essence, we see a smooth sine/cosine-like function for $S_x$ and $S_y$ in the rotation dominating regime, whereas in the spin-orbit dominating regime we see $S_x$ and $S_y$ becoming constants with sharp transitions over boundary lines (that correspond to the lines of vortices and the definition used in this paper for the domain boundary). The vortices of each component correspond to singularities in the
 $S_x$, $S_y$, $S_z$ components: pairs of upwards and downwards spikes.

\begin{figure}
\begin{center}
\includegraphics[scale=0.25]{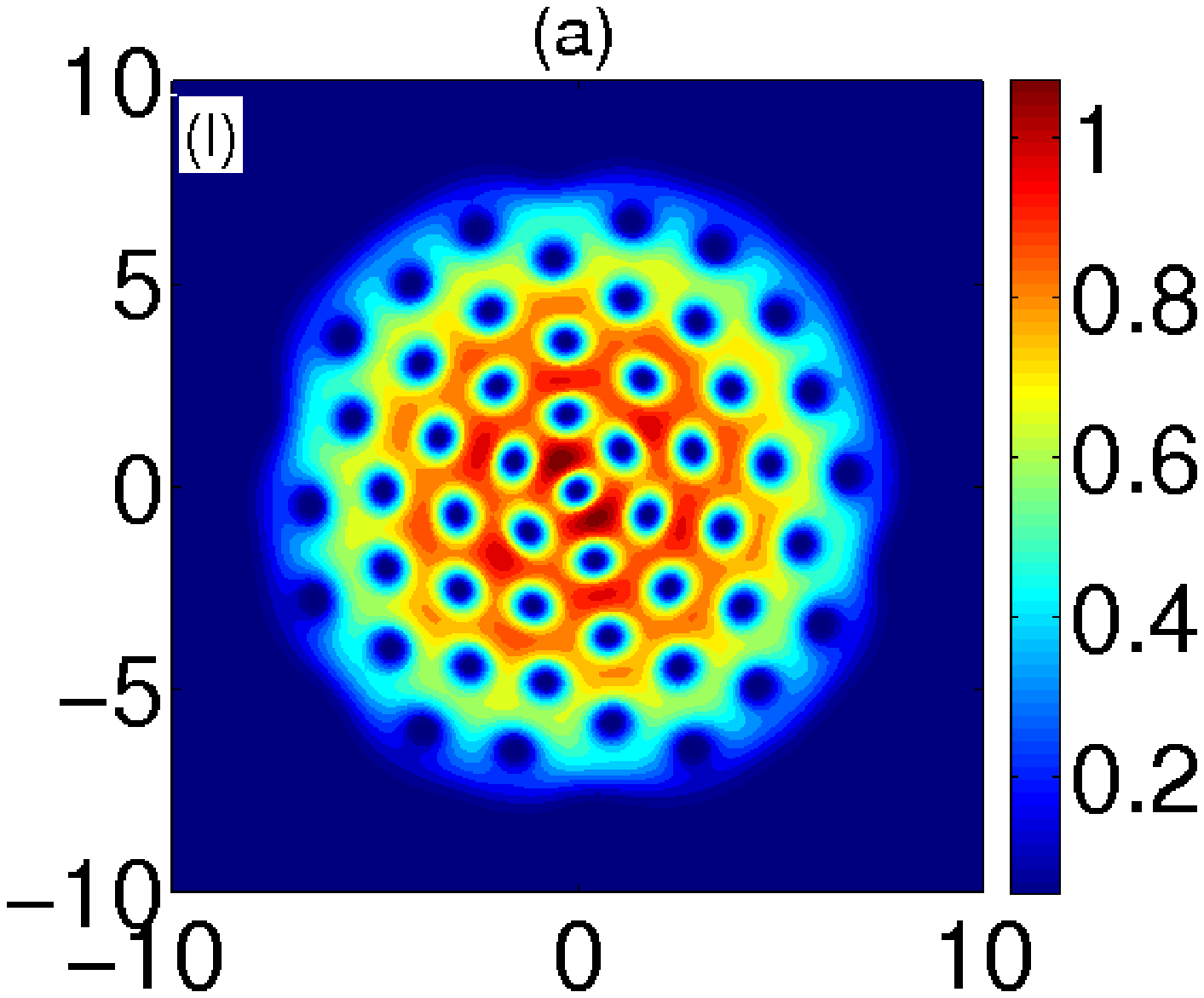}
\includegraphics[scale=0.25]{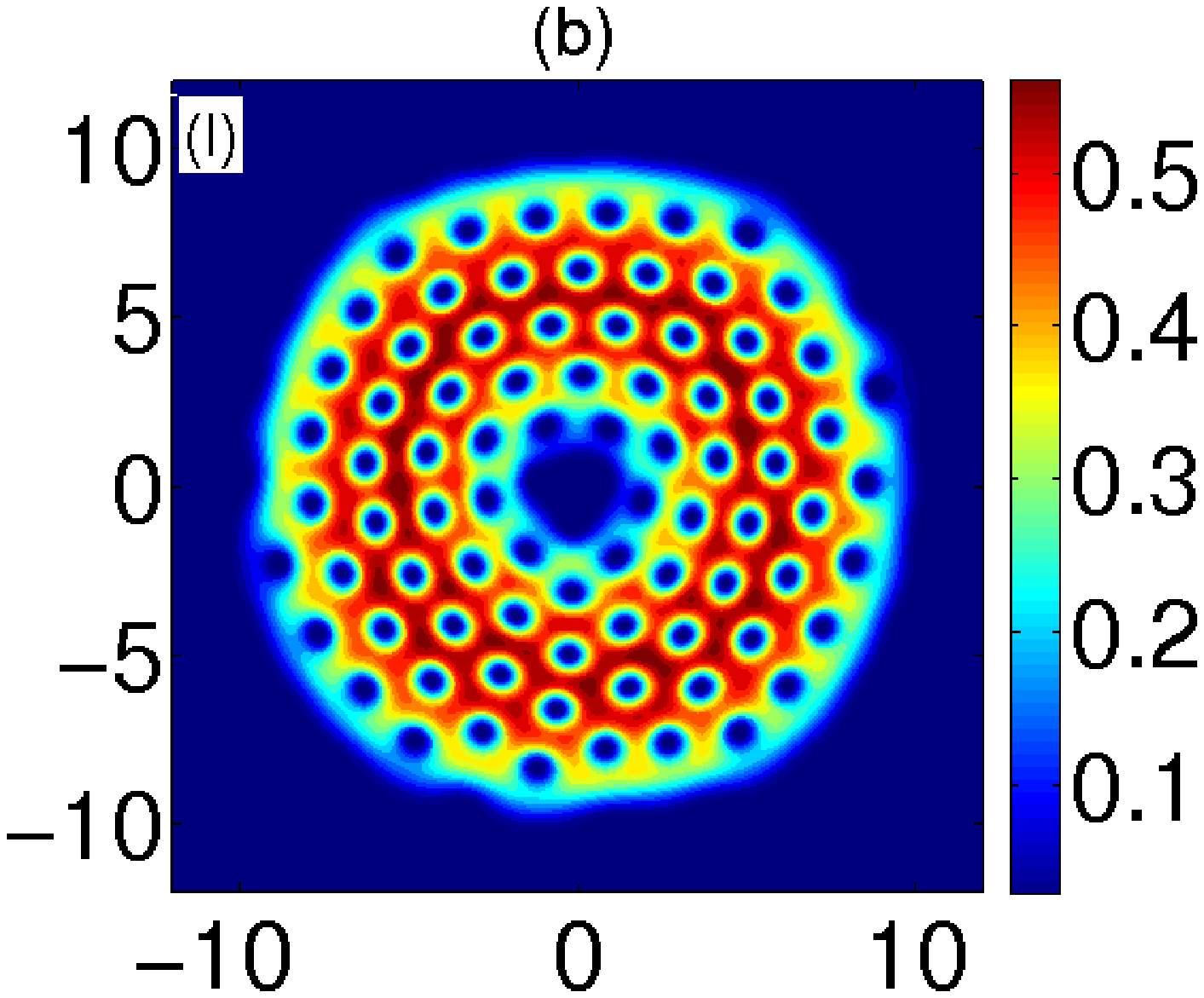}\\
\includegraphics[scale=0.25]{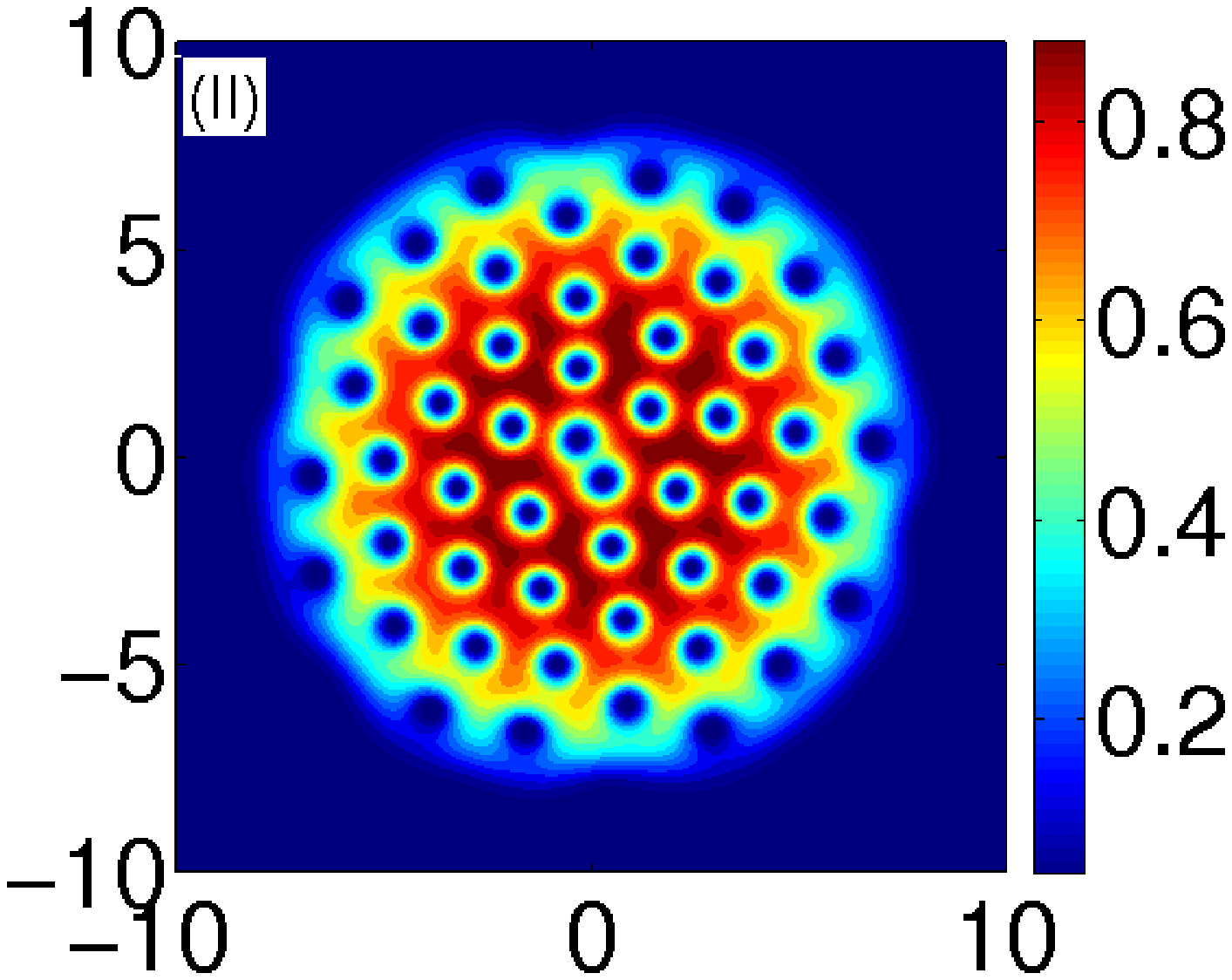}
\includegraphics[scale=0.25]{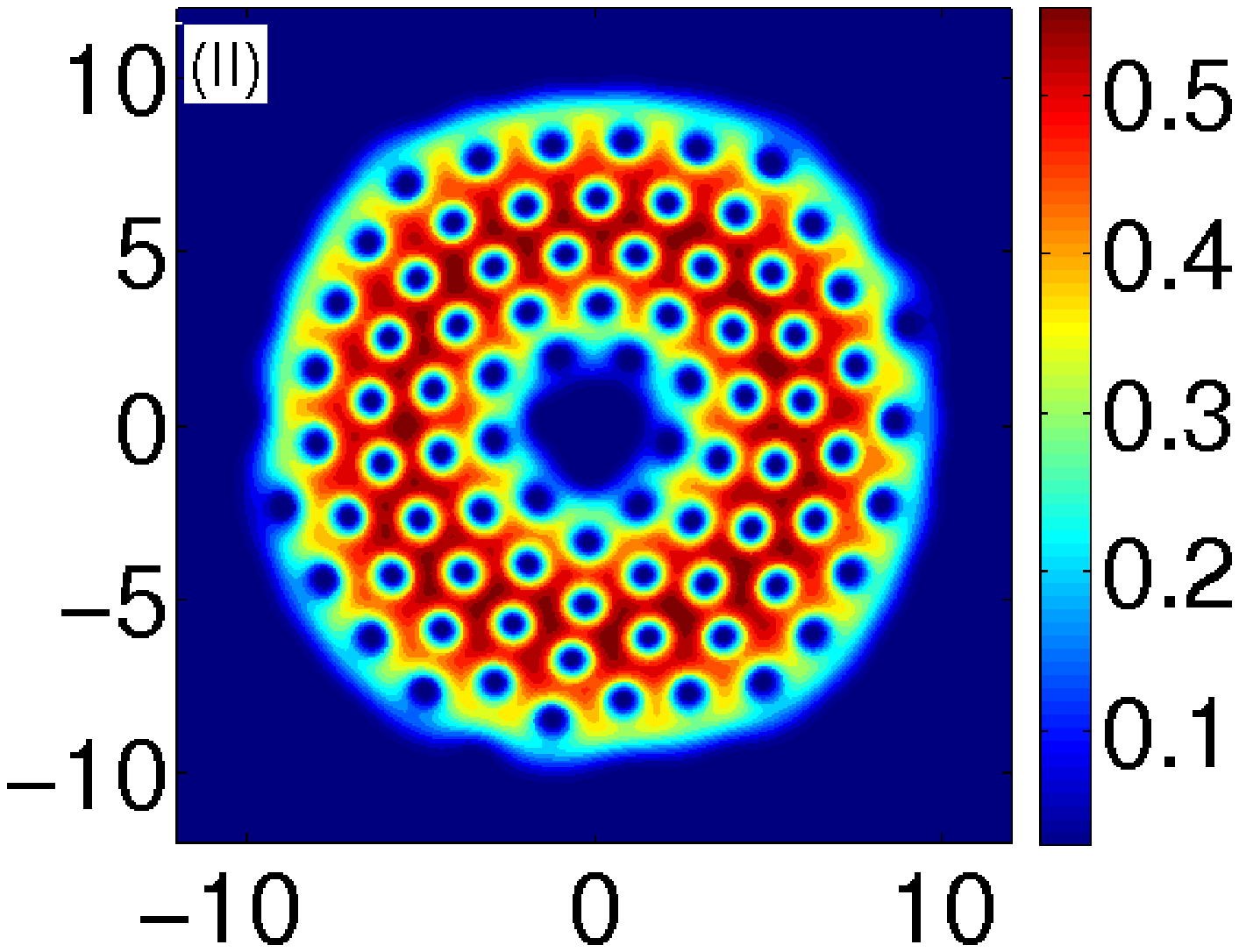}\\
\includegraphics[scale=0.25]{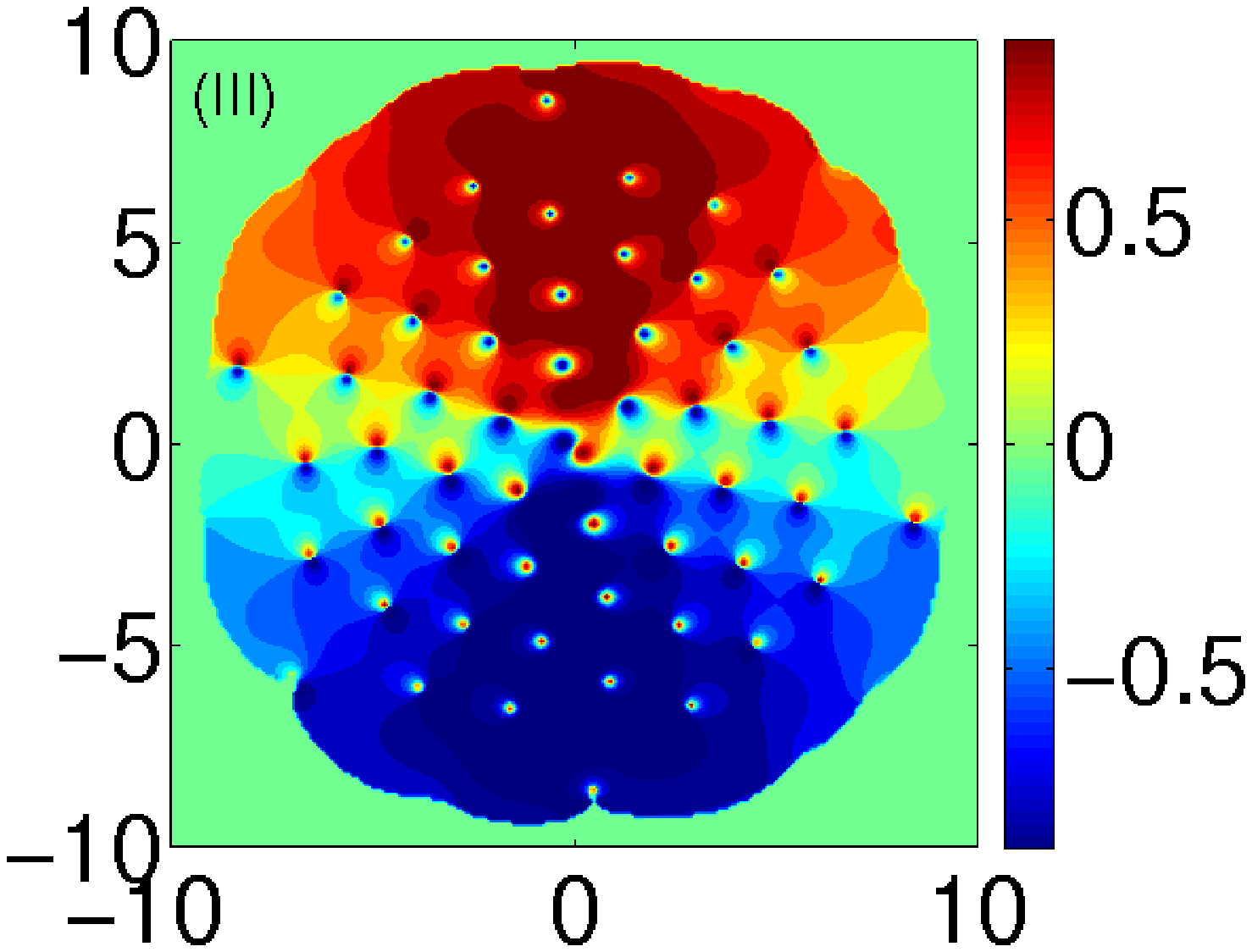}
\includegraphics[scale=0.25]{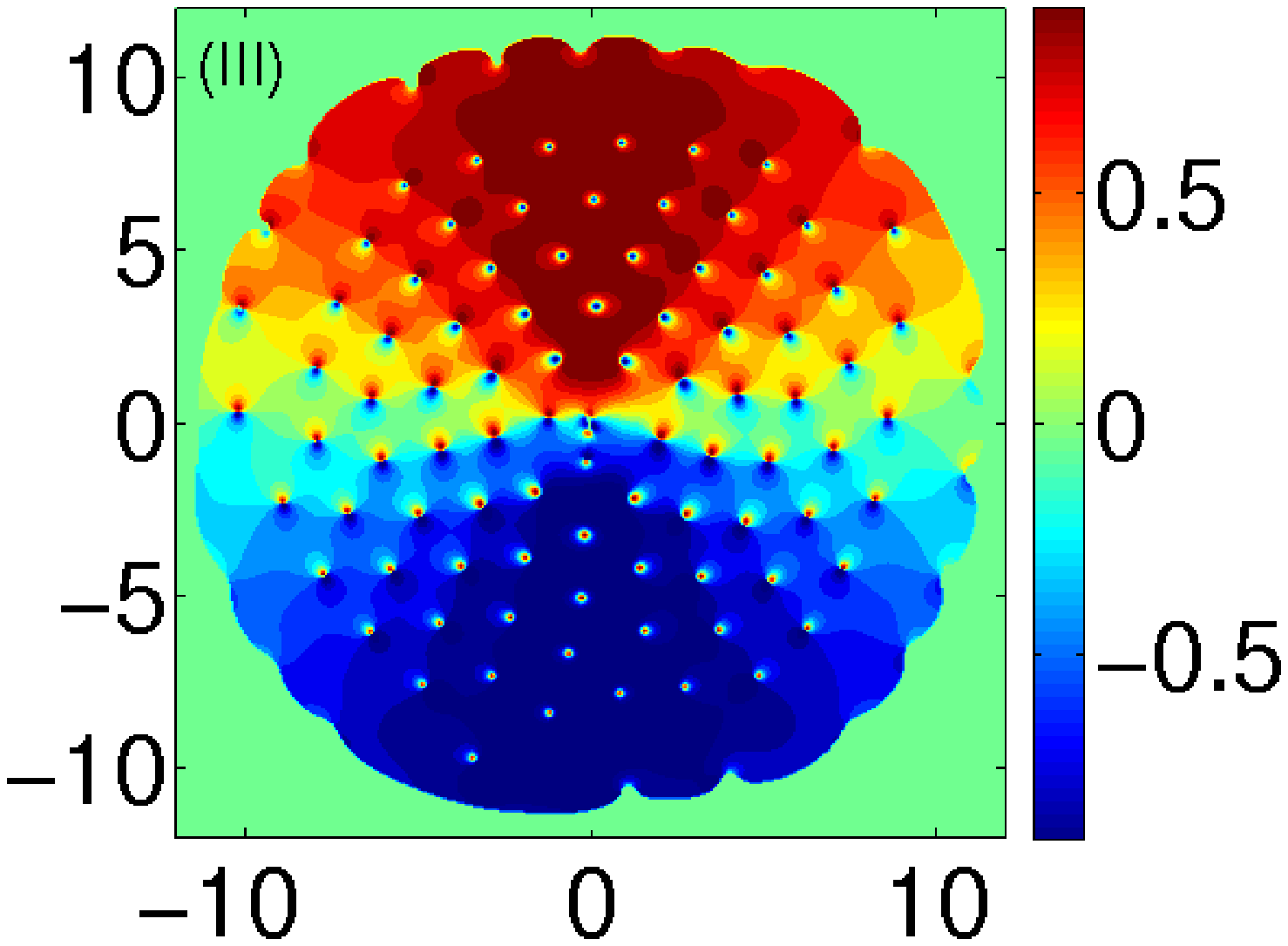}\\
\includegraphics[scale=0.25]{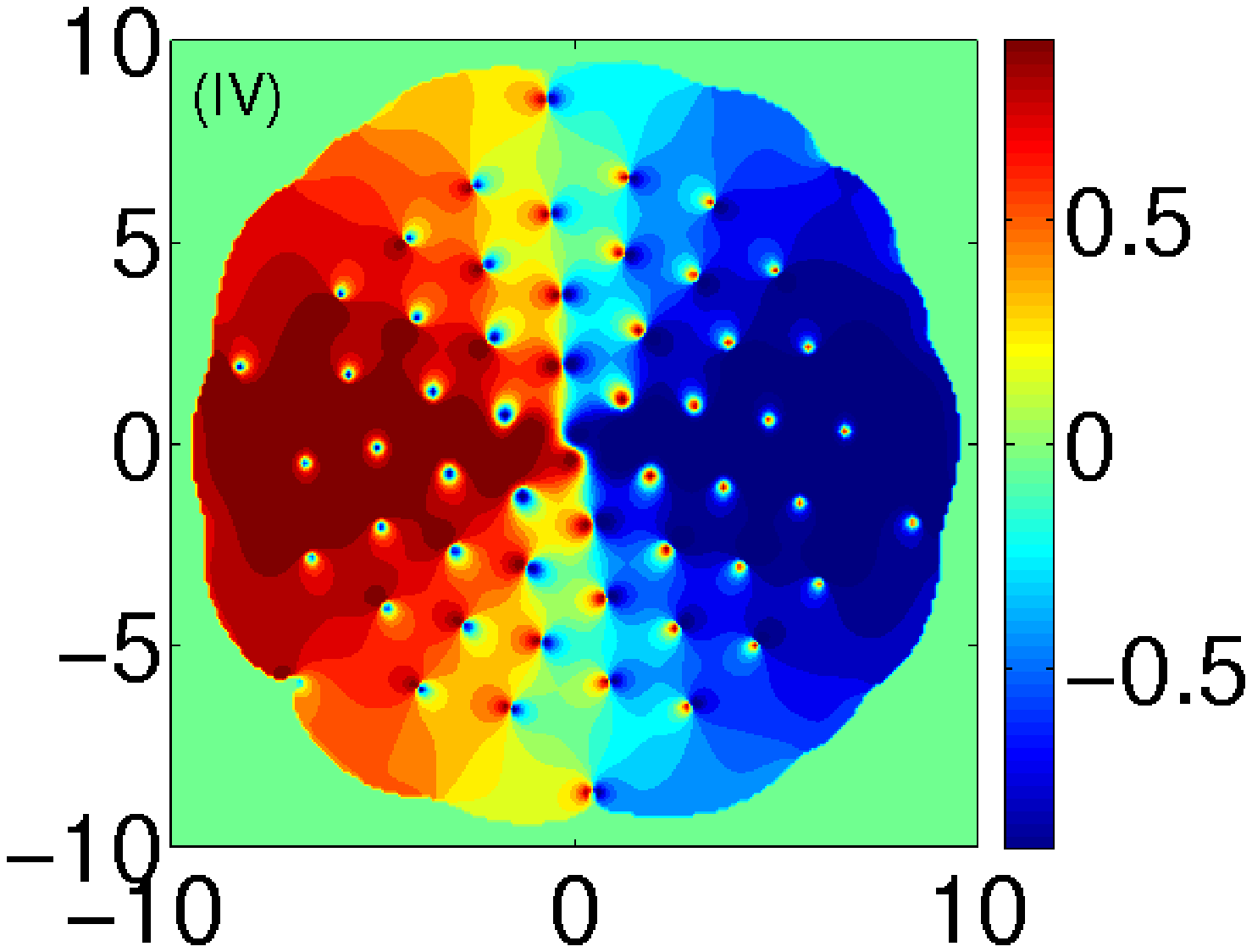}
\includegraphics[scale=0.25]{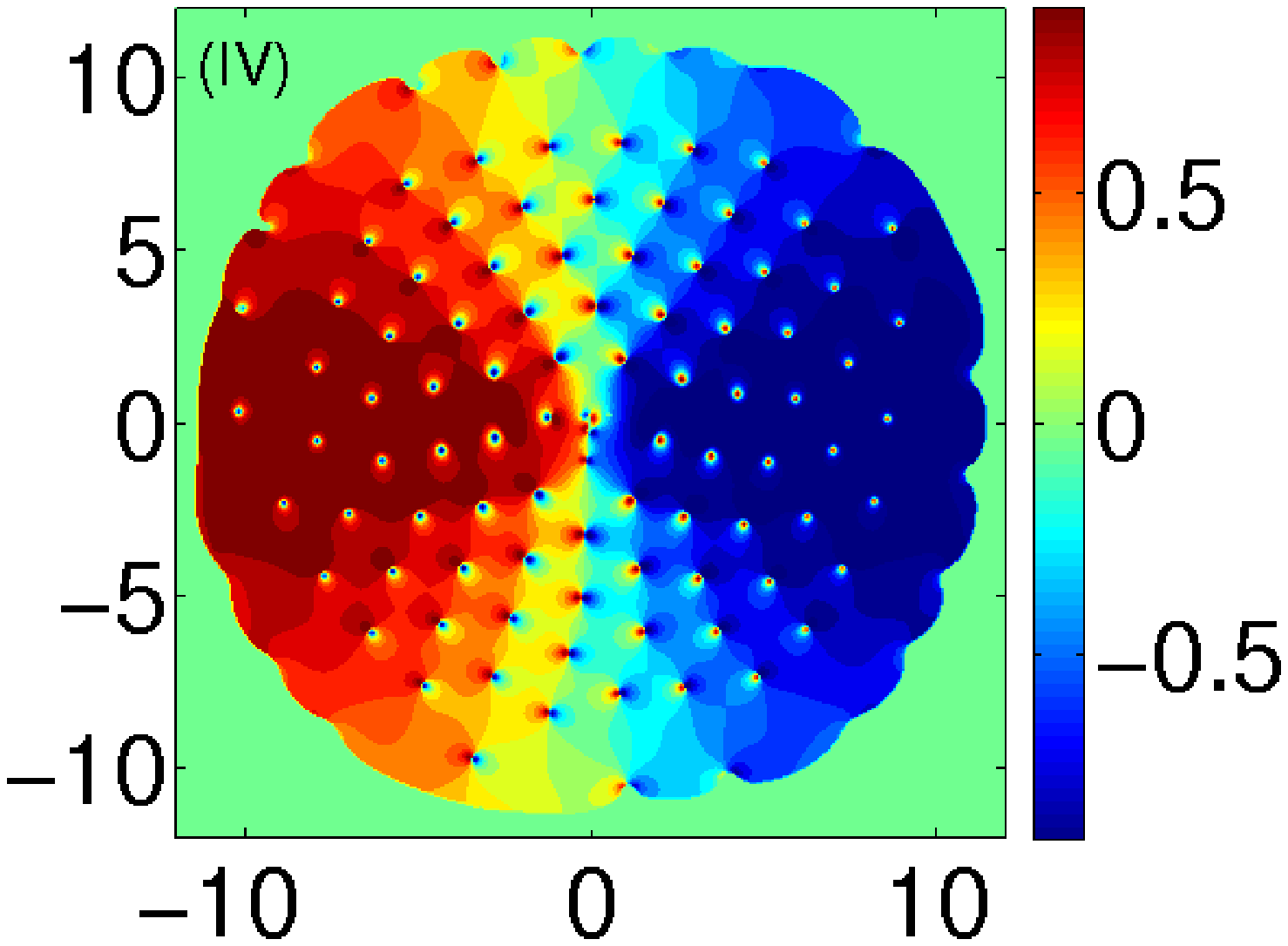}\\
\includegraphics[scale=0.25]{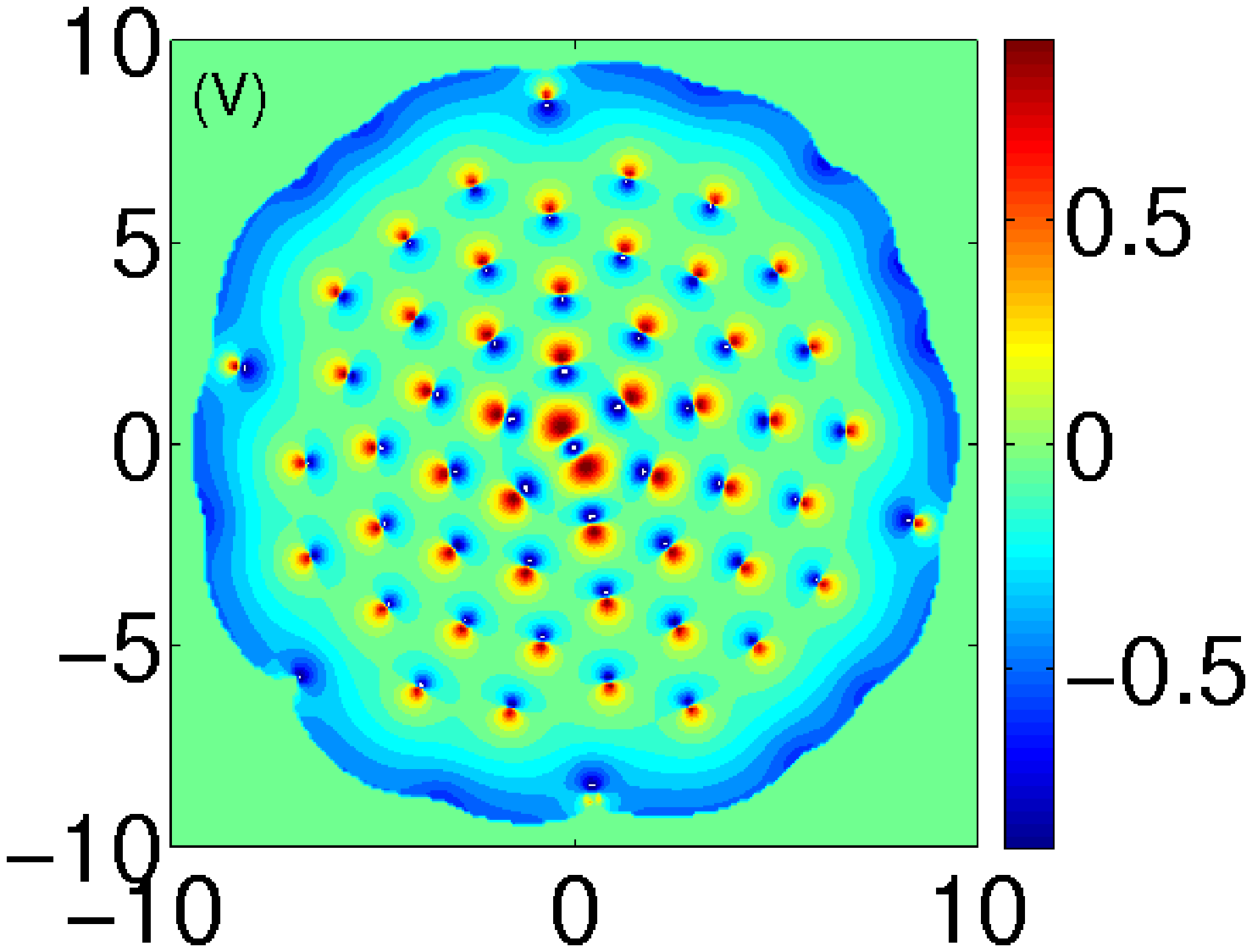}
\includegraphics[scale=0.25]{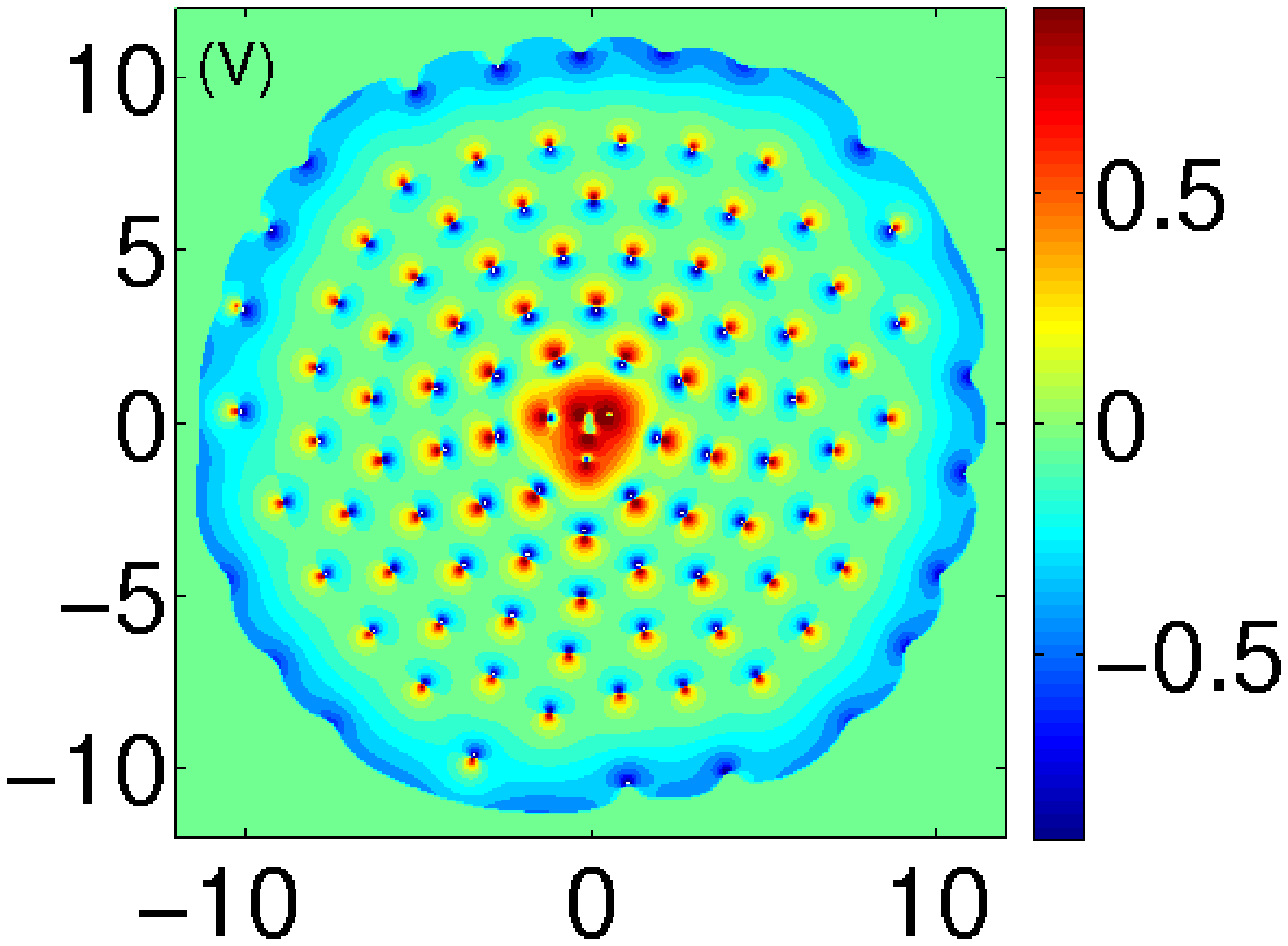}
\end{center}
\begin{picture}(0,0)(10,10)
\put(-53,23) {{$x$}}
\put(63,23) {{$x$}}
\put(-102,416) {{$y$}}
\put(-102,331) {{$y$}}
\put(-102,245) {{$y$}}
\put(-102,159) {{$y$}}
\put(-102,72) {{$y$}}
\end{picture}
\caption{(Color online) Numerical simulations  for regions (v) and (vii) of the $\kappa-\delta$ with $\Omega=0.9$ phase diagram of Fig. \ref{pd_omega0.9} for (a) $(\delta,\kappa)=(0.25,0.5)$ (left column) and (b) $(0.25,1.25)$ (right column) respectively. Density plots (frame (I), component-1, and (II), component-2) and spin density plots (frame (III), $S_x$, frame (IV), $S_y$ and frame (V), $S_z$).}
\label{prof_d_0.9}
\end{figure}

As $\kappa$ is increased, one can see the development of two annular components. These annular components still preserve the vortex lattices, and the sine/cosine-like form of the spin density - see Fig. \ref{prof_d_0.9}(b). In the next section we find an analytical expression for the critical parameters at which the geometry changes from two disks to two annuli. This analytical result (dashed line) can be compared to the numerical simulations (solid line) in Fig. \ref{pd_omega0.9}.

\subsection{D. $\Omega$-$\delta$ phase diagrams}

We have up to now only presented $\kappa$-$\delta$ phase diagrams with the value of the rotation held constant (either $\Omega=0$ for Fig. \ref{summ}, $\Omega=0.1$ for Fig. \ref{pd_omega0.1} or $\Omega=0.9$ for Fig. \ref{pd_omega0.9}). In addition to these we present two $\Omega$-$\delta$ phase diagrams with the value of $\kappa$ held constant: in Fig. \ref{pd_kap}(a) we take $\kappa=1$ (small) and in Fig. \ref{pd_kap}(b) $\kappa=8$ (large).

\begin{figure}
\begin{center}
\includegraphics[scale=0.3]{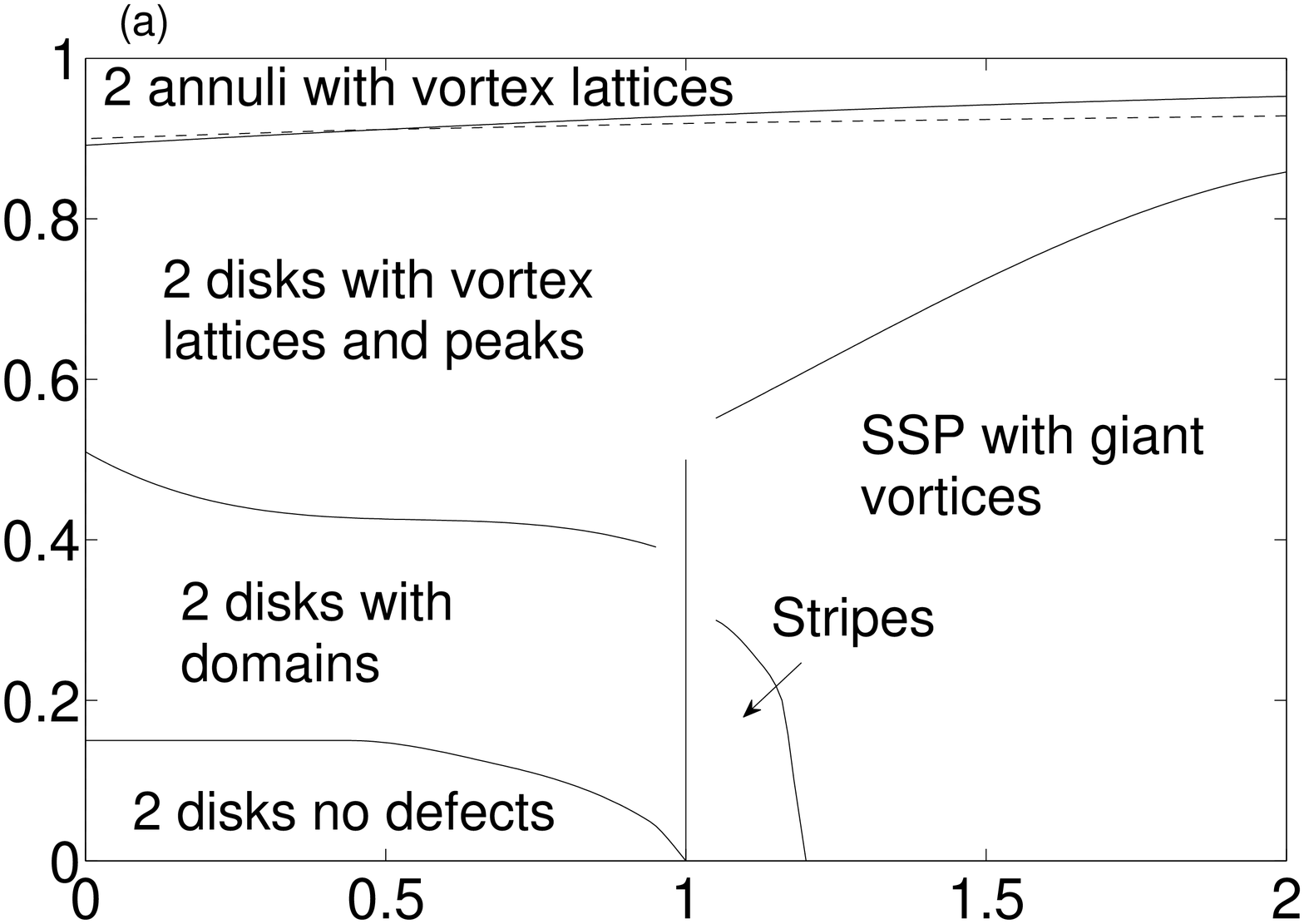}\\
\includegraphics[scale=0.3]{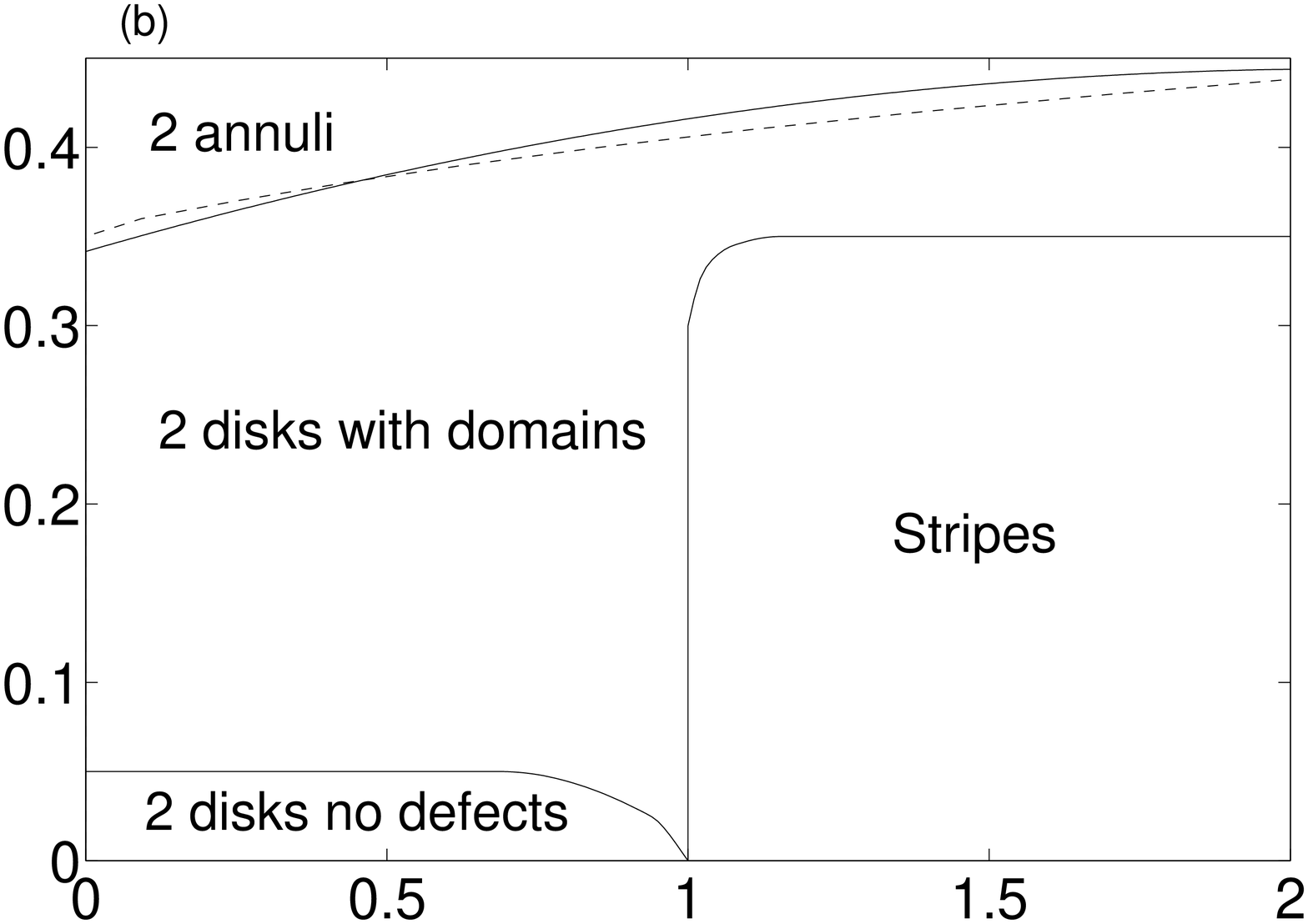}
\end{center}
\begin{picture}(0,0)(10,10)
\put(20,28) {{$\delta$}}
\put(-108,125) {{$\Omega$}}
\put(20,200) {{$\delta$}}
\put(-108,290) {{$\Omega$}}
\end{picture}
\caption{$\Omega-\delta$ phase diagrams with (a) $\kappa=1$ and (b) $\kappa=8$. The numerical parameters are taken as $g=4$ and $N=200$. There are six identified regions: (i) two disks with no defects, (ii) two disks with domains, (iii) segregated symmetry preserving (SSP) with a giant vortex, (iv) stripes, (v) two disks with vortex lattices and peaks and (vii) two annuli with vortex lattices. The boundary between regions (v) and (vii) is plotted according to the numerical simulations (solid line) and analytically (dashed line), calculated according to Eq. (\ref{crit_val}).}
\label{pd_kap}
\end{figure}


\section{IV. Analysis in the Thomas-Fermi regime}

In this Section we perform an analysis based on the Thomas-Fermi approximation to describe various features of the phase diagrams that we introduced in the previous section. In particular, we will concentrate on the symmetry preserving ground states (those featured in regions (i)-(iii), (v) and (vii) of the phase diagrams shown in Fig.'s \ref{summ}, \ref{pd_omega0.1} and \ref{pd_omega0.9}). Our starting point is the energy functional [Eq. (\ref{ener2})] written in terms of the total density and the spin density, for which we consider, as in the numerical simulations, $g_1=g_2$ which gives $c_1=0$:
\begin{equation}
		\label{ener2_2}
\begin{split}
 E=&\int\frac{1}{2}\left(\nabla\sqrt{\rho}\right)^2+\frac{\rho}{8}\left(\nabla\bm{S}\right)^2+\frac{\rho}{2}
 \left(\bm{v}_{\text{eff}}-\bm{\Omega}\times\bm{r}\right)^2\\
		&+\rho\kappa\left(\bm{S_{\perp}}\cdot \bm{v}_{\text{eff}}+\frac{1}{2}\bm{S}\cdot\nabla\times\bm{S}\right)\\
		&+\frac{\rho}{2}(1-\Omega^2)r^2+(c_0+c_2S_z^2)\frac{\rho^2}{2}\qquad d^2{r}.
\end{split}
\end{equation}
 Under the assumption that $gN$ is large, we are in the Thomas-Fermi limit, which allows us to make various approximations regarding the importance of the individual terms in Eq. (\ref{ener2_2}). We divide our analysis into looking at the cases of zero rotation and non-zero rotation which we further divide into low and high rotation.


\subsection{A. No Rotation}

{We assume that there is no rotation, $\Omega=0$. The phase diagram of Fig. \ref{summ} shows that $\delta=1$ is a critical value. We thus need to look at $\delta<1$ and $\delta>1$ separately.}

\subsubsection{1. $\delta<1$}Since $\delta<1$, that is $c_2>0$, we can assume in the energy (\ref{ener2_2}), that
$c_2S_z^2$ is negligible in front of $c_0$. The fact that $S_z$ is negligible (which can be seen in Fig. \ref{phase_00}),
 also implies that  $\bm{v}_{\text{eff}}\sim \nabla\Theta/2$ (we will see that $\nabla \Theta$ is of order 1). This leads to
\begin{equation}
		\label{ener_omega0}
\begin{split}
	 E=&\int\frac{1}{2}\left(\nabla\sqrt{\rho}\right)^2+\frac{\rho}{2}\left(\frac{1}{4}
\left(\nabla\bm{S}\right)^2+\kappa\bm{S}\cdot\nabla\times\bm{S}\right)\\		 &+\frac{\rho}{8}\left(\nabla\Theta\right)^2+\frac{\rho}{2}\kappa\bm{S_{\perp}}\cdot\nabla\Theta+
\frac{\rho}{2}r^2+c_0\frac{\rho^2}{2}\quad d^2{r}.
\end{split}
\end{equation}
This energy leads to two orders of magnitude, one for $\rho$ and the other for $\bm{S}$ and $\Theta$.
 We will see that the $\rho$ energy is of order $N^{3/2} \sqrt{c_0}$, which is large in
 the Thomas-Fermi limit, while the energy for $\bm{S}$ and $\Theta$ is of order $N \kappa^2$,
  which is much smaller than $N^{3/2} \sqrt{c_0}$, since $\kappa$ is of order 1.

  Thus, when $\kappa^2\ll \sqrt{N c_0}$, we can separately
 minimize
\begin{subequations}
\begin{equation}
	\label{rho_eq}
E_{\rho}=\int\frac{1}{2}\left(\nabla\sqrt{\rho}\right)^2+\frac{\rho}{2}r^2+c_0\frac{\rho^2}{2}\quad d^2{r},
\end{equation}
and
\begin{eqnarray}
E_{\bm{S},\Theta}&=&\int\frac{\rho}{2}\Bigg(\frac{1}{4}\left(\nabla\bm{S}\right)^2+\kappa\bm{S}\cdot\nabla\times\bm{S}\nonumber\\
&&\qquad+\frac{1}{4}\left(\nabla\Theta\right)^2+\kappa\bm{S_{\perp}}\cdot\nabla\Theta\Bigg)\quad d^2{r}.
\label{sth_eq}
\end{eqnarray}
\end{subequations} Eq. (\ref{rho_eq}) yields the Thomas-Fermi profile for $\rho$ that we will discuss below.
Eq. (\ref{sth_eq}) leads to two coupled problems for $\Theta$ and $\bm S$.

{\it{Thomas-Fermi Density Profiles}}. In the Thomas-Fermi approximation, the minimization of  (\ref{rho_eq}) yields
\begin{equation}
	\label{tf:omega0}
	\rho=\frac{1}{c_0}\left(\mu-\frac{1}{2}r^2\right),
\end{equation}
where $\mu$ is the  chemical potential.
This Thomas-Fermi density profile has a harmonic trapping potential and so the components will always be disk shaped. To complete the analysis we use the normalisation condition to obtain
\begin{subequations}
\begin{eqnarray}
	\label{radius_0}
	R&=&\left(\frac{4Nc_0}{\pi}\right)^{1/4},\\
	\mu&=&\left(\frac{Nc_0}{\pi}\right)^{1/2},
\end{eqnarray}
\end{subequations}
where $R$ is the Thomas-Fermi radius.
 This value fits very well with the numerical computations. We can check that the energy is thus of order $N^{3/2} \sqrt{c_0}$.

{\it{Equations for $\Theta$ and $\bm S$}}. The minimization of  (\ref{sth_eq}) leads to an equation for
 $\Theta$ written as
\begin{equation}\label{divtheta}
	\nabla\cdot\left(\rho\left(\nabla\Theta+2\kappa \bm{S_{\perp}}\right)\right)=0.
\end{equation} This is reminiscent of the  continuity equation written in \cite{XH}. For small
 $\kappa$, $\bm{S_{\perp}}$ can be written as a gradient so that
\begin{equation}
	\label{cond1}
	\nabla\Theta+2\kappa \bm{S_{\perp}}=0.
\end{equation}
{{In Fig. \ref{summ} with $\delta<1$ we have either $S_x=1$, $S_y=S_z=0$ or $|S_x|=|S_y|=1/\sqrt{2}$, $S_z=0$ (see Fig. \ref{phase_00}). In both cases, numerical computation of $\nabla\Theta$ gives that $\nabla\Theta=-2\kappa \bm{S_{\perp}}$ is satisfied everywhere.}}
Therefore, when Eq. (\ref{cond1}) is satisfied, the minimization of $E_{\bm{S},\Theta}$, given by (\ref{sth_eq}), becomes
\begin{equation}
	\begin{split}
E_{\bm{S},\Theta}=&\int\frac{\rho}{2}\Big(\frac{1}{4}\left(\nabla\bm{S}\right)^2+\kappa\bm{S}\cdot\nabla\times\bm{S}\\
&\qquad+\kappa^2(S_z^2-1)
 +c_2\rho S_z^2\Big)\quad d^2{r},
\end{split}
\end{equation}
since $\bm{S_{\perp}}^2=1-S_z^2$. The ground state of this energy for small $\kappa$ should {{prove}} to be
 close to $S_z=0$ and $\bm{S_{\perp}}$ constant.

\subsubsection{2. $\delta>1$}If $\delta>1$, that is $c_2<0$, the minimization of (\ref{ener2_2}) leads to $S_z^2\sim 1$, and at leading order, the density
 minimizes \begin{equation}
	\label{rho_eq2}
E_{\rho}=\int\frac{1}{2}\left(\nabla\sqrt{\rho}\right)^2+\frac{\rho}{2}r^2+(c_0+c_2)\frac{\rho^2}{2}\quad d^2{r}.
\end{equation} Note that because $g_1=g_2$, then $c_0+c_2=g_1$.
 In the Thomas-Fermi approximation, the minimization of  (\ref{rho_eq2}) yields
\begin{equation}
	\label{tf:omega02}
	\rho=\frac{1}{c_0+c_2}\left(\mu-\frac{1}{2}r^2\right),
\end{equation}
with
\begin{subequations}
\begin{eqnarray}
	\label{radius_02}
	R&=&\left(\frac{4N(c_0+c_2)}{\pi}\right)^{1/4},\\
	\mu&=&\left(\frac{N(c_0+c_2)}{\pi}\right)^{1/2},
\end{eqnarray}
\end{subequations}
where $R$ is the Thomas-Fermi radius.
 Note that at $\delta=1$, then $c_2=0$, so that both profiles (\ref{tf:omega0}) and (\ref{tf:omega02}) in $\rho$ are the same. The value (\ref{radius_02}) fits well with the numerical computations.

 In order to understand the skyrmion structure, we go back to the energy in $\psi_1$, $\psi_2$ and assume
  that $\psi_1=f(r) e^{in\theta}$, $\psi_2=g(r)$.
   Then, the spin orbit energy is equal to
   \begin{equation}E_{so}= -\kappa \int \left ( f g' e^{i\theta (n-1)}-f' g e^{-i\theta (n-1)}+\frac n r fg  e^{i\theta (n-1)}\right )\ d^2{r}.\end{equation}
   It follows that if $n\neq 1$, then this term is zero and having a skyrmion of order bigger than
   1 increases the energy. Therefore, the giant
    skyrmion is necessarily of degree 1, leading to a circulation of $2\pi$. This is similar to what \cite{HRPL,ram} find in the lowest Landau level with
    small interaction. Note that in the case of several annuli (Fig. \ref{prof2}. a.I), our analysis also yields that each annulus encompasses a degree 1, because the computation is valid per annulus. To check this numerically, we compute
    \begin{equation}\label{Cr} C(r)=\frac {i}{2|\psi_1(r)|^2}\int_{r=R} (\psi_1^*\nabla \psi_1 -\psi_1\nabla \psi_1^*)\ d^2{r}\end{equation} which is equal to $2\pi$ if the giant vortex is of degree 1.
    In Fig.'s \ref{circ1} and \ref{circ2} we plot $C(r)$ for $(\delta,\Omega)=(5,0)$ and for  $\kappa=1.5$ and $\kappa=3$. We check
    numerically in the case of a single annulus or multiple annuli that indeed the circulation is $2\pi$ per annulus of component 1.

\begin{figure}
\begin{center}
\includegraphics[scale=0.25]{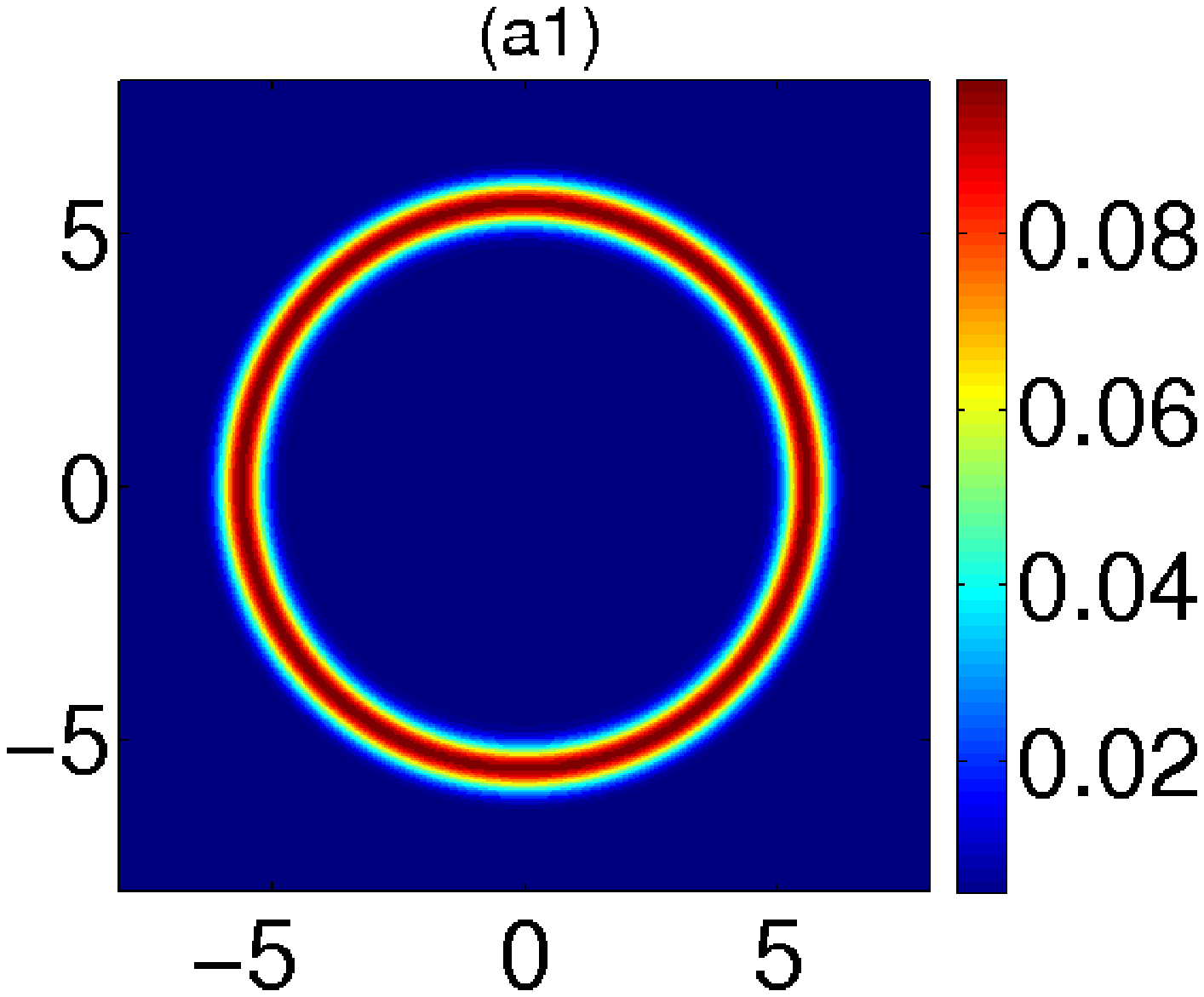}
\includegraphics[scale=0.25]{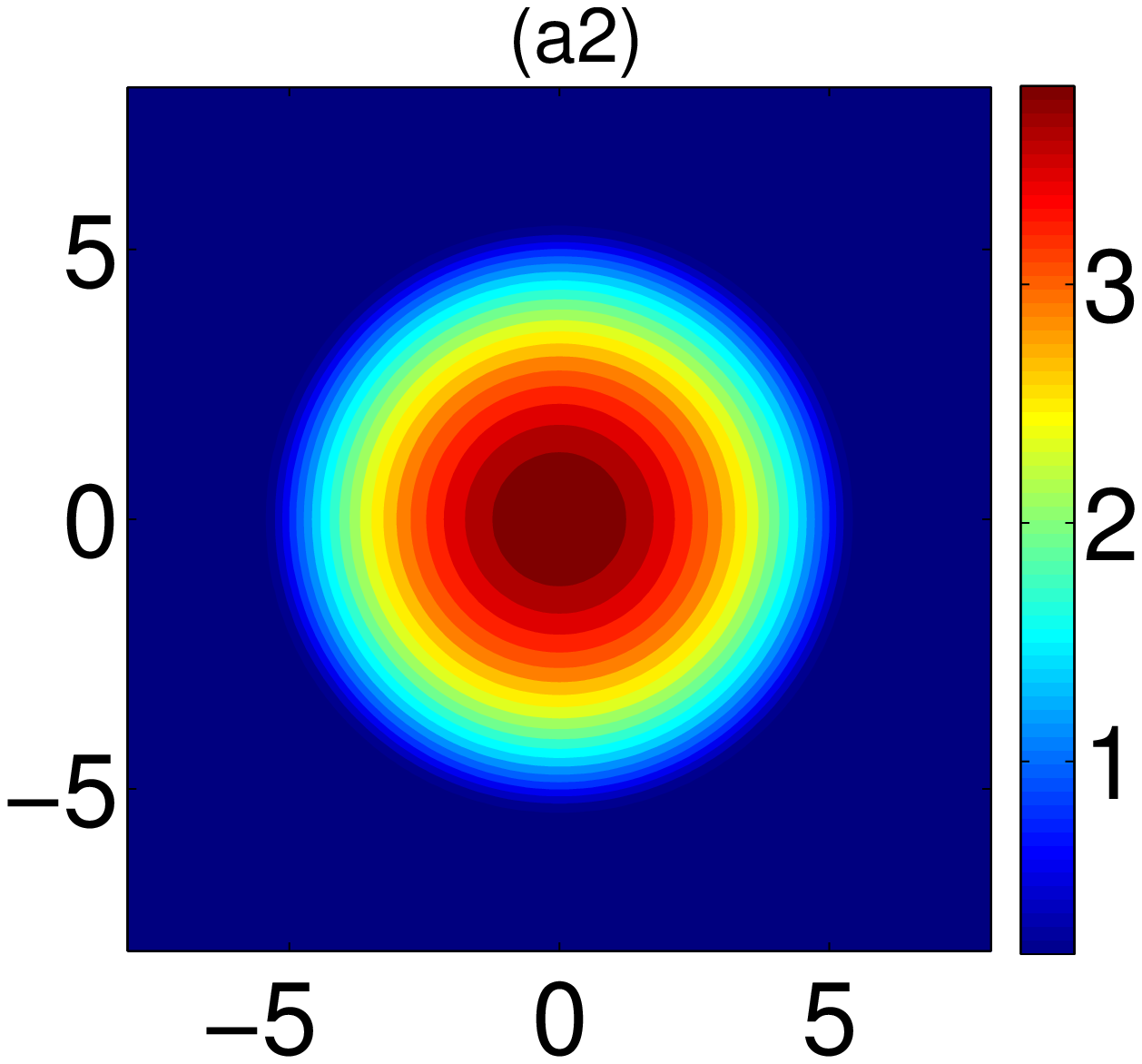}\\
\includegraphics[scale=0.3]{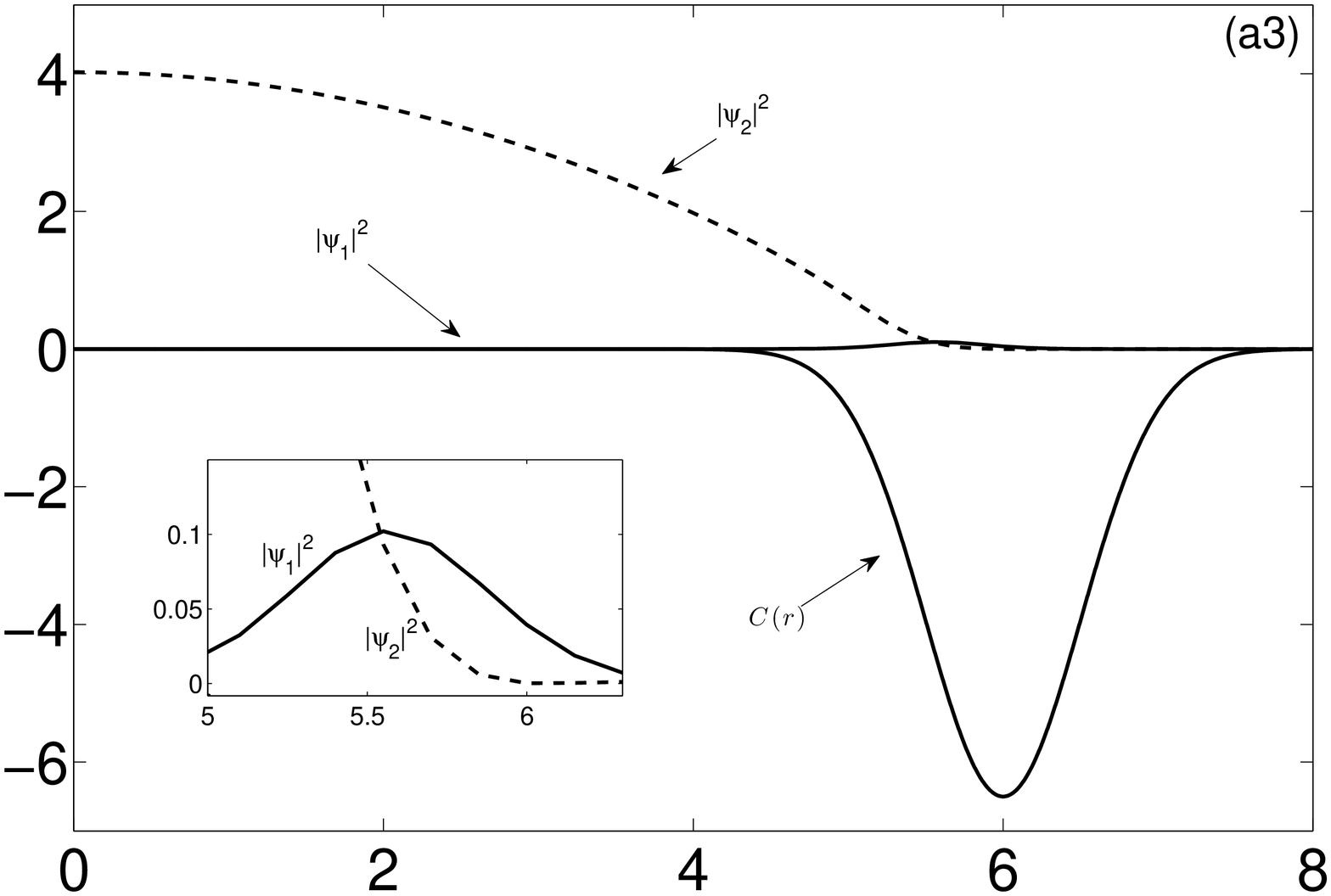}
\end{center}
\begin{picture}(0,0)(10,10)
\put(20,200)  {{$r$}}
\put(20,28)  {{$r$}}
\end{picture}
\caption{Density plots for component-1 (a1) and component-2 (a2) together with a plot of $C(r)$ for component-1. In (a3) we include the density curves for both wave functions. The inset shows a zoom in on the density profiles around the Thomas-Fermi radius. The parameters are $\delta=5$, $\Omega=0$ and $\kappa=1.5$.}
\label{circ1}
\end{figure}

\begin{figure}
\begin{center}
\includegraphics[scale=0.25]{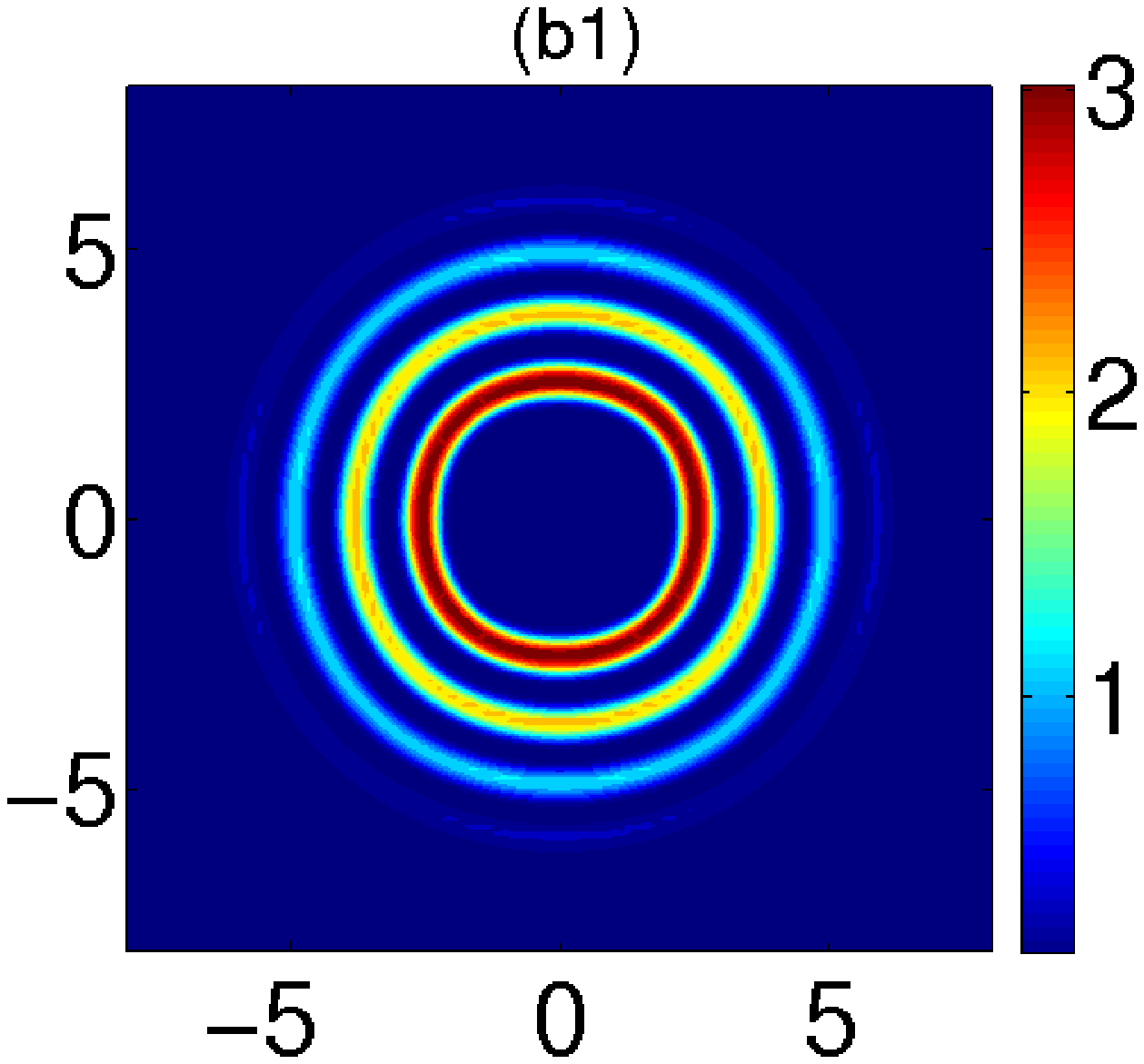}
\includegraphics[scale=0.25]{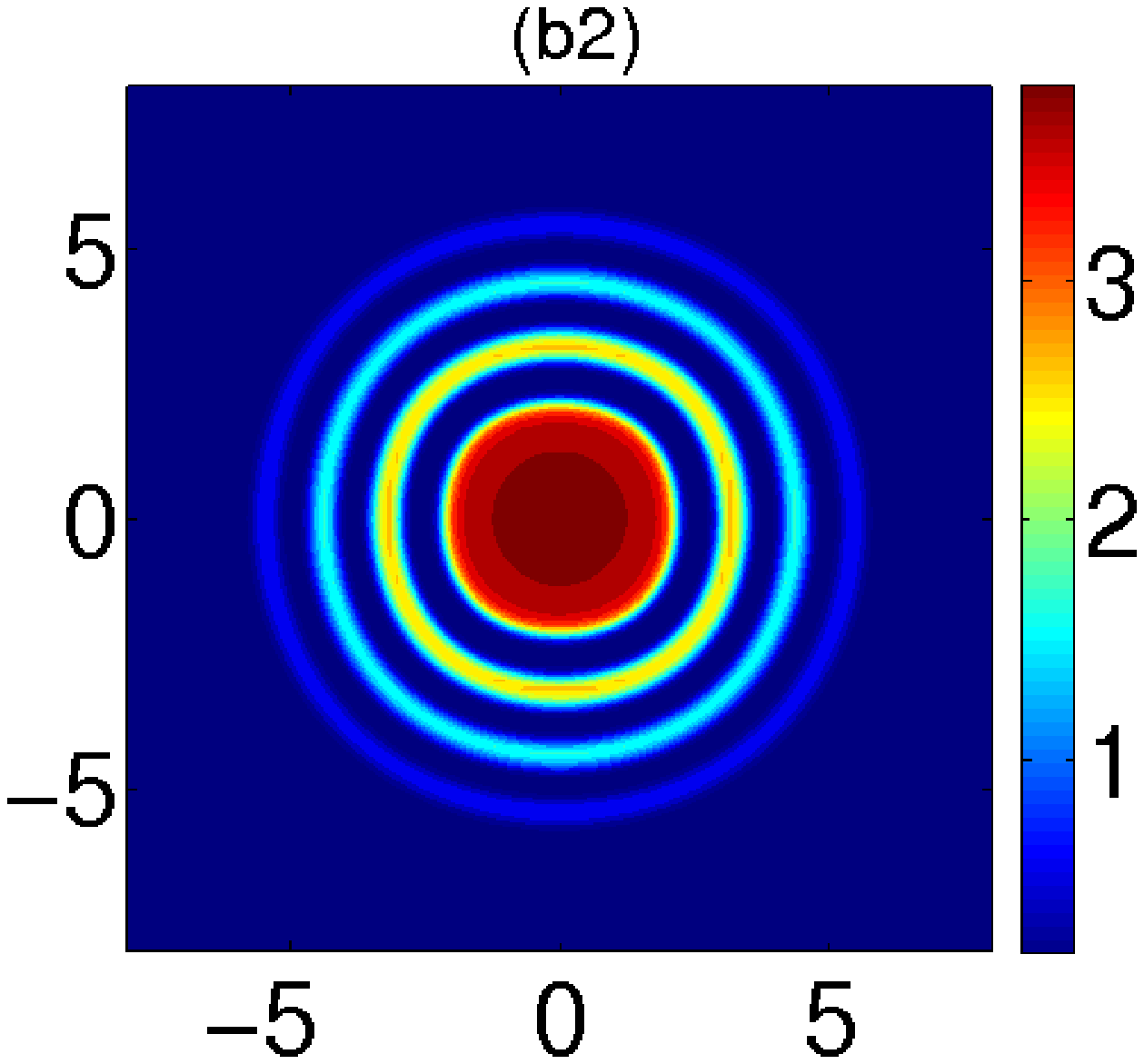}\\
\includegraphics[scale=0.3]{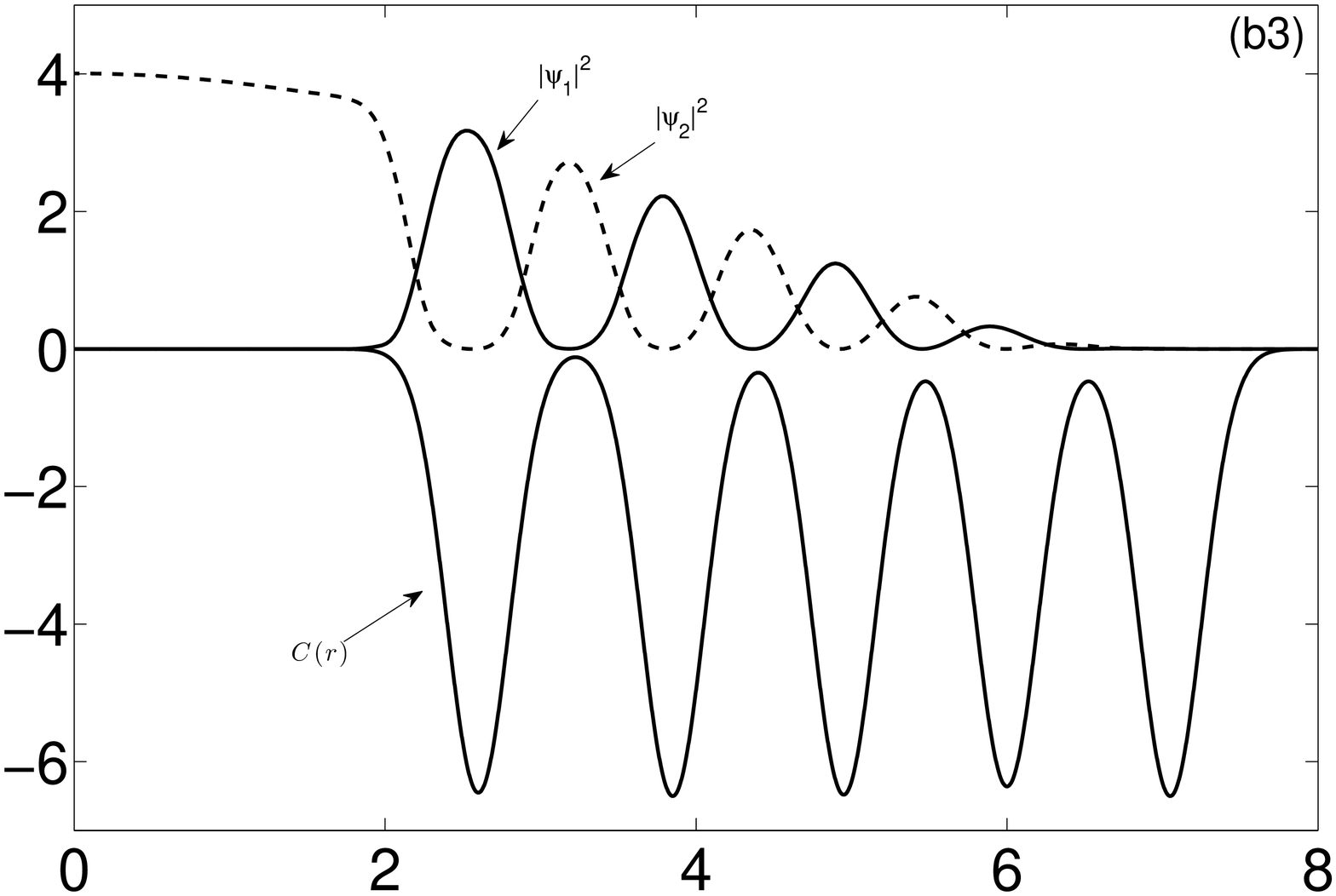}
\end{center}
\begin{picture}(0,0)(10,10)
\put(20,200)  {{$r$}}
\put(20,28)  {{$r$}}
\end{picture}
\caption{Density plots for component-1 (b1) and component-2 (b2) together with a plot of $C(r)$ for component-1. In (b3) we include the density curves for both wave functions. The parameters are $\delta=5$, $\Omega=0$ and $\kappa=3$.}
\label{circ2}
\end{figure}

    If $\delta $ is much bigger than 1, then
  $S_z^2\sim 1$, the components are segregated and only a boundary layer exists at the interface.
   Therefore, we can assume further that there exists a radius $R_0$ such that
    $\psi_1=\sqrt \rho e^{i\theta}{\bf 1}_{R_0\leq r\leq R}$ and $\psi_2=\sqrt \rho {\bf 1}_{ r\leq R_0}$.
    Then only the derivatives $f'$ and $g'$ produce a contribution which is a delta function at $r=R_0$ and
    then the spin orbit energy becomes \begin{equation}E_{so}\sim  -4\pi \kappa  \rho(R_0) .\end{equation}
    This dependence is again consistent with the numerical computations.

\subsection{B. Non-zero Rotation}

\subsubsection{1. Low Rotation}

In the case $\delta <1$, when the rotation is included into the problem, then the energy (\ref{ener2_2}) also leads to decoupled problems, one in $\rho$, and one for  $\bm{S}$ and $\Theta$:
\begin{subequations}
 \begin{equation}
 	\label{rho_eq_2}
 E_{\rho}=\int\frac{1}{2}\left(\nabla\sqrt{\rho}\right)^2+\frac{\rho}{2}r^2+c_0\frac{\rho^2}{2}\quad d^2{r},
 \end{equation}
 \begin{eqnarray} E_{\bm{S},\Theta}&=&\int\frac{\rho}{2}\Bigg(\frac{1}{4}\left(\nabla\bm{S}\right)^2+
 \kappa\bm{S}\cdot\nabla\times\bm{S}
 +\kappa^2 S_z^2+c_2 \rho S_z^2 \nonumber\\&&\qquad+\frac{1}{4}\left(\nabla\Theta+2\kappa \bm{S_{\perp}} \right)^2- \nabla\Theta \cdot \bm \Omega\times \bm r\Bigg)\quad d^2{r}.
\label{sth_eq_2}
 \end{eqnarray}
 \end{subequations} The Thomas-Fermi expression for $\rho$ does not change with respect to the $\Omega=0$ expression [Eq. (\ref{tf:omega0})]. We still have that (\ref{cond1})
  holds except on singularity lines. New singularities related to the rotation emerge, on the
   boundary of the domain regions of $\bm{S_{\perp}}$ as illustrated in Fig. \ref{prof_b_0.1}.

   In the case $\delta >1$, the coupling between spin orbit and rotation leads to a giant skyrmion for low $\kappa$,
    and then discontinuities in the outside annulus as illustrated in Fig. \ref{trans}.III. The Thomas Fermi approach may
    yield information on this behaviour. Numerically, though ${\bm S}_\perp$ is almost zero in the Thomas-Fermi radius
     it has a circulation which produces a circulation in $\Theta$.




\subsubsection{2. High Rotation}
When the rotation is increased, the energy becomes (see Appendix)
\begin{equation}
		\label{ener2high}
\begin{split}
 E=&\int\frac{1}{2}\left(\nabla\sqrt{\rho}\right)^2+\frac{\rho}{8}\left(\nabla\bm{S}\right)^2+\frac{\rho}{2}
 \left(\bm{v}_{\text{eff}}+\kappa \bm{S_{\perp}}-\bm{\Omega}\times\bm{r}\right)^2\\
		&+\rho\Omega\kappa(-yS_x+xS_y)+\frac{1}{2}\rho\kappa \bm{S}\cdot\nabla\times\bm{S}\\
		&+\frac 12 \rho \kappa^2 S_z^2+\frac{\rho}{2}(1-\Omega^2)r^2+(c_0+c_2S_z^2)\frac{\rho^2}{2}\qquad d^2{r}.
\end{split}
\end{equation} In the case $\delta <1$, in the Thomas-Fermi regime, this decouples into a problem for $\rho$, (for which the kinetic energy is negligible
 and $c_2 S_z^2$ can be neglected in front of $c_0$) and a problem for $\bm{v}_{\text{eff}}$ and $\bm S$.
This allows us to rewrite Eq. (\ref{ener2high}) as follows
\begin{equation}
		\label{ener_high}
	E_{\Omega}=\int\rho\Omega\kappa(-yS_x+xS_y)+\frac{\rho}{2}(1-\Omega^2)r^2+c_0\frac{\rho^2}{2}\quad d^2{r},
\end{equation}
\begin{equation}
		\label{ener2highveff}
\begin{split}
 E_{\bm{v}_{\text{eff}}, \bm S}=&\int\frac{\rho}{8}\left(\nabla\bm{S}\right)^2+\frac{\rho}{2}
 \left(\bm{v}_{\text{eff}}+\kappa \bm{S_{\perp}}-\bm{\Omega}\times\bm{r}\right)^2\\
		&+\frac{1}{2}\rho\kappa \bm{S}\cdot\nabla\times\bm{S}
		+\frac 12 \rho \kappa^2 S_z^2+\frac {c_2}2 \rho^2 S_z^2 \qquad d^2{r}.
\end{split}
\end{equation}
The minimization of (\ref{ener_high}) leads to two Euler-Lagrange equations:
\begin{subequations}
\begin{eqnarray}
	\label{elrho}
	\frac{1}{2}(1-\Omega^2)r^2+\kappa\Omega(-yS_x+xS_y)+c_0\rho&=&{\mu},\\
	\kappa\Omega\rho(y\mp x(1-S_x^2)^{-1/2}S_x)&=&0.
	\label{elsx}
\end{eqnarray}
\end{subequations}
We note that, at leading order, $S_x^2+S_y^2\approx1$, so that Eq. (\ref{elsx}) gives
\begin{equation}
	\label{S}
	\bm{S}\sim (\sin\theta,-\cos\theta,0).
	\end{equation}
	{{An analysis to the next order in $\bm{S}$, with the minimisation of (\ref{ener2highveff}), will lead to the vortex lattice.}}

Substituting (\ref{S}) into Eq. (\ref{elrho}) gives
\begin{equation}
	\label{rho}
	\rho=\frac{1}{c_0}\left({\mu}-\frac{1}{2}(1-\Omega^2)r^2+\kappa\Omega r\right).
\end{equation}
We can use the normalisation condition to find ${\mu}$. However first we must consider the two possible geometries: by our assumption that $S_z\approx0$ away from the defects, we expect the two components to share the same geometry. The numerical simulations presented in Fig. \ref{prof_d_0.9} (and also present in \cite{xuhan}) indicate that both components are either disks or annuli.


{\it Two disks}. When the components are both disks (we show later that this corresponds to ${\mu}>0$), then we can use the normalisation condition, integrating from $r=0$ to the outer boundary, $r=R$, to find that
$R$ is given as the solution of the quartic
\begin{equation}
	\label{disk}
	R^4-\frac{4\kappa\Omega}{3(1-\Omega^2)}R^3-\frac{4Nc_0}{\pi(1-\Omega^2)}=0,
\end{equation}
which then gives the chemical potential as
\begin{equation}
	{\mu}=\frac{1}{2}(1-\Omega^2)R^2-\kappa\Omega R.
\end{equation}


{\it Two annuli}. When the components are both annuli, we can analyse an effective potential, defined from Eq. (\ref{rho}) as $V_e(r)=\frac{1}{2}(1-\Omega^2)r^2-\kappa\Omega r$. The total density is zero when this effective potential is equal to the chemical potential: i.e. $V_e(r)={\mu}$. Thus
\begin{eqnarray}
	{\mu}&=&\frac{1}{2}(1-\Omega^2)r^2-\kappa\Omega r\nonumber\\
	\Rightarrow r&=&\frac{\kappa\Omega\pm\sqrt{\kappa^2\Omega^2+2(1-\Omega^2){\mu}}}{(1-\Omega^2)},
\end{eqnarray}
from which we identify
\begin{subequations}
\begin{eqnarray}
	R_1&=&\frac{\kappa\Omega-\sqrt{\kappa^2\Omega^2+2(1-\Omega^2){\mu}}}{(1-\Omega^2)},\\
	R_2&=&\frac{\kappa\Omega+\sqrt{\kappa^2\Omega^2+2(1-\Omega^2){\mu}}}{(1-\Omega^2)},
\end{eqnarray}
\end{subequations}
with the inner radius $R_1$ of the annulus and the outer radius $R_2$. For $R_1$ to exist we must have ${\mu}<0$, i.e. ${\mu}>0$ implies that both components are disks and ${\mu}<0$ implies that both components are annuli. We thus here assume that ${\mu}<0$. Note that
\begin{subequations}
\begin{eqnarray}
	R_1+R_2&=&\frac{2\kappa\Omega}{(1-\Omega^2)},\\
	R_2-R_1&=&\frac{2\sqrt{\kappa^2\Omega^2+2(1-\Omega^2){\mu}}}{(1-\Omega^2)}.\label{R2R1}
\end{eqnarray}
\end{subequations}

We can find ${\mu}$ from the normalisation condition as
\begin{equation}
 {\mu}=\frac{1}{2(1-\Omega^2)}\left(-\kappa^2\Omega^2+\left(\frac{3Nc_0(1-\Omega^2)^3}{4\pi\kappa\Omega}\right)^{2/3}\right),
\end{equation}
which gives $R_1$ and $R_2$ explicitly as
\begin{subequations}
\begin{eqnarray}
	\label{annulus}
	 R_1&=&\frac{\kappa\Omega}{(1-\Omega^2)}-\left(\frac{(\kappa_c\Omega_c)^4}{\kappa\Omega(1-\Omega^2_c)^3}\right)^{1/3},\\
	 R_2&=&\frac{\kappa\Omega}{(1-\Omega^2)}+\left(\frac{(\kappa_c\Omega_c)^4}{\kappa\Omega(1-\Omega^2_c)^3}\right)^{1/3},
\end{eqnarray}
\end{subequations}
which leaves the width of the annulus, $d=R_2-R_1$, as
\begin{subequations}
\begin{equation}
	d=\frac{\alpha}{(\kappa\Omega)^{1/3}},
	\end{equation}
where
\begin{equation}
	\alpha=2\frac{(\kappa_c\Omega_c)^{4/3}}{(1-\Omega^2_c)}
\end{equation}
\end{subequations}
is a constant, with the $\kappa_c$ and $\Omega_c$ critical values of $\kappa$ and $\Omega$ (and $g_{12}$) that can be found from setting ${\mu}=0$ and using the normalisation condition. This gives
\begin{equation}
	\label{crit_val}
	\frac{(\kappa_c\Omega_c)^4}{(1-\Omega_c^2)^3}=\frac{3Nc_0}{4\pi}.
\end{equation} In particular, for a sufficiently large $\kappa$, $d$ gets small, and a thin annulus with a large circulation can be created (as has been seen in Fig. 2 of Ref. \cite{xuhan}).

{\it Comparison to Phase Diagrams}. A phase diagram identifying the numerically determined ground states for the case of large rotation is given in Fig. \ref{pd_omega0.9}. We see the appearance of three regimes, of which the above analysis pertains to the regimes in which there exists two disks with vortex lattices and two annuli with vortex lattices. The boundary between these two regimes has been calculated analytically to be given by Eq. (\ref{crit_val}), which we include on the phase diagram of Fig. \ref{pd_omega0.9} (dashed line), together with the numerically determined boundary (solid line). One can see a good agreement between the theory and numerics.

The width of the annulus given by (\ref{R2R1}) becomes thinner as the product $\kappa\Omega$ increases. This scenario is reminiscent of the rotating single component condensate that is trapped by a harmonic plus quartic trapping potential \cite{bfs} in which an annulus develops as the rotation is increased, and in which the width of the annulus becomes smaller. In the case of a condensate held by a harmonic plus quartic trapping potential, there is no upper limit to the rotation since the quartic term acts to keep the condensate bounded for all $\Omega$. This increasing rotation leads to the development of a giant vortex (a large circulation) inside the annulus. At the same time the width of the annulus is decreasing such that the condensate can no longer support any vortices in the condensate bulk. A similar situation can occur if one considers a harmonic plus Gaussian trapping potential (a toroidal trap), although in this case the width of the annulus is dependent on additional factors, notably the strength and `waist' of the Gaussian term (which is generally taken to be centred at the origin) and an upper limit on the rotation which must be enforced to ensure the condensate stays bounded \cite{am1}.

\section{V. Conclusion}

We have provided phase diagrams in terms of the magnitude of rotation, the strength of the spin-orbit coupling and interactions.
 We have found that plotting the total phase and the components of the spin leads to an interesting classification of the ground states. In the case of coexisting condensates, the Thomas-Fermi approximation for the total density leads to a simplification
  of the energy. We are able to determine the boundary between regions of disks and annuli leading to vortex lattices at high
   rotation, and to derive a ferromagnetic energy. In the case of segregation, we analyze the giant skyrmion in the Thomas Fermi limit
    and find that it is of degree 1.

\section*{Acknowledgments}

The authors wish to thank J. Dalibard, R. Ignat  and P. \"Ohberg for useful discussions that took place during this work.
 They are very grateful to the anonymous referee for his remarks that lead to improvements of the paper.

\section*{Appendix: Derivation of the Energy Functional (\ref{ener2})}
	
\subsection*{1. First formulation}	
	
We show in this Appendix how to derive the energy functional
\begin{equation}
		\label{ener2old}
\begin{split}
 E=&\int\frac{1}{2}\left(\nabla\sqrt{\rho}\right)^2+\frac{\rho}{8}\left(\nabla\bm{S}\right)^2+\frac{\rho}{2}\left(\bm{v}_{\text{eff}}-\bm{\Omega}\times\bm{r}\right)^2\\
		&+\rho\kappa\left(\bm{S_{\perp}}\cdot \bm{v}_{\text{eff}}+\frac{1}2 \bm{S}\cdot \nabla \times \bm{S} \right)\\
		&+\frac{\rho}{2}(1-\Omega^2)r^2+(c_0+c_1S_z+c_2S_z^2)\frac{\rho^2}{2}\qquad d^2{r},
\end{split}
\end{equation}
in terms of the non-linear Sigma model. We start with the energy functional given in terms of the wave functions $\psi_k$ [Eq. \ref{er}], rewritten here
\begin{equation}
	\label{er_app}
	\begin{split}
	E&=\int\sum_{k=1,2}\Bigg(\frac{1}{2}|\nabla\psi_k|^2+\frac{1}{2}r^2|\psi_k|^2
-{\Omega}\psi^*_k L_z \psi_k\\
&\qquad+\frac{g_k}{2}|\psi_k|^4-\kappa\psi_k^*\left[i\frac{\partial\psi_{3-k}}{ \partial x}+(-1)^{3-k}\frac{\partial\psi_{3-k}}{\partial y}\right]\Bigg)\\
&\qquad+g_{12}|\psi_1|^2|\psi_2|^2\qquad d^2{r}.
\end{split}
\end{equation}
In \cite{am2}, we showed how to transform the energy functional of the rotating two-component condensate (i.e the above energy functional with $\kappa=0$) into one given in terms of the total density $\rho$, the total phase $\Theta$ and the spin density $\bm{S}$. So we split the above energy functional into terms independent of the spin-coupling and terms dependent on the spin-coupling, $E=E_i+E_{so}$, where \cite{am2}
\begin{equation}
	\begin{split}
			E_i&=\int\sum_{k=1,2}\left(\frac{1}{2}|\nabla\psi_k|^2+\frac{1}{2}r^2|\psi_k|^2
-{\Omega}\psi^*_k L_z \psi_k+\frac{g_k}{2}|\psi_k|^4\right)\\
&\qquad+g_{12}|\psi_1|^2|\psi_2|^2\qquad d^2{r}	\\
&=\int\frac{1}{2}\left(\nabla\sqrt{\rho}\right)^2+\frac{\rho}{8}\left(\nabla\bm{S}\right)^2+\frac{\rho}{2}\left(\bm{v}_{\text{eff}}-\bm{\Omega}\times\bm{r}\right)^2\\
		&\qquad+\frac{\rho}{2}(1-\Omega^2)r^2+(c_0+c_1S_z+c_2S_z^2)\frac{\rho^2}{2}\qquad d^2{r},
	\end{split}
\end{equation}
and
\begin{equation}
	\label{eqso}
	E_{so}=-\int\kappa\sum_{k=1,2}\psi_k^*\left(i\frac{\partial\psi_{3-k}}{ \partial x}+(-1)^{3-k}\frac{\partial\psi_{3-k}}{\partial y}\right)\quad d^2{r}.
\end{equation}
We work with Eq. (\ref{eqso}), and use the non-linear Sigma model where we have $\psi_k=\sqrt{\rho}\chi_k$, where $\rho=|\psi_1|^2+|\psi_2|^2$ and the $\chi_k$ are related to $\bm{S}$ by Eq.'s (\ref{seqs}). Note that $|\bm{S}|^2 = 1$ everywhere, so that one of the components of $\bm{S}$ is given in terms of the other two.

Upon substitution of $\psi_k=\sqrt{\rho}\chi_k$ in to Eq. (\ref{eqso}), we find that
\begin{equation}
	\label{eso1}
	\begin{split}
		E_{so}=&-\int\frac{\kappa}{2}(1-S_z^2)^{1/2}\sum_{k=1,2}e^{i(\Theta_{3-k}-\Theta_k)}\times\\
		&\Bigg[i\rho\left(i\frac{\partial\Theta_{3-k}}{\partial x}+(-1)^{3-k}\frac{\partial\Theta_{3-k}}{\partial y}\right)\\
		&\quad+\frac{(-1)^{k}\rho}{2(1+(-1)^{k}S_z)^{2}}\left(i\frac{\partial S_z}{\partial x}+(-1)^{3-k}\frac{\partial S_z}{\partial y}\right)\\
		&\quad+\frac{1}{2}\left(i\frac{\partial \rho}{\partial x}+(-1)^{3-k}\frac{\partial \rho}{\partial y}\right)\Bigg]\quad d^2{r},
\end{split}
\end{equation}
where we have written the spinor $\chi_k$ in terms of its amplitude and phase: $\chi_k=|\chi_k|\exp(i\Theta_k)$. Furthermore, we use the identities $S_z=|\chi_1|^2-|\chi_2|^2$ and $|\chi_1|^2+|\chi_2|^2=1$ to give $|\chi_k|^2=(1+(-1)^{3-k}S_z)/2$.

Next we note that $S_x=2|\chi_1||\chi_2|\cos(\Theta_1-\Theta_2)=(1-S_z^2)^{1/2}\cos(\Theta_1-\Theta_2)$ and $S_y=-2|\chi_1||\chi_2|\sin(\Theta_1-\Theta_2)=-(1-S_z^2)^{1/2}\sin(\Theta_1-\Theta_2)$, which allows us to write
\begin{equation}
	\label{com1}
	\begin{split}
	&-i(1-S_z^2)^{1/2}\sum_{k=1,2}e^{i(\Theta_{3-k}-\Theta_k)}\times\\
	&\qquad\left(i\frac{\partial\Theta_{3-k}}{\partial x}+(-1)^{3-k}\frac{\partial\Theta_{3-k}}{\partial y}\right)\\
	&=S_x\left(\frac{\partial}{\partial x}(\Theta_1+\Theta_2)+\frac{\partial}{\partial y}(\Theta_1-\Theta_2)\right)\\
	&\quad+S_y\left(\frac{\partial}{\partial y}(\Theta_1+\Theta_2)-\frac{\partial}{\partial x}(\Theta_1-\Theta_2)\right)\\
	&=\bm{S_{\perp}}\cdot\nabla\Theta-i\left(\frac{\partial S_x}{\partial x}+\frac{S_xS_z}{(1-S_z^2)}\frac{\partial S_z}{\partial x}\right)\\
	&\quad-\left(\frac{\partial S_x}{\partial x}+\frac{S_xS_z}{(1-S_z^2)}\frac{\partial S_z}{\partial x}\right),
\end{split}
\end{equation}
where the last line follows from
\begin{subequations}
	\begin{eqnarray}
		S_x\frac{\partial}{\partial y}(\Theta_1-\Theta_2)&=&-\frac{\partial S_y}{\partial y}-\frac{S_yS_z}{(1-S_z^2)}\frac{\partial S_z}{\partial y},\\
		S_y\frac{\partial}{\partial x}(\Theta_1-\Theta_2)&=&\frac{\partial S_x}{\partial x}+\frac{S_xS_z}{(1-S_z^2)}\frac{\partial S_z}{\partial x}.
	\end{eqnarray}
\end{subequations}

We move on to the second term of Eq. (\ref{eso1}), which, by noting that
\begin{subequations}
	\begin{align}
		\begin{split}
		&\sum_{k=1,2}\left(\frac{1+(-1)^{3-k}S_z}{1-(-1)^{3-k}S_z}\right)^{1/2}e^{i(\Theta_{3-k}-\Theta_k)}\\
		&\qquad=\frac{2}{(1-S_z^2)}(iS_y+S_xS_z),
	\end{split}\\
	\begin{split}
		 &\sum_{k=1,2}(-1)^{3-k}\left(\frac{1+(-1)^{3-k}S_z}{1-(-1)^{3-k}S_z}\right)^{1/2}e^{i(\Theta_{3-k}-\Theta_k)}\\
		&\qquad=\frac{2}{(1-S_z^2)}(iS_yS_z+S_x),
\end{split}
\end{align}
\end{subequations}
is equal to
\begin{equation}
	\label{com2}
	\begin{split}
		&-(1-S_z^2)^{1/2}\sum_{k=1,2}e^{i(\Theta_{3-k}-\Theta_k)}\times\\
		&\qquad\frac{(-1)^{k}\rho}{2(1+(-1)^{k}S_z)^{2}}\left(i\frac{\partial S_z}{\partial x}+(-1)^{3-k}\frac{\partial S_z}{\partial y}\right)\\
		&=\frac{1}{(1-S_z^2)}\left[(iS_xS_z-S_y)\frac{\partial S_z}{\partial x}+(iS_yS_z+S_x)\frac{\partial S_z}{\partial y}\right].
	\end{split}
\end{equation}
We combine Eq.'s (\ref{com1}) and (\ref{com2}) to get
\begin{equation}
	\bm{S_{\perp}}\cdot\nabla\Theta+\frac{1}{(1-S_z^2)}\left(S_x\frac{\partial S_z}{\partial y}-S_y\frac{\partial S_z}{\partial x}\right)-i\left(\frac{\partial S_x}{\partial x}+\frac{\partial S_y}{\partial y}\right).
\end{equation}
Finally, notice that
\begin{equation}
\begin{split}
	&-(1-S_z^2)^{1/2}\sum_{k=1,2}e^{i(\Theta_{3-k}-\Theta_k)}\frac{1}{2}\left(i\frac{\partial \rho}{\partial x}+(-1)^{3-k}\frac{\partial \rho}{\partial y}\right)\\
		&=-i\left(S_x\frac{\partial\rho}{\partial x}+S_y\frac{\partial\rho}{\partial y}\right).
\end{split}
\end{equation}
Thus, Eq. (\ref{eqso}) becomes
	\begin{equation}
		\begin{split}
		 E_{so}&=\int\frac{\kappa\rho}{2}\Bigg(\bm{S_{\perp}}\cdot\nabla\Theta\\
		&\quad+\frac{1}{(1-S_z^2)}\left(S_x\frac{\partial S_z}{\partial y}-S_y\frac{\partial S_z}{\partial x}\right)\Bigg)\quad d^2{r}.
	\end{split}
\end{equation}
The last step is to notice that
\begin{equation}
	\bm{S}\cdot\nabla\times\bm{S}=\frac{1}{(1-S_z^2)}\left( S_x\frac{\partial S_z}{\partial y}-S_y\frac{\partial S_z}{\partial x}-S_z\bm{S_{\perp}}\cdot\bm{R}\right),
	\end{equation}
so that	
\begin{equation}
	\label{so1}
		 E_{so}=\int\rho\kappa\left(\bm{S_{\perp}}\cdot \bm{v}_{\text{eff}}+\frac{1}2 \bm{S}\cdot \nabla \times \bm{S} \right)\quad d^2{r}
	\end{equation}
	since
\begin{equation}
	\bm{v}_{\text{eff}}=\frac{1}{2}\left(\nabla\Theta+\frac{S_z\bm{R}}{(1-S_z^2)}\right),
\end{equation}	
	and thus Eq. (\ref{er_app}) yields Eq. (\ref{ener2old}).

\subsection*{2. Alternative Forms for the Energy Functional (\ref{ener2})}	

Equation (\ref{ener2old}) can be decomposed into its constituent parts, i.e. we can write $E=E_{KE}+E_{PE}+E_I$, where
\begin{subequations}
	\begin{align}
	E_{KE}&=\int\frac{1}{2}\left(\nabla\sqrt{\rho}\right)^2+\frac{\rho}{8}\left(\nabla\bm{S}\right)^2\quad d^2{r},\\
	\begin{split}
	 E_{PE}&=\int\frac{\rho}{2}\left(\bm{v}_{\text{eff}}-\bm{\Omega}\times\bm{r}\right)^2+\rho\kappa\Big(\bm{S_{\perp}}\cdot \bm{v}_{\text{eff}}\\
	&\quad+\frac{1}2 \bm{S}\cdot \nabla \times \bm{S} \Big)+\frac{\rho}{2}(1-\Omega^2)r^2\quad d^2{r},
	\end{split}\\
	E_{I}&=\int (c_0+c_1S_z+c_2S_z^2)\frac{\rho^2}{2}\quad d^2{r}.
\end{align}
\end{subequations}
In the following we give two alternative expressions for $E_{PE}$; (i) note that
	\label{all:1}
\begin{equation}
	\begin{split}
	\left(\bm{v}_{\text{eff}}-\bm{\Omega}\times\bm{r}\right)^2&=\left(\bm{v}_{\text{eff}}+\kappa \bm{S_{\perp}}\right)^2-\left(\kappa^2(1-S_z^2)-\Omega^2r^2\right)\\
	&\quad-2\kappa \bm{S_{\perp}}\cdot \bm{v}_{\text{eff}}-2\bm{\Omega}\times\bm{r}\cdot \bm{v}_{\text{eff}},
	\end{split}
\end{equation}
so that we are able to write $E_{PE}$ as
\begin{equation}
		\label{alt1}
\begin{split}
		 E_{PE}=&\int\frac{\rho}{2}\left(\bm{v}_{\text{eff}}+\kappa\bm{S_{\perp}}\right)^2+\frac{\rho\kappa}{2}\bm{S}\cdot\nabla\times\bm{S}\\
		&-\frac{\rho}{2}\left(\nabla\Theta+\frac{S_z\bm{R}}{(1-S_z^2)}\right)\cdot\bm{\Omega\times\bm{r}}\\
&\quad+\frac{\rho}{2}(\kappa^2(S_z^2-1)+r^2)\quad d^2{r}.
\end{split}
\end{equation}
(ii) Similarly, we note that
	\label{all:2}
\begin{equation}
	\begin{split}
	 \left(\bm{v}_{\text{eff}}-\bm{\Omega}\times\bm{r}\right)^2&=\left(\bm{v}_{\text{eff}}-\bm{\Omega}\times\bm{r}+\kappa \bm{S_{\perp}}\right)^2-\kappa^2(1-S_z^2)\\
	&\quad+2\kappa \bm{S_{\perp}}\cdot\bm{\Omega}\times\bm{r}-2\kappa \bm{v}_{\text{eff}}\cdot \bm{S_{\perp}},
	\end{split}
\end{equation}
so that we are able to write $E_{PE}$ as
\begin{equation}
		\label{alt2}
\begin{split}
		 E_{PE}=&\int\frac{\rho}{2}\left(\bm{v}_{\text{eff}}-\bm{\Omega}\times\bm{r}+\kappa\bm{S_{\perp}}\right)^2+\frac{\rho\kappa}{2}\bm{S}\cdot\nabla\times\bm{S}\\
		&\quad+\rho\kappa\Omega(-yS_x+xS_y)\\
		&\qquad+\frac{\rho}{2}\left(\kappa^2(S_z^2-1)+(1-\Omega^2)r^2\right)\quad d^2{r}.
\end{split}
\end{equation}
Some of these formulations of the energy are related to some computations in \cite{HLL} or the hydrodynamic formulation in
  \cite{XH}.


\begin{thebibliography}{0}
\expandafter\ifx\csname natexlab\endcsname\relax\def\natexlab#1{#1}\fi
\expandafter\ifx\csname bibnamefont\endcsname\relax
  \def\bibnamefont#1{#1}\fi
\expandafter\ifx\csname bibfnamefont\endcsname\relax
  \def\bibfnamefont#1{#1}\fi
\expandafter\ifx\csname citenamefont\endcsname\relax
  \def\citenamefont#1{#1}\fi
\expandafter\ifx\csname url\endcsname\relax
  \def\url#1{\texttt{#1}}\fi
\expandafter\ifx\csname urlprefix\endcsname\relax\def\urlprefix{URL }\fi
\providecommand{\bibinfo}[2]{#2}
\providecommand{\eprint}[2][]{\url{#2}}

\end{thebibliography}


\begin{thebibliography}{99}
	
\bibitem{jacob}A. Jacob, P. \"Ohberg, G. Juzeli\={u}nas, and L. Santos, New Journal of Physics {\bf10}, 045022 (2008).
\bibitem{juz}G. Juzeli\={u}nas, J. Ruseckas, and J. Dalibard, Phys. Rev. A \textbf{81}, 053403 (2010).
\bibitem{campbell}D. I. Campbell, G. Juzeli\={u}nas, and I. B. Spielman, Phys. Rev. A \textbf{84}, 025602 (2011).
\bibitem{DGJO} J. Dalibard, F. Gerbier, G. Juzeli\={u}nas, and P. \"{O}hberg, Rev. Mod. Phys. {\bf83}, 1523 (2011).
\bibitem{rashba1} Y.-J. Lin, R. L. Compton, K. Jim\'{e}nez-Garcia, J. V. Porto, and I. B. Spielman, Nature (London) \textbf{462}, 628 (2009).	
\bibitem{rashba2} Y.-J. Lin, K. Jim\'{e}nez-Garcia, and I. B. Spielman, Nature (London) \textbf{471}, 83 (2011).
\bibitem{OB}T. Ozawa and G. Baym, Phys. Rev. A \textbf{85}, 013612 (2012).
\bibitem{zhai} C. Wang, C. Gao, C.-M. Jian, and H. Zhai, Phys. Rev. Lett. \textbf{105}, 160403
(2010).
\bibitem{WuMZ} C. Wu, I. Mondragon-Shem, and X.-F. Zhou, Chin. Phys. Lett. \textbf{28}, 097102 (2011).
\bibitem{galitski} T. D. Stanescu, B. Anderson, and V. Galitski, Phys. Rev. A \textbf{78}, 023616 (2008).
\bibitem{HZ}T.-L. Ho and S. Zhang, Phys. Rev. Lett. {\bf107}, 150403 (2011).
\bibitem{HLL}P.-S. He, R. Liao, and W.-M. Liu, Phys. Rev. A \textbf{86}, 043632 (2012).
\bibitem{yip}S.-K. Yip, Phys. Rev. A \textbf{83}, 043616 (2011).
\bibitem{ZMZ}Y. Zhang, L. Mao, and C. Zhang, Phys. Rev. Lett. {\bf108}, 035302 (2012).
\bibitem{HRPL}H. Hu, B. Ramachandhran, H. Pu, and X.-J Liu, Phys. Rev. Lett. {\bf108}, 010402 (2012).
\bibitem{OB2}T. Ozawa and G. Baym, Phys. Rev. A \textbf{85}, 063623 (2012).
\bibitem{KMM} T. Kawakami, T. Mizushima, and K. Machida, Phys. Rev. A \textbf{84}, 011607 (2011).
\bibitem{XKYU}Z. F. Xu, Y. Kawaguchi, L. You, and M. Ueda, Phys. Rev. A \textbf{86}, 033628 (2012).
\bibitem{spin-2} Z. F. Xu, R. Lu, and L. You, Phys. Rev. A \textbf{83}, 053602 (2011).
\bibitem{CZ}Z. Chen and H. Zhai, Phys. Rev. A \textbf{86}, 041604(R) (2012).
\bibitem{ram}B. Ramachandhran, B. Opanchuk, X.-J Liu, H. Pu, P. D. Drummond, and H. Hu, Phys. Rev. A \textbf{85}, 023606 (2012).
\bibitem{JZ} C.-M. Jian and H. Zhai, Phys. Rev. B \textbf{84}, 060508(R) (2011).
\bibitem{subh} S. Subhasis, R. Nath, and L. Santos, Phys. Rev. Lett. {\bf107}, 270401 (2011).
\bibitem{fialko}O. Fialko, J. Brand, and U. Z\"ulicke, Phys. Rev. A \textbf{85}, 051605(R) (2012).
\bibitem{xu} Y. Xu, Y. Zhang, and B. Wu, Phys. Rev. A, {\bf87}, 013614 (2013).
\bibitem{su}S.-W. Su, L.-K. Liu, Y.-C. Tsai, W. M. Liu, and S.-C. Gou, Phys. Rev. A \textbf{86}, 023601 (2012).
\bibitem{KMNM}T. Kawakami, T. Mizushima, M. Nitta, and K. Machida, Phys. Rev. Lett. {\bf109}, 015301 (2012).
\bibitem{RHM}E. Ruokokoski, J. A. M. Huhtam\"aki, and M. M\"ott\"onen, Phys. Rev. A \textbf{86}, 051607(R) (2012).
\bibitem{LL}C.-F. Liu and W. M. Liu, Phys. Rev. A \textbf{86}, 033602 (2012).
\bibitem{radic}  J. Radi\'c, T. A. Sedrakyan, I. B. Spielman, and V. Galitski, Phys. Rev. A {\bf84}, 063604 (2011).
\bibitem{LFZWL}C.-F. Liu, H. Fan, Y.-C. Zhang, D.-S. Wang, and W.-M. Liu, Phys. Rev. A \textbf{86}, 053616 (2012).
\bibitem{zzwu} X. F. Zhou, J. Zhou, and C. Wu, Phys. Rev. A, {\bf84}, 063624 (2011).
\bibitem{xuhan} X.-Q. Xu, J. H. Han, Phys. Rev. Lett. \textbf{107}, 200401 (2011).
\bibitem{cornell} V. Schweikhard, I. Coddington, P. Engels, S. Tung,
and E. A. Cornell, Phys. Rev. Lett. \textbf{93}, 210403 (2004).
\bibitem{MHo} E. J. Mueller and T.-L. Ho, Phys. Rev. Lett.
\textbf{88}, 180403 (2002).
\bibitem{ueda} K. Kasamatsu, M. Tsubota, and M. Ueda, Phys. Rev. A \textbf{71}, 043611 (2005).
\bibitem{ktu} K. Kasamatsu, M. Tsubota, and M. Ueda, Int. J. Modern Phy. B {\bf19}, 11(1835-1904) (2005).
\bibitem{am2} P. Mason and A. Aftalion, Phys. Rev. A. {\bf84}, 033611 (2011).
\bibitem{XH}X.-Q Xu and J. H. Han, Phys. Rev. Lett. {\bf108}, 185301 (2012).
\bibitem{bfs}A. L. Fetter, B. Jackson, and S. Stringari, Phys. Rev. A. {\bf71}, 013605 (2005).
\bibitem{am1} A. Aftalion and P. Mason, Phys. Rev. A. {\bf81}, 023607 (2010).

\end{thebibliography}
\end{document}